\documentclass[12pt]{article}

\usepackage[english]{babel}

\usepackage[a4paper,top=1.8cm,bottom=2cm,left=2.5cm,right=2.5cm,marginparwidth=1.75cm]{geometry}

\usepackage{amsmath}
\usepackage{amssymb}
\usepackage{graphicx}
\usepackage[colorlinks=true, allcolors=blue]{hyperref}
\usepackage[round]{natbib} 

\usepackage{subfig}
\usepackage{adjustbox}
\usepackage{tabto}
\usepackage{eucal}
\usepackage{titlesec}
\usepackage{float}
\usepackage{bm}
\usepackage{bbm}
\usepackage{multirow}

\usepackage{amsthm}
\usepackage{mathtools}
\usepackage{makecell} 

\theoremstyle{definition}

\theoremstyle{remark}

\usepackage{graphicx}

\usepackage{authblk}

\title{Flood risk estimation via geometric extremal graphical models}
\author[1]{Kristina Grolmusova}
\author[2]{Jennifer L. Wadsworth}
\author[3]{Thordis L. Thorarinsdottir}
\affil[1]{STOR-i Centre for Doctoral Training, Lancaster University, Lancaster, United Kingdom}
\affil[2]{School of Mathematical Sciences, Lancaster University, Lancaster, United Kingdom}
\affil[3]{Department of Mathematics, University of Oslo, Oslo, Norway}

\date{}

\usepackage{lipsum}                     
\usepackage{xargs}                      
\usepackage[pdftex,dvipsnames]{xcolor}  

\begin{document}
\maketitle

\begin{abstract}
    \noindent We exploit the new framework of multivariate geometric extreme value theory for the statistical analysis of river flow extremes at multiple locations on a river network. Current methodologies within the geometric framework are limited to a relatively low number of dimensions. This is insufficient for the purposes of flood risk estimation, since the number of gauging stations on a river network is often of the order $10-20+$. In order to create a parsimonious model in higher dimensions, we translate recent theoretical work on geometric extremal graphical models into statistical practice. We define the gauge function, a key object in geometric extremes, in a structured way using block graphs, which are a natural way of expressing the river network. We introduce both simple models, and more complex ones that can accommodate both simultaneous and non-simultaneous flows, and apply them to extreme flows at 10 locations on a river network around Preston, in north-west England. The models are shown to fit well and indicate strong extrapolation performance. We also introduce a correction coefficient for the geometric framework to address potential over- or under-estimation of marginal probabilities. The overall utility of our approach is illustrated through calculation of probabilities of simultaneous flooding at four locations on the network.
\end{abstract}

\section{Introduction}
The importance of flood risk estimation is indisputable. For example, according to \cite{UKHSA2023HECC}, approximately 6.1 million people in the UK currently live in flood prone areas, which is around 10\% of the total UK population. Moreover, the number of people in the UK significantly at risk of flooding is projected to increase between 61\% to 118\% by 2050.
According to findings of \cite{rentschler_flood_2022}, 1.81 billion people, around 23\% of the global population, live in locations that are directly exposed to a significant level of flood risk (1-in-100-year floods) with South and East Asia being the most affected. Approximately 89\% of the population that are exposed to flooding are in low- and middle-income countries.
Understanding the risks of floods is crucial to mitigate negative impacts that they have on individuals, society and environment. 

The challenge of estimating flood risk through statistical models has been addressed in several different ways. 
Many of the methods used by UK hydrologists are based on the Flood Estimation Handbook (FEH) \citep{FEH1999}, which is considered to be an industry standard for flood risk estimation. It comprises two main approaches -- statistical flood frequency estimation and rainfall-runoff methods. 
In the statistical approach, which has been updated since in reports by \cite{Kjeldsen2008} and \cite{VesuvianoGriffin2025}, a flood frequency curve is estimated for a site of interest from which the risk is subsequently calculated. If the site is ungauged, does not have data measurements, or if the site does not have a long record or has a lot of  missing data, data from other hydrologically similar sites are used to estimate the frequency curve. While the FEH provides a very comprehensive guide to UK flood risk estimation, the statistical model only focuses on the estimation at each site individually. However, in order to understand flood risk at multiple locations, dependence modelling is crucial. 

Approaches to dependence modelling that accommodate both simultaneous and non-simultaneous extremes across sites are typically based on conditional extreme value theory \citep{heffernan_conditional_2004}. Extensions of this framework allow for temporal dependence and missing data \citep{keef_spatial_risk_2009}, and have been used to study river flow dependence at regional and national scales, including applications across Great Britain, Austria and the UK \citep{keef_spatial_dependence_2009, schneeberger_generation_2018, keef_estimating_prob_2013}. Subsequent developments address high-dimensional residual structure and substantial missingness \citep{towe_modelbased_2019}, incorporate river-network graphical structure to reduce the effective parameter dimension and enable higher-dimensional analyses \citep{farrell_conditional_2024}, and support large-scale, model-based flood event catalogues in data-sparse regions \citep{quinn_spatial_2019, wing_toward_2020, olcese_use_2022,olcese_developing_2024}. These methods have been used to estimate joint flood risk, population exposure and related exceedance probabilities.
As shown through some of these applications, the conditional model is  relatively easy to fit in high dimensions. The disadvantage is that a conditioning site has to be chosen, and there is usually no natural choice. In that case, multiple models can be fitted for each conditioning site in turn. However, the risk at multiple sites can be calculated using a number of these models, so that we must make a non-trivial choice about which model to use, or how to combine estimates. 

\cite{asadi_extremes_2015} modelled joint river flow extremes using max-stable processes, modified for flood risk estimation by incorporating river distance and hydrological distance into the dependence structure. While this model avoids issues with the single site conditioning, it entails an assumption that floods may occur simultaneously at all sites, which does not always hold. \cite{engelke_graphical_2020} developed extremal graphical models based around multivariate Pareto distributions. Like the max-stable approach of \cite{asadi_extremes_2015}, Pareto models also entail that all sites may record simultaneous extremes. Despite this limitation, their model was shown to fit relatively well when applied to flow data on a small portion of the Danube river. 

\cite{wadsworth_statistical_2024} introduced statistical inference for the tail of a multivariate distribution based on the shape of the underlying data. This framework, known as the geometric approach to multivariate extreme value analysis, avoids several of the issues encountered in the approaches mentioned above. Only one model fit is required to estimate the risk at different combinations of sites in different regions of space. Together with the ability to capture both simultaneous and non-simultaneous extremes and the potential to capture complex dependence structures such as floods occurring at all sites at the same time, or at two specific sites, but not at others etc., the geometric approach provides an excellent choice for the problem of estimating the risk of widespread floods.
Owing to the recent nature of the geometric approach, current methodologies within this framework are limited to relatively low number of dimensions. Parametric and semi-parametric approaches have been fit in up to four dimensions in the literature \citep{wadsworth_statistical_2024, majumder_semiparametric_2025, simpson_estimating_2024,campbell_piecewise-linear_2024}.
Two deep-learning approaches have been introduced that go up to eight \citep{murphy-barltrop_deep_2024} and 10 dimensions  \citep{de_monte_generative_2025}. However, these can lack interpretability and the model by \cite{de_monte_generative_2025}, when applied to a 10-dimensional problem, gives less accurate results than in lower dimensions.

Many river networks have $10-20+$ sites, suggesting that new methods are needed. This is the contribution of the present paper. In order to create a parsimonious model in higher dimensions, we incorporate the river structure into our model via the statistical exploitation of \emph{geometric extremal graphical models}, introduced in a theoretical setting by \cite{papastathopoulos_geometric_2026}. To our knowledge, this is the first statistical implementation of this approach. We fit the geometric graphical models to extreme flows at 10 locations on a river network around Preston, in the north west of England. Both \citet{de_monte_generative_2025}, and in a spatial setting \citet{kakampakou_geometric_2025}, can fit models in such dimensions, but in both cases performance is sub-optimal. In contrast, our model fit diagnostics appear strong.
Various extremal dependence structures can be achieved within the setting of geometric extremal graphical models. The simplest model that we present ultimately entails that the most extreme events cannot occur simultaneously.
We therefore propose an extension to this simpler model, where the choice of simultaneous or non-simultaneous extremes is determined during the fitting procedure. Finally, we introduce a correction coefficient for the geometric framework to address potential over- or underestimation of marginal probabilities.

The rest of the paper proceeds as follows.
Section~\ref{sec:river_flow_data_and_pre-processing} introduces the river flow data used for the analysis and the pre-processing steps applied to the data. Section~\ref{sec:geometric_approach} provides background on geometric extreme value theory, and proposes the new correction coefficient. Section~\ref{sec:incorporating_river_structure_into_the_model} introduces geometric extremal graphical models, and how these can be leveraged for inference in the geometric setting.
The modelling of the river flow data and the performance of the model is illustrated in Section~\ref{sec:modelling_the_river_flow_data}. An extension to the model, which allows inference on whether connected sites exhibit simultaneous extremes, is proposed in Section~\ref{sec:extension_of_the_model}.
We conclude with a discussion in Section~\ref{sec:discussion}.

\section{River flow data and pre-processing}
\begin{figure}[t!]
    \centering
    \includegraphics[width=0.35\linewidth]{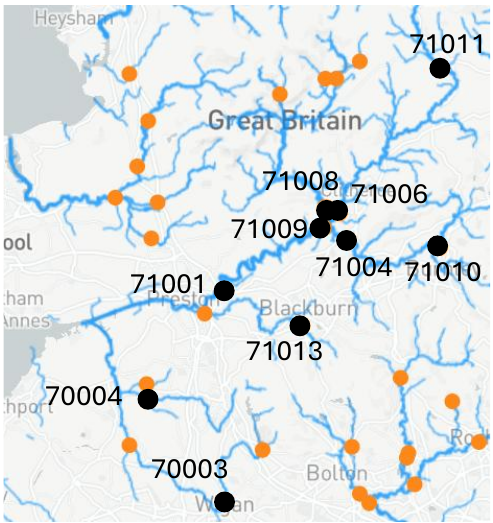}
    \includegraphics[width=0.35\linewidth]{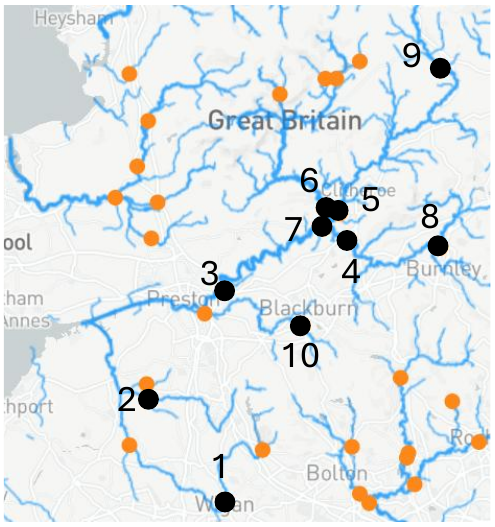}
    \caption{Partial map of the UK showing the rivers as blue lines with gauging stations as orange dots \citep{NRFA}. Black numbered dots represent the gauging stations used for the analysis with station IDs (left) and numbering used throughout the paper (right).}
    \label{fig:Preston_river_network}
\end{figure}

\label{sec:river_flow_data_and_pre-processing}

\subsection{Data}
We obtained data from the National River Flow Archive \citep{NRFA} at gauging stations on a river network in north west England. We consider gauged daily flow (GDF) data, which is the mean river flow in cubic metres per second in a water-day (09.00 to 08.59 GMT).  
The selected river network is close to the city of Preston, and consists of rivers Ribble and Yarrow and their tributaries. Data from all currently operating gauging stations with natural flows and GDF data available for more than 30 years were chosen for the analysis, with any rows of missing data removed. This resulted in a data set with 43 years worth of data between 1980 and 2023 for 10 gauging stations, which are shown in Figure~\ref{fig:Preston_river_network}.
Gauging stations (3)~-~(9) are interconnected with each other -- water from one of these stations flows directly into another one of these stations: we say they are flow-connected stations. Water from stations (1), (2) and (10) does not pass through any other gauging stations: we say these are flow-unconnected stations. 
We also define the term flow-connected pairs, which are pairs of gauging stations $(i,j)$, where water from station ($i$) eventually flows to station ($j$) or vice versa. If there is a direct flow between station ($i$) and ($j$), we say these are adjacent flow-connected pairs. Flow-unconnected pairs are any pairs of gauging stations $(i,j)$, where water from station ($i$) does not flow to station ($j$) or vice versa.

\subsection{Pre-processing}
\begin{figure}[t!]
    \centering
    \includegraphics[width=0.32\linewidth]{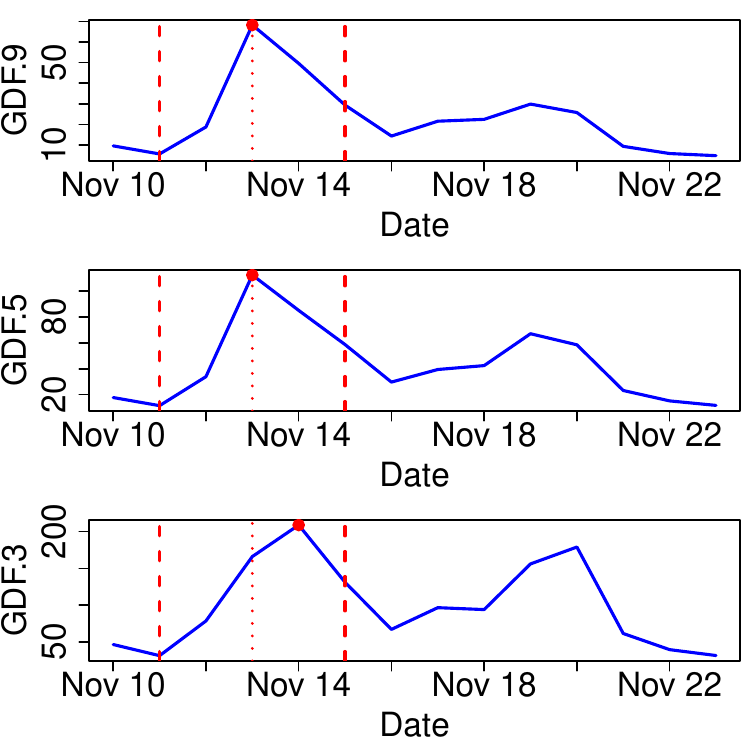}
    \includegraphics[width=0.32\linewidth]{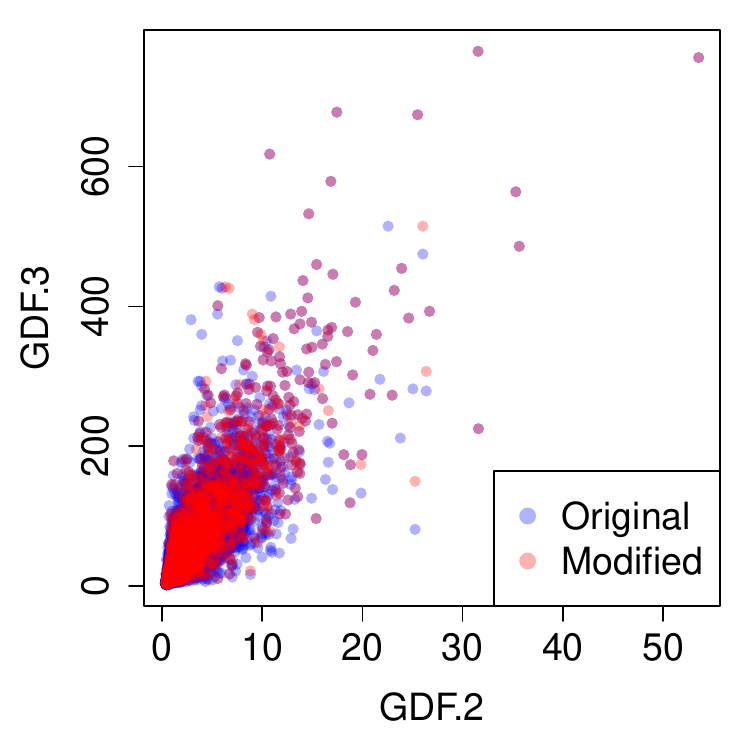}
    \includegraphics[width=0.32\linewidth]{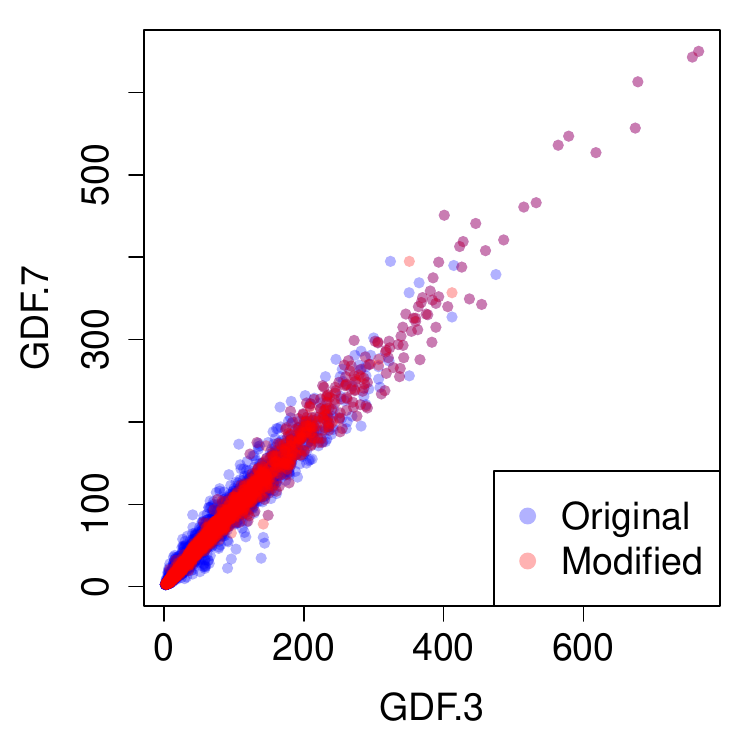}
    \caption{Left: Plot of the GDF at gauging stations (9), (5) and (3) with an extreme event on 13/11/1994 at (9) and (5), and on 14/11/1994 at (3) shown as red dot. Dashed red vertical lines show the window of $\pm 2$ days. Middle and right: Plot of the original (blue) and matched (red) dataset for a pair of flow-unconnected and adjacent flow-connected gauging stations.}
    \label{fig:time_lag_illustration_and_original_vs_matched_dataset}
\end{figure}

Despite the fact that the river network is contained in a relatively small area, time lags can occur between the most extreme recorded flows. This may happen due to differences in runoffs at the different gauging stations, or when the river distances between gauging stations are large, or the river flows are slow. 
In the Preston dataset, this time lag can be illustrated by looking at the extreme river flows at gauging stations (9), (5) and (3). The water from gauging station (9) flows directly to (5) and eventually to (3). The distance between stations (3) and (9) is around 95~km. Dates of events with the most extreme river flows at the station (9) were recorded and GDF values on the dates around the recorded event were investigated at the other gauging stations. 
The aim was to track the peak flow event and investigate whether these peak flows are also observed on the same day. The left panel of Figure~\ref{fig:time_lag_illustration_and_original_vs_matched_dataset} shows the GDF against the date at the three selected gauging stations. It shows that a peak flow event occurred on 13th November 1994 at gauging station (9) which reached station (5) on the same day, but station (3) on the next one. Due to the largest values being recorded on different days, the extreme river flows would be considered as separate events in the data. However, this is not desirable, since we would like to accurately represent a single flood event through one multivariate data point. In order to obtain a dataset of unique flood events, we pre-processed the data by matching the peak flows.

The matching of the peak flows is based on a method by \cite{asadi_extremes_2015}, where a time window of size $\pm~p$ is constructed around the highest ranked event and the peak flows within the time window at each station are taken as the values of river flows for the highest ranked event. Only this event is added to a new dataset and the remaining events within the time window are deleted. This procedure is repeated until no events are left in the dataset. Further details of this procedure are given in Appendix~A.1.
For the Preston dataset, $p = 2$ was selected based on the time lag analysis discussed in the previous section. 
This resulted in a matched dataset of 2,582 points, which is approximately 17\% of the original 15,459 observations. Despite the number of data points being significantly reduced, it was found to be sufficient for further analysis.
Plot in the middle and right of Figure~\ref{fig:time_lag_illustration_and_original_vs_matched_dataset} demonstrates the effect of the pre-processing on a pair of flow-unconnected and adjacent flow-connected stations. In both cases, the matched dataset exhibits stronger dependence, which is expected since many of the river flow values are matched with higher values in the matched dataset compared to the original one.

A few flow values at gauging station (7) were significantly lower than the sum of the flow values at (4), (5) and (6), which are the stations directly upstream of (7). This suggests that the inflow is a lot higher than the outflow, which means the water was lost before reaching (7), or that there is some discrepancy between the measurements at the different stations. For that reason, we removed any data points where 80\% of the sum of the river flows at (4), (5) and (6) was higher than the river flows at (7). This step was performed after the pre-processing and 23 points were removed. None of these points correspond to river heights close to bankfull, and so are not highly informative on the extremes. 

Seasonality both in the original and matched dataset was assessed by obtaining a plot of river flows at a given day of the year for each gauging station. Figure~S.1 in Appendix~A.2 shows an example for gauging station (1). There appears to be a seasonal pattern in the data with higher and lower river flows in the winter and summer months, respectively. In common with several previous studies \citep[e.g.][]{keef_spatial_risk_2009,keef_spatial_dependence_2009,keef_estimating_prob_2013}, we decided not to address the seasonality in the data and proceeded with the matched dataset. Discussion on the reasons for this can be found in Appendix~A.2. 
\section{Geometric approach to multivariate extremes}
\label{sec:geometric_approach}
We now introduce the framework of geometric extreme value theory. We describe the background theory, statistical estimation, and extrapolation from the fitted model. 

\subsection{Convergence of the scaled sample cloud onto the limit set}
\label{sec:convergence_scaled_sample_cloud}
\begin{figure}[t!]
    \centering
    \includegraphics[width=\linewidth]{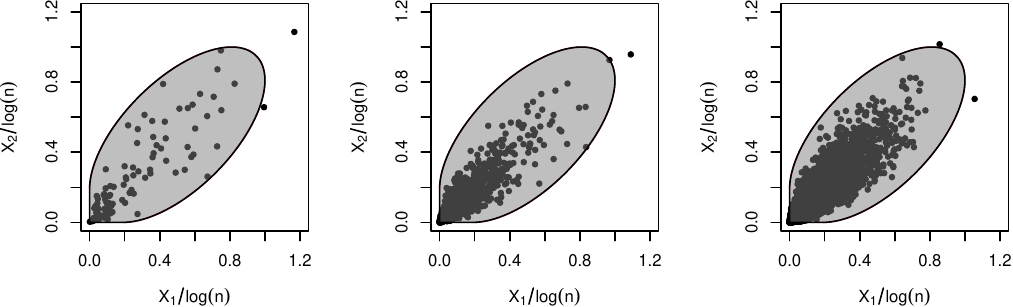}
    \caption{Illustration of the convergence of scaled sample cloud $N_n$ (black points) onto limit set $G$ (grey region) described by the Gaussian gauge function with $\rho = 0.9$. From left to right, sample sizes are $n$ = 100, 1000, 10000.}
    \label{fig:convergence_to_sample_cloud}
\end{figure}

The geometric approach to extreme value theory is based on the convergence of scaled sample clouds of light-tailed random vectors onto a limit set. To disentangle the effects of margins and dependence, a common marginal form is adopted, which is possible statistically via estimation and transformation. We assume standard exponential margins.
The scaled sample cloud is defined as $N_n = \left\{\bm{X}_1/\log(n), \ldots, \bm{X}_n/\log(n)\right\}$, where $\bm{X}_i \in \mathbb{R}^d, i=1,\ldots,n$ are iid random vectors with exponential margins. Under mild conditions, as $n \rightarrow \infty$, the scaled sample cloud $N_n$ converges onto a limit set, defined by $G = \{\bm{x} \in \mathbb{R}_+^d: g(\bm{x}) \leq 1\}$, where $g(\bm{x})$ is termed the gauge function. The gauge function $g(\bm{x})$ is 1-homogeneous, i.e., $g(c\bm{x}) = cg(\bm{x})$ for any $c > 0$, and describes the shape of the boundary of the limit set $G$. The set $G$ is star-shaped, which means that for any $t \in (0,1)$, if $\bm{x}$ is in $G$, then $t\bm{x}$ is also in $G$. Taking the scaling sequence as $\log(n)$ means that the coordinate-wise supremum of $G$ is $(1, \ldots, 1)$.
Figure~\ref{fig:convergence_to_sample_cloud} illustrates the notion of convergence of $N_n$ onto $G$ for random samples from a bivariate Gaussian vector with exponential margins, and the limit set defined by the corresponding gauge function $g_G(x_1,x_2) =(x_1 + x_2 - 2\rho\sqrt{x_1 x_2})/(1-\rho^2)$. With increasing sample size, the black points start to form the shape of the grey region.

The shape of $G$ is informative about the extremal dependence structure. \cite{nolde_linking_2022} linked the geometry of $G$ to several existing dependence concepts, including the key parameters of the conditional extreme value model. \cite{campbell_piecewise-linear_2024} and \cite{papastathopoulos_geometric_2026} use the intersections of the limit set with its boundary box to determine which variables exhibit simultaneous extremes.
For $D = \{1, \ldots, d\}$ and $A \subseteq D$, the variables in group $A$ are simultaneously extreme while the rest of the variables are of smaller order, if the intersection of the boundary of the limit set $G$ with the unit box $[0,1]^d$ occurs at $x_j = 1$ for all $j \in A$ and $x_j = \gamma_j$ for all $j \in D\backslash A$ for some $\gamma_j \in [0,1)$. \cite{papastathopoulos_geometric_2026} termed the full collection of sets $A$ \emph{geometric extremal directions}. In Figure~\ref{fig:convergence_to_sample_cloud}, the boundary of the limit set $G$ intersects the unit box $[0,1]^2$ at $(1,0.9^2)$ and $(0.9^2,1)$, which means the variables do not take their most extreme values simultaneously.

\subsection{Truncated gamma model}
\cite{wadsworth_statistical_2024} introduced a model for the extreme observations, which is rooted in this geometric representation.
The multivariate random vector $\bm{X}$ with standard exponential margins can be expressed in terms of radii and angles, $R = \sum_{j=1}^d X_j$ and $\bm{W} = \bm{X}/R$, respectively. Under certain assumptions, they showed that the conditional distribution of radii greater than a high, angle-dependent, radial threshold $r_0(\bm{w})$, asymptotically follows the gamma distribution with some shape parameter $\alpha>0$ and rate parameter given by the gauge function $g(\bm{w})$.
That is, we assume 
\begin{equation}
        R|\{\bm{W}=\bm{w},R>r_0(\bm{w})\} \; \dot\sim\; \text{truncGamma}(\alpha, g(\bm{w})).
    \label{eq:truncated_gamma_model}
\end{equation}

\noindent This model was shown to be applicable in many different cases by \cite{wadsworth_statistical_2024}, and can be assessed through model diagnostics. The shape parameter is often theoretically equal to the dimension $d$, but estimated for extra flexibility.

\subsubsection{Fitting}
\label{sec:fitting_the_truncated_gamma_model}
To fit the truncated gamma model, the data need to be transformed to exponential margins. We first transform to uniform margins by applying the rank transform to the body of the data and the generalised Pareto distribution (GPD) fitted to the tail \citep{coles_modelling_1991}, then apply the standard exponential quantile function.
We then perform the radial-angular transformation, and consider estimation of $r_0(\bm{w})$. Here we obtain this using a method based on kernel density estimation, introduced by \cite{campbell_piecewise-linear_2024}. 
The threshold is defined as the $\tau$ quantile of the univariate radial distribution $F(r|\bm{w})$. Since $F(r|\bm{w})$ is not known, kernel based estimates are used to get an estimate for $F(r|\bm{w})$, $\hat{F}(r|\bm{w})$. The high radial threshold $r_0(\bm{w})$ is then obtained by numerical inversion of the $\hat{F}(r|\bm{w}) = \tau$, for $\tau$ near 1. 
This method can return a value of $r_0(\bm{w})$ at any angle $\bm{w}$, which is particularly advantageous in high dimensions, where more empirical methods struggle to obtain $r_0(\bm{w})$ in the regions in the angular space that contain very few points.

The truncated gamma model~(\ref{eq:truncated_gamma_model}) can be fitted to the exceedances above $r_0(\bm{w})$ using maximum likelihood estimation. The maximum likelihood estimates (MLEs) from the fitted model are $(\hat{\alpha}, \hat{\bm{\theta}})$, where $\hat{\bm{\theta}}$ are the parameters in the gauge function.
In two dimensions, there are a few different choices for the gauge function: standard parametric models such as logistic, Gaussian, square or inverted logistic, which can also be additively mixed to obtain more flexible shapes \citep{wadsworth_statistical_2024, lee_geometric_2025}, and semi-parametric options such as piecewise-linear \citep{campbell_piecewise-linear_2024}, or Bézier splines \citep{majumder_semiparametric_2025}.

As we increase the number of dimensions, there are fewer options for the standard gauge functions and the number of parameters increases significantly. For example, the $d$-dimensional Gaussian gauge has $d(d-1)/2$ parameters. Both additively mixed and piecewise-linear gauge functions require even higher number of parameters.
In order to apply the methodologies within the geometric framework for flood risk estimation along a river network, reduction in the number of parameters is essential. To obtain a parsimonious model, we incorporate the river structure into the gauge function via the graphical modelling framework. The details of how this is done are described in Section~\ref{sec:incorporating_river_structure_into_the_model}.

\subsubsection{Extrapolation from the model}
\label{sec:extrapolation_and_probability}
Fitting a model to the extreme observations allows extrapolation beyond the region of the data and calculation of probabilities of extreme events far into the tail.
\citet{wadsworth_statistical_2024} perform extrapolation by simulating from the estimated distribution of $\bm{X}|\{R>r_0(\bm{W})\}$. 
This is done first by drawing $\bm{w}^\star$ from the distribution of $\bm{W}|\{R>r_0(\bm{W})\}$. In this case, the samples of $\bm{w}^\star$ are taken from the empirical distribution of the angles. Secondly, conditional upon $\bm{w}^\star$, $r^\star$ is drawn from the distribution of $R|\{\bm{W} = \bm{w}^\star, R > r_0(\bm{w}^\star)\}$, which is the fitted truncated gamma distribution at the sampled angle $\bm{w}^\star$. The new observation is then $\bm{x}^\star = r^\star\bm{w}^\star$.

Once the set of new observations is obtained, we seek to estimate $\Pr(\bm{X} \in B)$, where $B$ is extreme in some way. To do this, we use the decomposition
\begin{equation}
    \Pr(\bm{X} \in B) = \Pr(\bm{X} \in B|R> r_0(\bm{W})) \Pr(R> r_0(\bm{W})) + \Pr(\bm{X} \in B, R<r_0(\bm{W})).
\label{eq:prob_in_region_B_k=1}
\end{equation}
Both of the probabilities in the first term of the equation~(\ref{eq:prob_in_region_B_k=1}) are obtained empirically. The quantity $\Pr(\bm{X} \in B|R > r_0(\bm{W}))$ is equated to the fraction of simulated points in the region $B$, while $\Pr(R > r_0(\bm{W}))$ is the fraction of data points where $R> r_0(\bm{W})$, which will be approximately, but rarely exactly, $1-\tau$. The probability $\Pr(\bm{X} \in B,R<r_0(\bm{W}))$ is calculated empirically using the data as the fraction of non-exceedances in the region $B$. If the set $B$ lies entirely in the region for which $\{R>r_0(\bm{W})\}$, $\Pr(\bm{X} \in B,R<r_0(\bm{W})) = 0$.

It is possible to simulate points in more extreme regions where $R > kr_0(\bm{W})$, for $k > 1$; this is useful for extrapolating further into the tail. In Section~\ref{sec:probability_of_floods_at_3_4_5_7}, we do this to calculate probabilities of simultaneous floods. Details on how this is done can be found in \cite{wadsworth_statistical_2024}.

\subsubsection{Correction coefficient}
\begin{figure}[b!]
    \centering
    \includegraphics[width=0.4\linewidth]{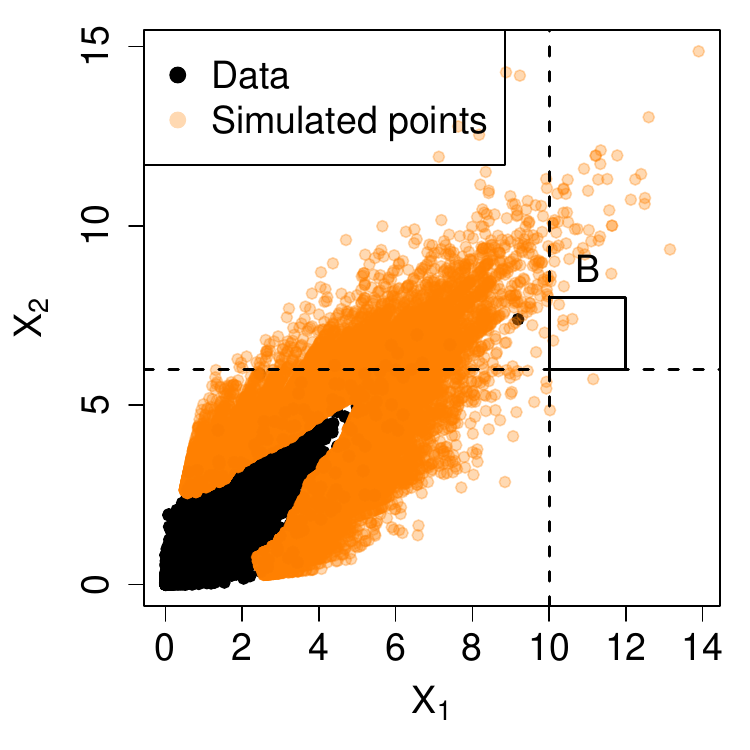}
    \includegraphics[width=0.4\linewidth]{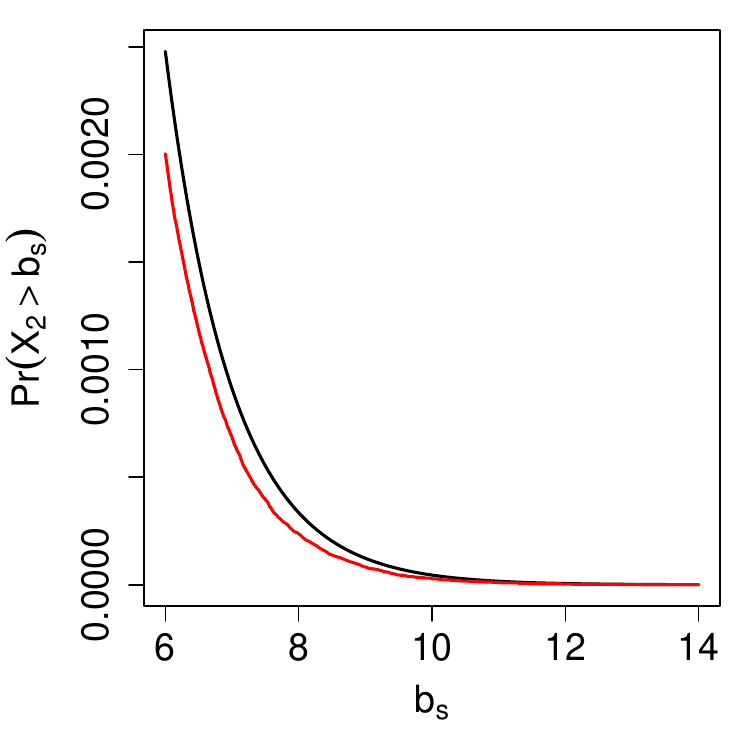}
    \caption{Left: Plot of the data (black) and simulated points (orange) with the region of interest $B=(10,12)\times(6,8)$ (black box). Dashed black lines represent lower boundary of $B$ for calculation of the marginal distribution. Right: Distribution of the marginal probability $\Pr(X_2>b_s)$ for $b_s \in [6,14)$ calculated using the simulated data (red) and using the exponential distribution (black).}
    \label{fig:correct_coeff_discrepancy}
\end{figure}

The data that we are working with has an underlying marginal standard exponential distribution. The method of simulating new points, described in Section~\ref{sec:extrapolation_and_probability}, does not enforce that the simulated points exactly follow a standard exponential distribution. This can be seen by comparing the marginal probability obtained using simulated points with using the exponential distribution directly. The plot on the left of Figure~\ref{fig:correct_coeff_discrepancy} shows a random sample from the logistic distribution, along with points simulated according to the method described in Section~\ref{sec:extrapolation_and_probability}, and an example region of interest $B = (b_{11},b_{12})\times (b_{21},b_{22}) = (10,12)\times(6,8)$. Using the simulated data, we obtain the distribution of marginal probabilities for $X_2$ for values $x \in [b_{21},\infty)$ using simulated points with equation~(\ref{eq:prob_in_region_B_k=1}). This is compared to the probabilities obtained using $\exp{(-x)}$, shown in the plot on the right of Figure~\ref{fig:correct_coeff_discrepancy}. We can see that in this case, the probability is underestimated when calculated using the simulated points.

This discrepancy can lead to systematic over- or underestimation of the probability $\Pr(\bm{X} \in B)$.
In order to correct for this, we multiply the estimate of $\Pr(\bm{X} \in B)$ by a correction coefficient $C_{\mathrm{corr}}$, based on the ratio of the true to estimated marginal probabilities of the lower endpoints of $B$. Numerical experiments show that margins tend to be under/overestimated to a similar degree. We thus take the average of these ratios to get
\begin{equation}
    C_{\mathrm{corr}} = \frac{1}{|S|}\sum_{s \in S}\frac{\exp(-b_s)}{\widehat{\Pr}(X_s > b_s)},
    \label{eq:correction_coefficient}
\end{equation}
where $b_s$ is the lower boundary of $B$ in exponential margins for variable $s$, $\widehat{\Pr}(X_s > b_s)$ is the marginal estimate, and $S \subseteq \{1,\ldots, d\}$, such that only dimensions are included for which $b_s \neq 0$. In the above example, the correction coefficient becomes
\begin{equation*}
     C_{\mathrm{corr}} = \frac{1}{2}\left\{\frac{\exp(-b_1)}{\widehat{\Pr}(X_1 > b_1)}+\frac{\exp(-b_2)}{\widehat{\Pr}(X_2 > b_2)}\right\}, \qquad b_1 = b_{11} = 10,\quad b_2 = b_{21} = 6.
\end{equation*}

\section{Incorporating river structure into the model}
\label{sec:incorporating_river_structure_into_the_model}
In this section, we outline the necessary basics of graphical modelling and introduce geometric extremal graphical models. We then detail the form of these that is suitable for the stations on the Preston river network. 

\subsection{Graphical modelling and terminology}
\begin{figure}[h!]
    \centering
    \includegraphics[width=0.45\linewidth]{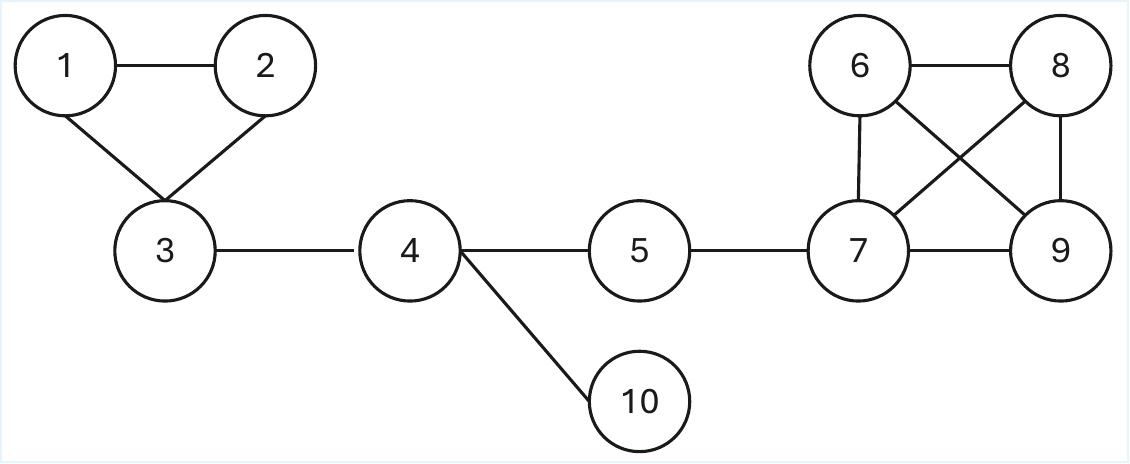}
    \caption{An example of a block graph.}
    \label{fig:example_graph}
\end{figure}
A graph $\mathcal{G}$ is defined as $\mathcal{G} = (\mathcal{V},\mathcal{E})$, where $\mathcal{V} = \{1,\ldots,d\}$ is a set of vertices and $\mathcal{E} \subset \mathcal{V} \times \mathcal{V}$ is a set of edges between pairs of distinct vertices.
We focus on an undirected graph, meaning that for two vertices $i,j \in \mathcal{V}$, $(i,j) \in \mathcal{E}$ implies $(j,i) \in \mathcal{E}$. This means that there is not a distinct edge when going from $i$ to $j$ and when going from $j$ to $i$. 
A special class of graphs, termed decomposable graphs, can be described in terms of $(\mathcal{C},\mathcal{D})$, where $\mathcal{C}$ is the set of cliques, which are maximal fully connected subsets of vertices, and $\mathcal{D}$ is the set of separators, which are intersections of cliques. Block graphs are a sub-type of decomposable graphs, where the separators are all singletons. Figure~\ref{fig:example_graph} shows an example of a block graph with $\mathcal{V} = \{1,\ldots,10\}$, $\mathcal{E} =\{ (1,2), (1,3), (2,3), (3,4), (4,5), (4,10), (5,7), (6,7), (6,8), (6,9), (7,8),(7,9), (8,9)\}$, $\mathcal{C} = \{(1,2,3), (3,4), (4,5), (4,10), (5,7), (6,7,8,9)\}$ and $\mathcal{D} = \{3,4,4,5,7\}$.
A tree is a sub-type of block graph in which any two vertices are only connected by one path, which means all cliques are of size two. In Figure~\ref{fig:example_graph} the graph formed by vertices 3, 4, 5, 7 and 10, for example, is a tree. 
For further details on decomposable graphs see Section 2.3 and Appendix A in \cite{engelke_graphical_2020} and Chapter 2 in \cite{lauritzen_graphical_1996}.

\subsection{Geometric extremal graphical models}
\cite{papastathopoulos_geometric_2026} introduced geometric graphical models on block graphs, which are defined through the gauge function of the limit set and take the following form
\begin{equation}
         g(\bm{x}) = \sum_{C \in \mathcal{C}} g_C(\bm{x}_C) - \sum_{D \in \mathcal{D}} x_D.
         \label{eq:graphical_gauge_function_block}
\end{equation}
The gauge function for a tree can also be expressed through $\mathcal{E}$ and $\mathcal{V}$, where $\mathcal{E} = \mathcal{C}$, by
\begin{equation}
         g(\bm{x}) = \sum_{(i,j)\in \mathcal{E}} g_{\{i,j\}}(x_i,x_j) - x_i - x_j + \sum_{k\in\mathcal{V}} x_k.
         \label{eq:graphical_gauge_function_tree}
\end{equation}

They showed some useful properties of the extremal graphical models in relation to joint extremes, proving that joint extremes cannot occur between groups of variables that cross cliques, without including the separators. 
Using the block graph in Figure~\ref{fig:example_graph} as an example, vertices 2 and 5 cannot have joint extremes on their own without including vertices 3 and 4. They also showed that all cliques exhibiting joint extremes is equivalent to all variables exhibiting joint extremes. These properties are useful when interpreting the results from the fitted graphical models.

\cite{nolde_linking_2022} showed that in general marginal gauge functions can be obtained from the full $d$-dimensional gauge function through the following minimisation:
\begin{equation}
    g_J(\bm{x}_J) = \min_{x_i \in [0,\infty): i\notin J} g(\bm{x}),
    \label{eq:marginal_gauge_min}
\end{equation}
where $\bm{x}_J = (x_j:j \in J)$ for any index set $J \subseteq \{1,\ldots,d\}$, and $g_J$ is the marginal gauge function for the variables in $J$. This is especially useful for visualisation purposes as we can project higher dimensional limit sets to two or three dimensional margins.
In the case of extremal graphical models, there are some simplifications of the minimisation, which can be computationally costly. For example, the marginal gauge function of a clique is $g_C$.
\cite{papastathopoulos_geometric_2026} showed that for a block graph the marginal bivariate gauge function $g_{\{i,j\}}(x_i,x_j)$ can be expressed in terms of a chain graph with vertices lying on the shortest path between $i$ and $j$. 
They also showed that the marginal three-dimensional gauge function $g_{\{i,\pi,j\}}(x_i,x_{\pi},x_j)$, where $\pi$ is the penultimate vertex in the shortest path from $i$ to $j$ can also be expressed in terms of a chain graph such that $g_{\{i,\pi,j\}}(x_i,x_{\pi},x_j) = g_{\{i,\pi\}}(x_i,x_{\pi}) + g_{\{\pi,j\}}(x_{\pi},x_j) - x_{\pi}$.
Depending on the graph and the vertices $i$ and $j$ that we are interested in, minimisation may or may not be required.

\subsection{Model for the Preston river network}
\begin{figure}[b!]
    \centering
    \includegraphics[width=0.4\linewidth]{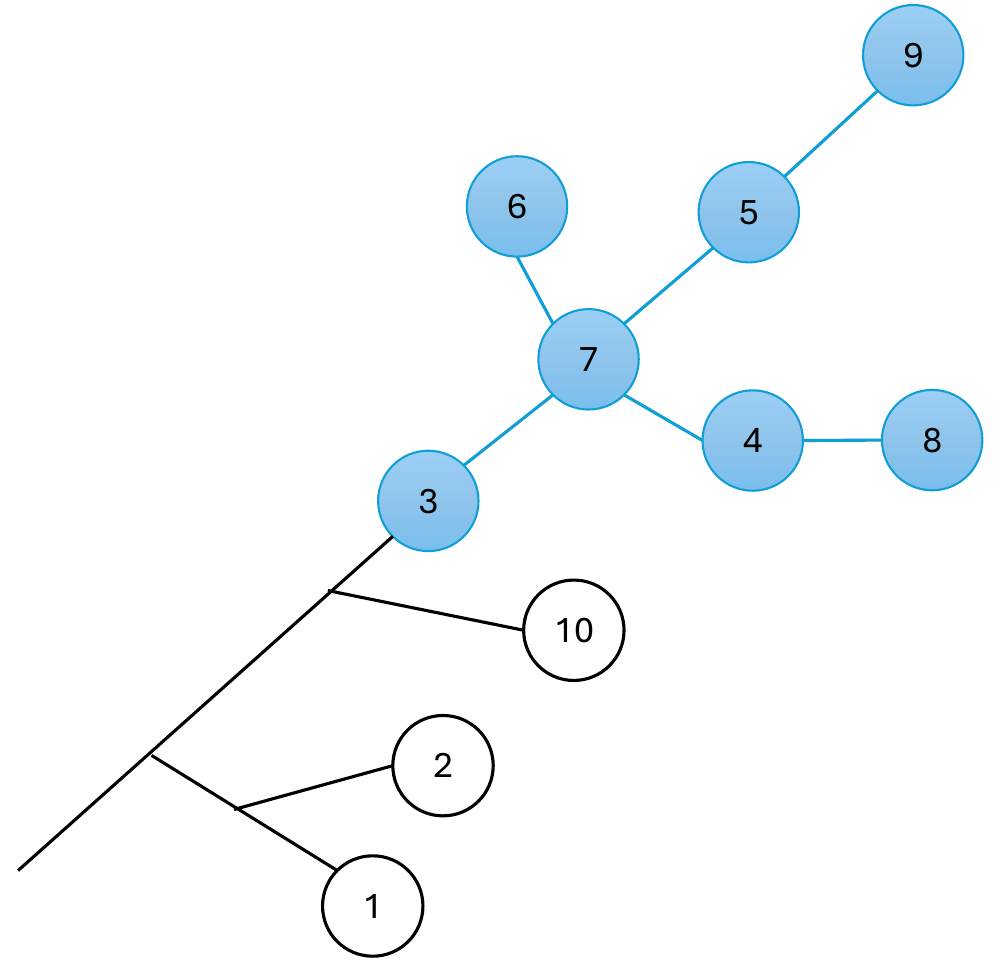}
    \includegraphics[width=0.4\linewidth]{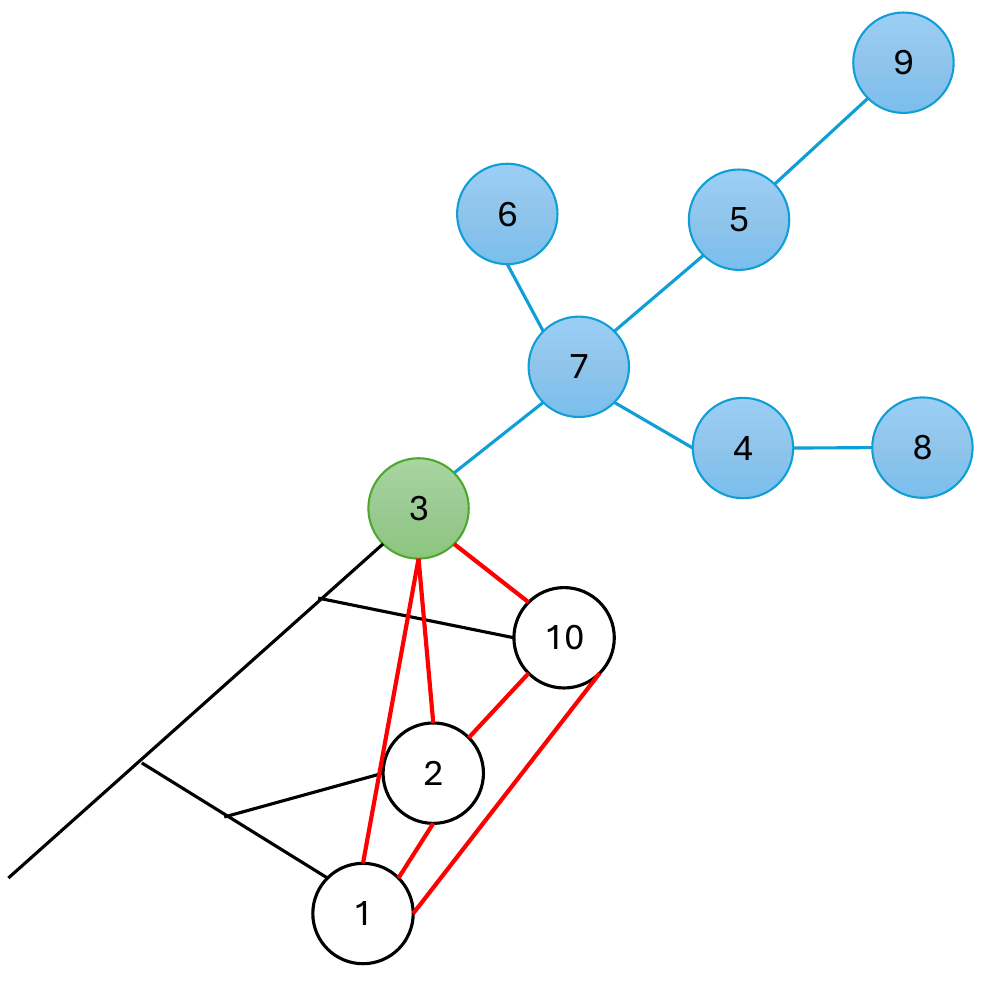}
    \caption{Left: River structure of the gauging stations of the Preston river network. The flow-connected stations are shown as filled blue circles and flow-unconnected as unfilled black circles. Right: Block graph of these stations, where the red lines show the extra edges between the flow-unconnected stations and the closest separator on the tree of flow-connected stations (filled green circle) to create a clique of vertices (1), (2), (3) and (10).}
    \label{fig:Preston_tree_graph}
\end{figure}

Flow gauging stations along a river network can naturally be expressed as a block graph with the flow-connected stations forming a tree.
Figure~\ref{fig:Preston_tree_graph} on the left shows the structure of the gauging stations of the Preston river network shown in Figure~\ref{fig:Preston_river_network}. The flow-connected stations (3), (4), $\ldots$, (9) form a tree with $\mathcal{V} = \{3,\ldots,10\}$ and $\mathcal{E} = \{ (3,7), (4,7), (4,8), (5,7), (5,9), (6,7) \}$.
The tree extremal graphical model~(\ref{eq:graphical_gauge_function_tree}) is used for these stations.
In order to model the whole river network, the flow-unconnected stations (1), (2) and (10) were assumed to form a clique on their own, connected to the rest of the graph via station (3), the closest flow-connected gauging station. 
There is no natural tree structure from the river flow between the flow-unconnected stations since the water is not flowing from one station to another, but there is still a dependence between these stations: due to their close geographical location, these gauging stations are likely to experience similar rainfall events. Figure~\ref{fig:Preston_tree_graph} on the right shows the additional edges between the gauging stations (1), (2), (3) and (10) that join the flow-unconnected stations to the tree of flow-connected stations with station (3) being a new separator.

Applying expression~(\ref{eq:graphical_gauge_function_block}) to the block graph of the Preston river network in Figure~\ref{fig:Preston_tree_graph}, the graphical gauge function for this river network is
\begin{align}
    g_{\{1,\ldots, 10\}}&(x_1,\ldots, x_{10}) = g_{\{3,7\}}(x_3, x_7) + g_{\{4,7\}}(x_4, x_7) + g_{\{4,8\}}(x_4, x_8) + g_{\{5,7\}}(x_5, x_7) \notag \\
    &+ g_{\{5,9\}}(x_5, x_9) + g_{\{6,7\}}(x_6, x_7) + g_{\{1,2,3,10\}}(x_1,x_2,x_3,x_{10}) - x_3 -x_4 - x_5 - 3x_7.
    \label{eq:10d_graphical_gauge_function}
\end{align}
The 10 dimensional gauge function in this case is expressed as a sum of six bivariate gauge functions coming from the adjacent flow-connected pairs and one four-dimensional gauge function coming from the clique formed when joining the flow-unconnected stations to the tree of flow-connected stations.

\section{Modelling the river flow data}
\label{sec:modelling_the_river_flow_data}

\subsection{Fitting the extremal graphical model to the river flow data}
The pre-processed river flow data was transformed to exponential margins as described in Section~\ref{sec:fitting_the_truncated_gamma_model}, where we took the body of the data to correspond to observations less than the 0.95 quantile. In order to obtain the high radial threshold $r_0(\bm{w})$, we need to select $\tau$. In general, $\tau$ should be as high as possible since we want to consider only extreme values, but at the same time it needs to result in enough exceedances for the fitting of the truncated gamma model~(\ref{eq:truncated_gamma_model}). In this case, $\tau = 0.85$ was chosen since it gave the best diagnostics in a threshold sensitivity analysis.

To fit the graphical gauge function using maximum likelihood estimation, choices for the bivariate and four-dimensional gauge functions had to be made. These choices impact the modelling and conclusions on whether gauging stations experience joint extremes or not.
To begin with, we fitted each of four bivariate gauge functions to all adjacent flow-connected pairs: logistic, Gaussian, inverted logistic and square gauge functions. For each pair, the two best candidates were chosen and different combinations of these were considered in the higher-dimensional graphical gauge functions. It was found that the simple choice of Gaussian, $g_G(x_1,x_2) =(x_1 + x_2 - 2\rho\sqrt{x_1 x_2})/(1-\rho^2), \; \rho \in [0,1]$, performed consistently best or near best for all pairs, and so we proceed with this choice. In Section~\ref{sec:extension_of_the_model}, we also introduce an exponential-Gaussian gauge function, which provides an extension to the Gaussian gauge function that allows joint extremes for some or all adjacent flow-connected pairs.

A multivariate Gaussian gauge function was chosen for the clique of flow-unconnected stations as there are not many other flexible candidates that could be used in four dimensions. It has the form
$ g(x_1,x_2,x_3,x_4) = (x_1^{1/2},x_2^{1/2},x_3^{1/2},x_4^{1/2}) \Sigma^{-1} (x_1^{1/2},x_2^{1/2},x_3^{1/2},x_4^{1/2})^\top$,
where $\Sigma$ is positive definite with diagonal elements equal to 1 and off-diagonal elements equal to $\theta_1,\ldots,\theta_6$ where $\theta_i \in [0,1)$ for $i=1,\ldots,6$.
The choice of Gaussian gauge function for all cliques allowed us to obtain the marginal gauge functions more easily using equation~(\ref{eq:marginal_gauge_min}), since the minimisation is known explicitly in this Gaussian case. This is useful for the assessment of the model fit.

This graphical gauge function has 12 parameters that need to be estimated, six from the bivariate Gaussian gauge functions of the adjacent flow-connected pairs and six from the four-dimensional Gaussian gauge function of the clique formed by joining the flow-unconnected stations to the tree of flow-connected stations. The gauge function is thus written as a function of parameters $\bm{\theta} = \allowbreak (\rho_{3,7}, \allowbreak \rho_{4,7}, \allowbreak\rho_{4,8}, \allowbreak \rho_{5,7}, \allowbreak \rho_{5,9}, \allowbreak \rho_{6,7}, \allowbreak \theta_{1,2}, \allowbreak \theta_{1,3}, \allowbreak \theta_{1,10},  \allowbreak \theta_{2,3}, \allowbreak \theta_{2,10}, \allowbreak \theta_{3,10})$.

\subsection{Results}
\subsubsection{Parameter estimates}
When fitting the truncated gamma model~(\ref{eq:truncated_gamma_model}), the shape parameter $\alpha$ is estimated together with the parameters of the graphical gauge function. We express $\alpha = ad$ and estimate $a$. This way all the parameters are of the same order of magnitude, which simplifies the optimisation. 

\begin{table}[h!]
\centering
\caption{Estimated parameters with 90\% CIs for the truncated gamma model with the graphical gauge function.}
\label{tab:parameter_estimates_Gaussian}
\resizebox{\textwidth}{!}{\begin{tabular}{@{}lrrrrrrrrrrrrrr@{}}
\hline
 & \multicolumn{1}{c}{$a$} & \multicolumn{1}{c}{$\rho_{3,7}$} & \multicolumn{1}{c}{$\rho_{4,7}$} & \multicolumn{1}{c}{$\rho_{4,8}$} & \multicolumn{1}{c}{$\rho_{5,7}$} & \multicolumn{1}{c}{$\rho_{5,9}$} &  \multicolumn{1}{c}{$\rho_{6,7}$}
 
 & \multicolumn{1}{c}{$\theta_{1,2}$} & \multicolumn{1}{c}{$\theta_{1,3}$} & \multicolumn{1}{c}{$\theta_{1,10}$} & \multicolumn{1}{c}{$\theta_{2,3}$} & \multicolumn{1}{c}{$\theta_{2,10}$} & \multicolumn{1}{c}{$\theta_{3,10}$} \\
\hline
{Estimate} & 0.33 & 0.98 & 0.95 & 0.96 & 0.94 & 0.97 & 0.89 & 0.97 & 0.75 & 0.71 & 0.76 & 0.74 & 0.73 \\ 
{Lower CI} & 0.33 & 0.93 & 0.88 & 0.87 & 0.83 & 0.84 & 0.77 & 0.93 & 0.30 & 0.27 & 0.35 & 0.32 & 0.40 \\
{Upper CI} & 0.61 & 1.00 & 0.99 & 0.99 & 1.00 & 0.99 & 0.99 & 0.99 & 0.91 & 0.92 & 0.92 & 0.95 & 0.88 \\
\hline
\end{tabular}}
\end{table}

Table~\ref{tab:parameter_estimates_Gaussian} shows the estimated parameters for this model with their 90\% confidence intervals (CIs). The CIs were obtained through block-bootstrap sampling, since the pre-processed data retains some temporal dependence. When looking at the ACF plots of the matched data, the dependence disappeared around time lag 10. Based on this, and a small sensitivity analysis, the CIs were obtained from 1000 block-bootstrap samples with block size 10.
Kernel density estimation was performed on the block-bootstrap samples in order to provide CIs representing the highest density of the bootstrap distribution. We selected this technique because some bootstrap distributions had long tails, even though the majority of support lay in a small region. We selected to provide 90\% CIs as a summary for the same purpose. The full bootstrap distributions are provided in Figure~S.2 and Figure~S.3 in Appendix~B.1. 
Estimates of the Gaussian parameters for the adjacent flow-connected pairs are high, with pair $(6,7)$ having slightly lower value compared to the rest of them. Estimates of the Gaussian parameters for the flow-unconnected pairs are lower and the corresponding CIs are wider with the exception of pair $(1,2)$. This is natural because we expect flow-connection to give rise to higher dependence. The flow-unconnected stations exhibit weaker dependence but are geographically still relatively close to one another and very likely affected by similar rainfall events.

\subsubsection{Assessment of model fit}
The model fit is assessed via a transformed probability-probability (PP) plot. A PP plot compares the empirical cdf to the cdf of the fitted model, $\hat{F}_{\mathrm{tg}}$. 
It is defined as $\{i/(n+1),u_{(i)}\}$, where $u_i = \hat{F}_{\mathrm{tg}}[r_i;\bm{w}_i,r_0(\bm{w}_i)]$ for threshold exceedances $i = 1, \ldots, n$, and $u_{(1)} \leq \ldots \leq u_{(n)}$ are the order statistics. In order to see clearly whether the fit is within the tolerance intervals, the difference between the cdf of the fitted model and the empirical cdf was plotted, resulting in the transformed PP plot.
If the model is describing the data well, we would expect the points to lie close to the line $y~=~0$ and be within the tolerance intervals, which are obtained from the beta distribution of order statistics. Such intervals are appropriate for independent data, and would likely be slightly wider for dependent data.

\begin{figure}[t!]
    \centering
    \includegraphics[width=0.32\linewidth]{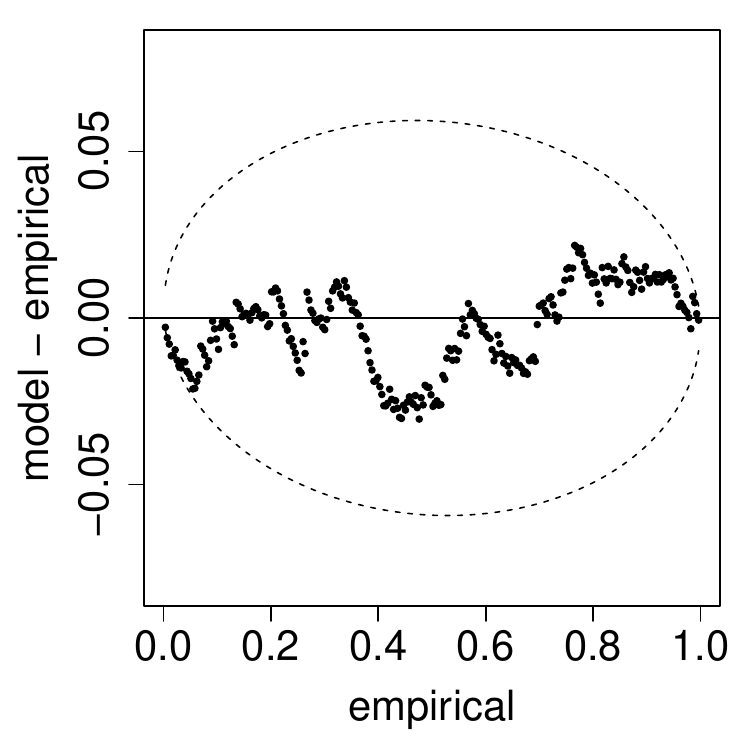}
    \includegraphics[width=0.32\linewidth]{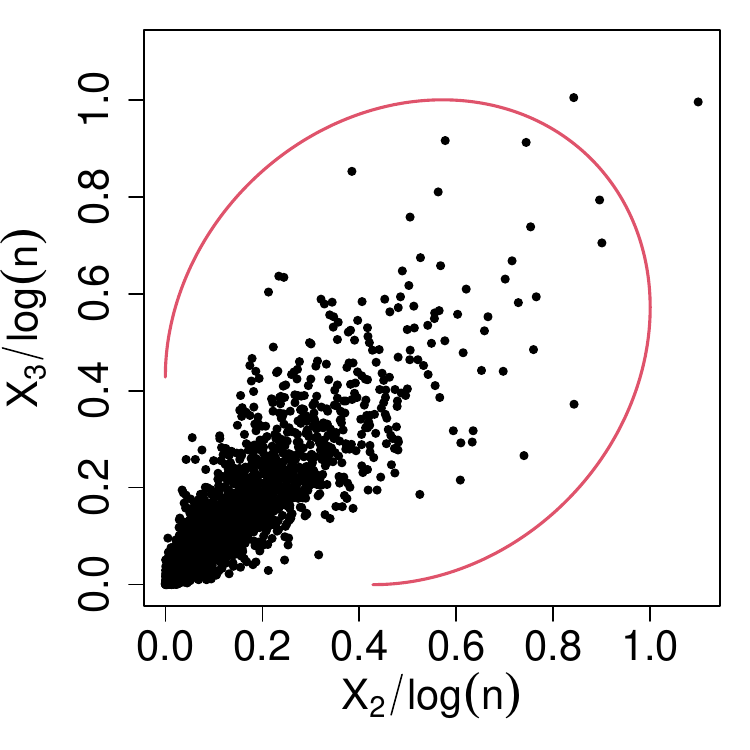}
    \includegraphics[width=0.32\linewidth]{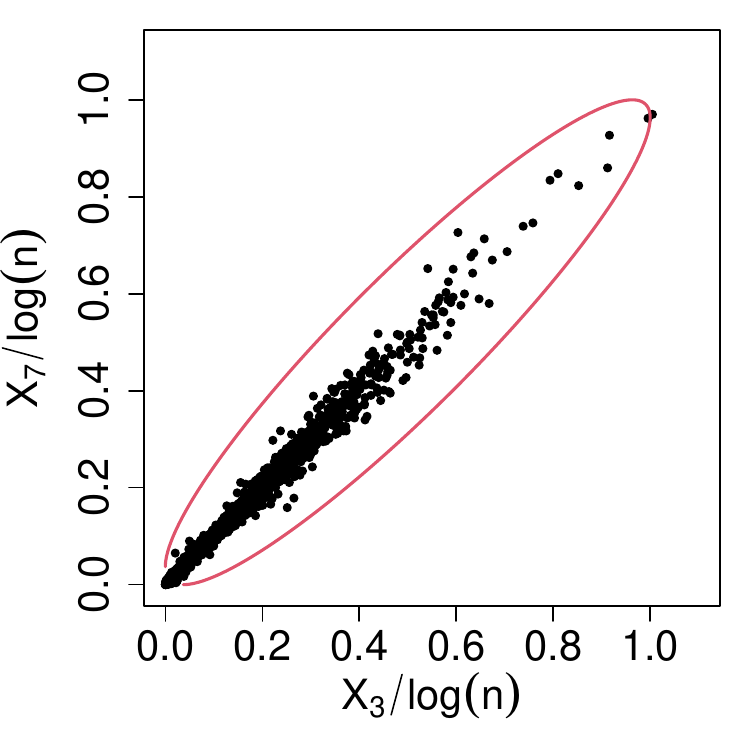}
    \caption{Transformed PP plot for the truncated gamma model with the graphical gauge function for the Preston river network (left). Plots of the bivariate projections of the limit set (red line) with corresponding scaled sample clouds (black points) for a flow-unconnected pair $(2,3)$ (middle) and an adjacent flow-connected pair (right).}
    \label{fig:model_fit_assessment}
\end{figure}

The left plot of Figure~\ref{fig:model_fit_assessment} shows the transformed PP plot for the fitted model to all 10 gauging stations, indicating a good fit. 
The other two plots in Figure~\ref{fig:model_fit_assessment} display bivariate projections of unit level sets from the graphical gauge function, together with the scaled data for a flow-unconnected and an adjacent flow-connected pair. Additional bivariate projections for the adjacent flow-connected pairs can be found in Figure~S.7 in Appendix~C.1.
Such plots give an indication of whether the fitted graphical gauge function represents a plausible limit set for the scaled sample cloud. 
If the model fits the data well, we would expect a broad agreement between the shape of the limit set and of the scaled sample cloud with the points being largely contained within the limit set; this appears to be the case here. For the pair $(6,7)$, the shape of the limit set is a little wider than the scaled sample cloud. In Appendix~B.3, we describe a sensitivity analysis, which suggests that the position of a single point is responsible for this result. We also observe that in the case of the flow-unconnected pair the dependence is weaker, which is reflected in a wider shape of the projection of the fitted limit set.

\subsubsection{Simulated points and comparison of model-based and empirical $\chi_S(u)$}
\label{sec:simulated_points_and_chi}
A main objective of fitting extreme value models is to extrapolate beyond the range of the data. In order to check the suitability of the model for this purpose, we begin by simulating points from the fitted model, and visually comparing them to the observed data. We first simulate 5,000 points on the exponential scale, as described in Section~\ref{sec:extrapolation_and_probability}, then reverse the marginal transformation. Points above the 0.95 quantile are back-transformed via the fitted GPD, while those below are transformed via the inverse empirical cdf, obtained by joining the points in the empirical cdf by a straight line.

\begin{figure}[t!]
    \centering
    \includegraphics[width=0.4\linewidth]{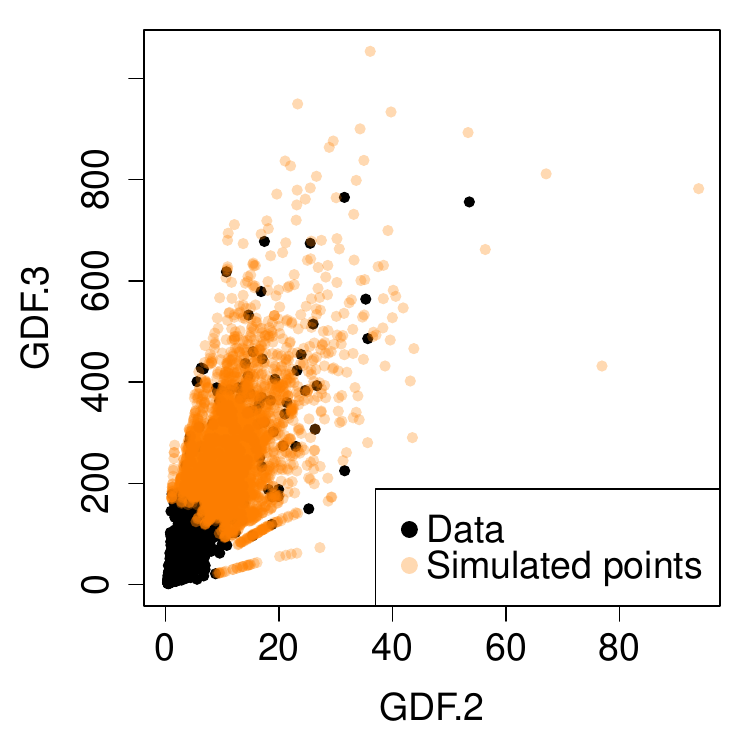}
    \includegraphics[width=0.4\linewidth]{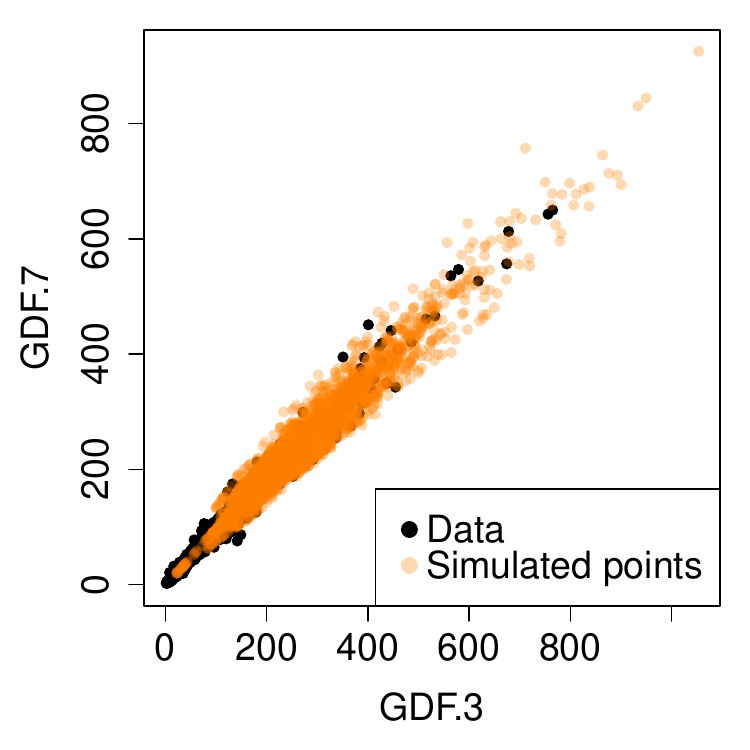}
    \caption{Plot of matched data (black) and simulated points (orange) for flow-unconnected (left) and adjacent flow-connected (right) pair.}
    \label{fig:simulated_points_2_3_and_3_7_original_scale}
\end{figure}

Figure~\ref{fig:simulated_points_2_3_and_3_7_original_scale} illustrates data at two pairs of stations, showing both the simulated points from the 10-dimensional model and the data on the original scale for a flow-unconnected and an adjacent flow-connected pair of gauging stations. It can be seen that the simulated points mimic the shape of the data, and provide plausible extrapolation. On the left hand side plot, trails of points can be seen, which is due to the fact that the empirical distribution $\bm{W}|\{R > r_0(\bm{W})\}$ was used to sample new $\bm{w}^\star$. 

In addition to the plot of simulated points, we used simulation to calculate a well-known tail dependence measure $\chi_S(u)$, defined as
\begin{equation*}
    \chi_{S}(u) = \frac{1}{1-u}\Pr[F_j(X_j) > u, \forall j \in S], \quad u \in (0,1), \quad S \subseteq\{1,\ldots,d\}.
\end{equation*}
We compare empirical estimates of $\chi_S(u)$ with model-based values, calculated using simulated points for a range of quantile levels $u \in (0,1)$ and $S \subseteq \{1,\ldots,d\}$.
This dependence measure gives an indication of the tendency of all variables to be large conditional on one variable being large.
The probability $\Pr[F_j(X_j) > u, \forall j \in S]$ is obtained as the probability $\Pr(\bm{X} \in B)$ where $B$ is the cross product of intervals $(- \log(1- u), \infty)$ for variables that are within the subset $S$, and $(0, \infty)$ for the variables that are not. 
In the case of the empirical $\chi_S(u)$, the probability was obtained as a fraction of data points lying in the region $B$. 
The probability $\Pr(\bm{X} \in B)$ for the model-based $\chi_S(u)$ was obtained using equation~(\ref{eq:prob_in_region_B_k=1}), where $\Pr(\bm{X} \in B|R>r_0(\bm{w}))$ was calculated from 100,000 simulated points, which ensured that there were enough points in the region $B$ for $u \rightarrow 1$. We used 1,000,000 points for the zoomed-in versions of the $\chi_S(u)$ plots for the original fit to produce a smoother curve.  Alternatively, a fewer points could be simulated in region where $R>kr_0(\bm{W})$. The correction coefficient was obtained using equation~(\ref{eq:correction_coefficient}), where $b_s = - \log(1- u), \forall s \in S$. The probability $\Pr(X_s > b_s)$ was calculated using equation~(\ref{eq:prob_in_region_B_k=1}) where $B$ is $(- \log(1- u), \infty)$ for variable $s$ and $(0, \infty)$ for the rest of the variables.

Block-bootstrap sampling with block size 10 was used to generate 1000 samples for both the empirical and model-based $\chi_S(u)$ separately. The empirical $\chi_S(u)$ was calculated directly from these samples. For the model-based $\chi_S(u)$, the truncated gamma model was fitted to each sample and new points were simulated for each sample. The model-based $\chi_S(u)$ was calculated from these. 
Since we are taking the bootstrap samples from the matched data in exponential margins, we need to account for the fact that the bootstrap samples might not precisely follow standard exponential distribution. This has an impact on the correction coefficient, since we assume the data follows standard exponential distribution in its calculation. Instead of making this assumption, we fit an exponential distribution to each bootstrap sample and use the fitted exponential distribution in the calculation. The 95\% CIs were obtained for both empirical and model-based $\chi_S(u)$ plots by taking the pointwise 2.5 and 97.5\% quantiles of the respective $\chi_S(u)$ samples.

\begin{figure}[t!]
    \centering
    \includegraphics[width=0.4\linewidth]{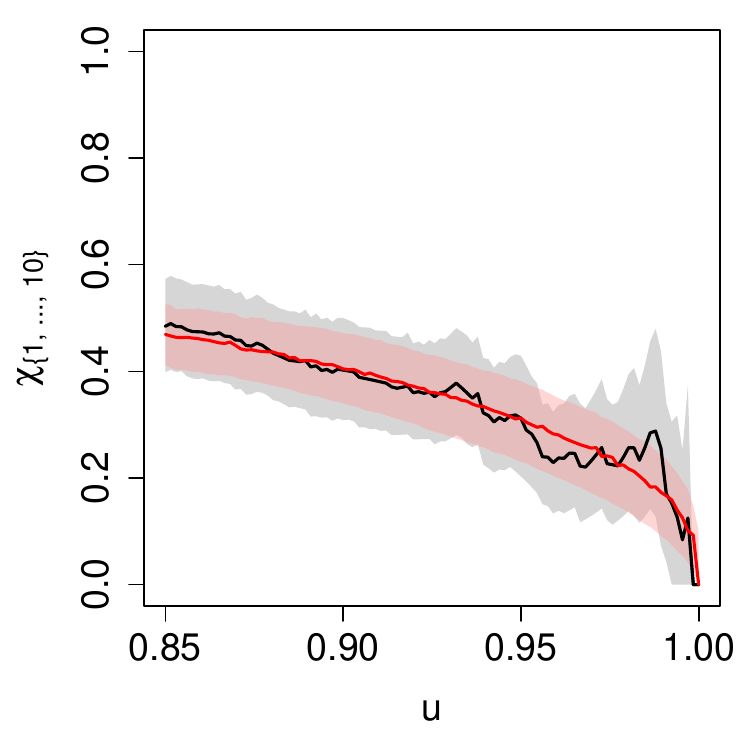}
    \includegraphics[width=0.4\linewidth]{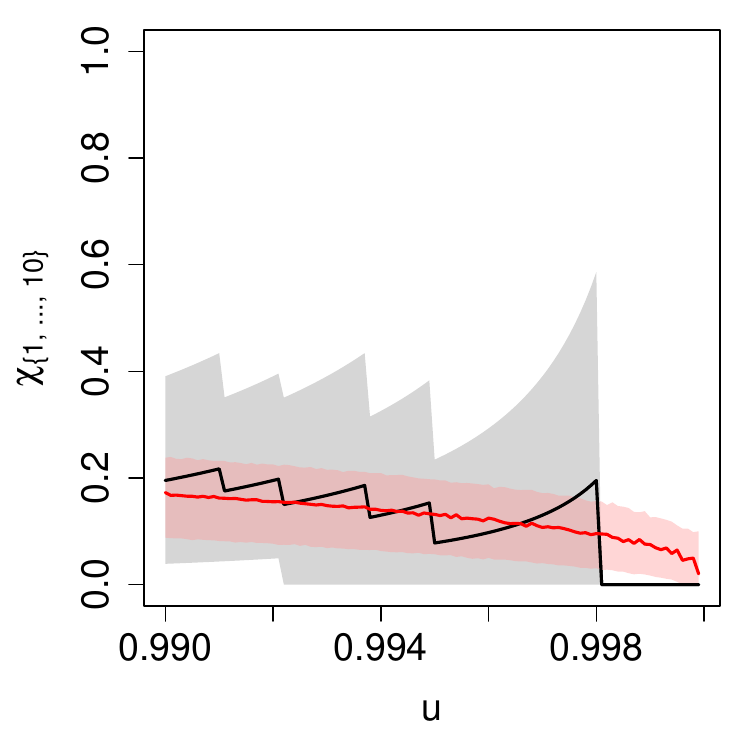}
    \includegraphics[width=0.4\linewidth]{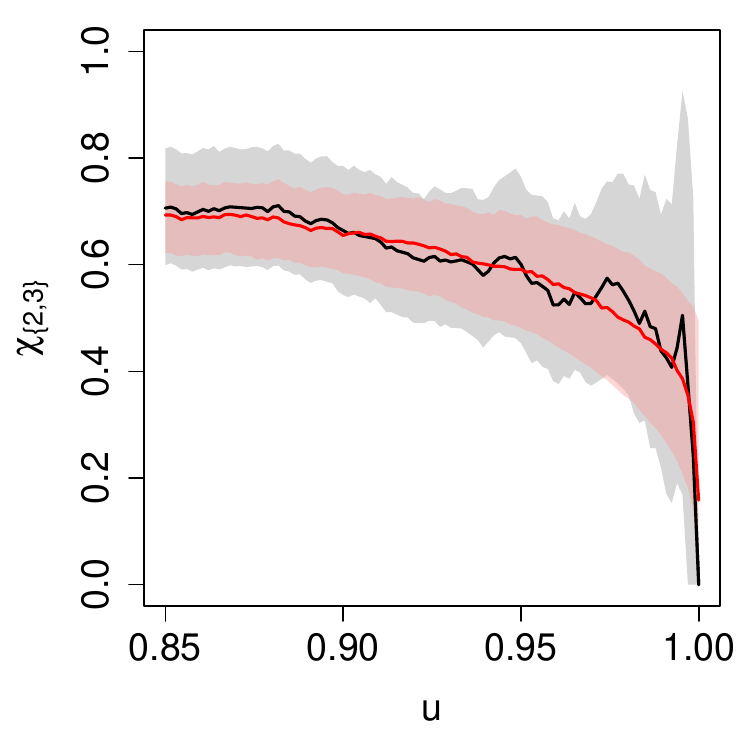}
    \includegraphics[width=0.4\linewidth]{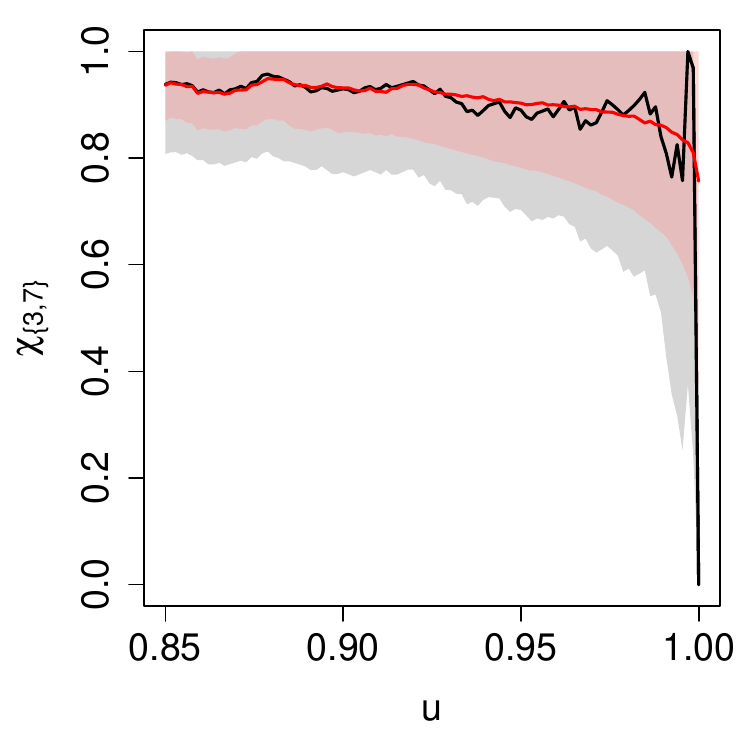}
    \caption{Plot of empirical (solid black) and model-based $\chi_S(u)$ (solid red) with empirical (grey region) and model-based (light red region) 95\% CIs with $S = \{1,\ldots,10\}$ (top), $\{2,3\}$ (bottom left) and $\{3,7\}$ (bottom right). The top right-hand side plot zooms in on $u \rightarrow 1$ values.}
    \label{fig:chi_plot_examples}
\end{figure}

The top plots in Figure~\ref{fig:chi_plot_examples} show the comparison of the empirical and model-based $\chi_S(u)$ with $S = \{1,\ldots,10\}$. It can be seen that the model-based $\chi_S(u)$ agrees well with the empirical one. The empirical $\chi_S(u)$ is mostly within the model-based CIs. Non-zero values for the model-based $\chi_S(u)$ can be observed for $u$ close to 1 when the empirical is zero. This highlights the ability of the model to extrapolate beyond the range of the data. The bottom plots show the comparison of bivariate empirical and model-based $\chi_S(u)$ for a flow-unconnected pair $S = \{2,3\}$ and an adjacent flow-connected $S = \{3,7\}$. Additional $\chi_S(u)$ plots can be found in Figure~S.8 in Appendix~C.2.1. The model-based $\chi_S(u)$ agrees well with the empirical one in different cases. 
The CIs widen as $u \rightarrow 1$, which is especially noticeable for the flow-connected pairs.

The $\chi_S(u)$ plots were obtained in three different ways: without $C_{\mathrm{corr}}$, with $C_{\mathrm{corr}}$ assuming standard exponential distribution, and with $C_{\mathrm{corr}}$ using a fitted exponential distribution. It was found that the model-based $\chi_S(u)$ was closer to the empirical $\chi_S(u)$ when the probability was multiplied by $C_{\mathrm{corr}}$, especially for $u \rightarrow 1$. The CIs were narrower when the probability was multiplied by $C_{\mathrm{corr}}$. When a standard exponential distribution was assumed for $C_{\mathrm{corr}}$, the CIs were very narrow whereas fitting the exponential distribution resulted in wider CIs, but ones that likely better reflect the true uncertainty of the model. An example of a comparison of these intervals can be seen in Figure~S.9 in Appendix~C.2.2.

\subsubsection{Probability of floods at gauging stations (3), (4), (5) and (7)}
\label{sec:probability_of_floods_at_3_4_5_7}
In order to illustrate practical utility of the model, we calculate the annual probability of floods occurring at gauging stations (3), (4), (5) and (7) simultaneously. We chose to focus on this particular probability because there is an indication that these gauging stations can experience simultaneous extremes, which is further discussed in Section~\ref{sec:extension_of_the_model}.
In this case a flood is defined as river flow values at each location exceeding their corresponding river flow value at bankfull. This value is obtained through the statistical equation used to obtain river flows from river heights that are measured. This equation, together with the river height corresponding to bankfull, were obtained from the information on each gauging station from \cite{NRFA}. 
The bankfull values at stations (3), (4), (5) and (7) are $v_3 = 860$, $v_4 = 318$, $v_5 = 580$ and $v_7 = 900$ $\mathrm{m^3/s}$, respectively, which on exponential margins are $v_{\mathrm{exp}, 3} = 8.6$, $v_{\mathrm{exp}, 4} = 9.3$, $v_{\mathrm{exp}, 5} = 13.0$ and $v_{\mathrm{exp}, 7} = 9.7$. There are no observations exceeding bankfull at any of these locations, hence the probability cannot be calculated empirically. 

There are two approaches to calculating this probability. One approach is based on the idea of event set generation that has previously been used with the conditional approach, see for example \cite{keef_estimating_prob_2013}. The event set generation method with our framework is described in more detail in Appendix~D. 
This approach can be computationally expensive if $B$ is very far in the tail. We thus focus on a more direct and efficient way to calculate this probability.
One million points were simulated from the fitted model with the graphical gauge function using methods described in Section~\ref{sec:extrapolation_and_probability}, but in a region where $R>kr_0(\bm{W})$ with $k>1$, since the river flow values corresponding to bankfull are far in the tail. This entails a modified version of equation~(\ref{eq:prob_in_region_B_k=1}) to calculate the probability $\Pr(\bm{X}\in B)$, as outlined in \citet{wadsworth_statistical_2024}. In this case $B =(0,\infty)^2\times(v_{\mathrm{exp},3},\infty)\times(v_{\mathrm{exp},4},\infty)\times(v_{\mathrm{exp},5},\infty)\times(0,\infty)\times(v_{\mathrm{exp},7},\infty)\times(0,\infty)^3$.
To minimise Monte Carlo uncertainty, we want to select $k$ as high as possible, subject to the high radial threshold $kr_0(\bm{W})$ remaining below the marginal values $v_{\mathrm{exp},i}$ for $i=3,4,5,7$. 
The value of $k$ was obtained using a heuristic method introduced by \cite{campbell_analysing_2026}, where a set of values $k' \in \{1,1.1,\ldots,4\}$ was considered. For each $k'$, the non-exceedances were scaled by $k'$ and the number of these scaled non-exceedances in the region $\max_{i = \{3,4,5,7\}} X_i > \min_{i = \{3,4,5,7\}}v_{\mathrm{exp},i}$ was obtained. The value of $k$ was chosen as the highest value in the range that gave zero scaled non-exceedances, leading to $k=2.7$
The resulting exceedance probability per event, with 95\% CIs obtained from the 1000 block-bootstrap samples by taking the 2.5 and 97.5\% quantile, is $2.9 [1.3, 3.9]\cdot 10^{-6}$. The annual exceedance probability is obtained by multiplying this estimate by the average number of events per year in which there is no missing data --- recalling our pre-processing --- which gives $1.8 [0.8, 2.4]\cdot 10^{-4}$. This simple multiplication is highly accurate given the very small probability.
The annual probability of floods occurring at stations (3), (4), (5) and (7) is therefore very low, corresponding to a return period of over 5000 years. 
This was an example used for illustration purposes. However, using the fitted model with the graphical gauge function, we are in theory able to calculate probabilities in the whole region of space. This is an advantage compared to the framework of conditional extremes, where multiple models conditioning on different stations in turn have to be fitted.
For example, if we wanted to calculate the same probability using the conditional approach, we would have to use a model conditioning on station (3), (4), (5) or (7). This gives four different models, where each one of them might give slightly different result or the models might have to be combined using a method with importance weights \citep{wadsworth_higher-dimensional_2022}, for example. Moreover, none of these models could be used if we were interested in calculating the probability of flooding at station (1) and (2), for example. This means that in order to be able to calculate probabilities for all different possible combinations, $d$ different models conditioning on each of the stations in turn would have to be fitted, each with $2(d-1)$ parameters, where in this case $d=10$. The appropriate models would then have to be chosen depending on the probabilities of interest and potentially combined together.

\section{Extension of the model}
\label{sec:extension_of_the_model}
Using the Gaussian gauge function for the adjacent flow-connected pairs entails an assumption that the pairs do not experience joint extreme river flows at the very highest levels. This is because the boundary of the limit set does not intersect the unit square at $(1,1)$, as discussed in Section~\ref{sec:convergence_scaled_sample_cloud}. For greater flexibility, we propose an extension to the model that modifies the bivariate Gaussian gauge function in a way that allows for the possibility of joint extremes between these pairs. This gauge function corresponds to that of the model introduced by \cite{huser_modeling_2019} for interpolating between asymptotic dependence and asymptotic independence.
Following \citet{ lee_geometric_2025} we term this the exponential-Gaussian gauge function, with form
\begin{equation*}
    g_{EG}(x_1,x_2) = \begin{dcases*}
\min_{s \in [0,{\min(x_1,x_2)}/{\gamma}]} \{s+g_G(x_1-\gamma s, x_2 - \gamma s)\}, \quad \gamma \in (0,1] \\
\min_{s \in [0,\min(x_1,x_2)]}\{s+g_G(\gamma x_1-\gamma s, \gamma x_2 - \gamma s)\}, \quad \gamma > 1,
\end{dcases*}
\end{equation*}
where $g_G$ is the bivariate Gaussian gauge and $\gamma$ is an extra parameter that controls whether there is a point at (1,1) or not. When $\gamma \leq 1/g_G(1,1)$, the shape of the Gaussian gauge function is unaffected. When $\gamma \in (1/g_G(1,1),1)$, the shape alters but there is still no point at (1,1). When $\gamma \geq 1$, the resulting gauge function has a point at (1,1).
This means that for $\gamma<1$ the pair does not experience joint extremes and for $\gamma\geq1$ it does.
This value is estimated when fitting the model, hence no prior assumption has to be made about the occurrence of joint extremes between the pairs. In addition, joint extreme river flows for certain pairs creates a higher order group of joint extreme river flows, which links to the property showed by \cite{papastathopoulos_geometric_2026}, whereby all cliques exhibiting joint extremes is equivalent to all variables exhibiting joint extremes. This property is still applicable to a subgraph involving sites (3), (4), (5) and (7).

Fitting the model with exponential-Gaussian gauge function for all adjacent flow-connected pairs resulted in the 10-dimensional gauge function having 18 parameters. The likelihood corresponding to this model was found to be relatively flat, leading to difficulties in optimisation and parameter estimates with large uncertainties. To simplify the optimisation, we decided to apply the exponential-Gaussian gauge function only to the pairs that showed an indication of joint extremes in the full exponential-Gaussian model. These pairs were $(3,7)$, $(4,7)$, and $(5,7)$. Results for this model follow.

\subsection{Results}
\begin{table}[t]
\caption{Estimated parameters with 90\% CIs for the truncated gamma model with the graphical gauge function with exponential-Gaussian gauge function used for pairs $(3,7)$, $(4,7)$, and $(5,7)$.}
\label{tab:parameter_estimates_exp-Gauss}
\begin{center}
\begin{tabular}{@{}lrrrrrrrrrrr@{}}
\hline
 & \multicolumn{1}{c}{$a$} & \multicolumn{1}{c}{$\gamma_{3,7}$} & \multicolumn{1}{c}{$\rho_{3,7}$} & \multicolumn{1}{c}{$\gamma_{4,7}$} & \multicolumn{1}{c}{$\rho_{4,7}$} & \multicolumn{1}{c}{$\rho_{4,8}$} & \multicolumn{1}{c}{$\gamma_{5,7}$}& \multicolumn{1}{c}{$\rho_{5,7}$} & \multicolumn{1}{c}{$\rho_{5,9}$} &  \multicolumn{1}{c}{$\rho_{6,7}$} \\
\hline
{Estimate} & 0.36 & 1.06 & 0.96 & 1.27 & 0.86 & 0.96 & 1.00 & 0.85 & 0.97 & 0.89 \\  
{Lower CI} & 0.34 & 0.83 & 0.83 & 0.81 & 0.67 & 0.87 & 0.67 & 0.16 & 0.86 & 0.78 \\ 
{Upper CI} & 0.63 & 1.81 & 1.00 & 1.93 & 0.99 & 0.99 & 1.97 & 1.00 & 0.99 & 0.99 \\[1pt]

& \multicolumn{1}{c}{$\theta_{1,2}$} & \multicolumn{1}{c}{$\theta_{1,3}$} & \multicolumn{1}{c}{$\theta_{1,10}$} & \multicolumn{1}{c}{$\theta_{2,3}$} & \multicolumn{1}{c}{$\theta_{2,10}$} & \multicolumn{1}{c}{$\theta_{3,10}$} \\
\hline
{Estimate} & 0.97 & 0.77 & 0.71 & 0.77 & 0.74 & 0.74\\  
{Lower CI} & 0.93 & 0.32 & 0.29 & 0.35 & 0.34 & 0.43 \\ 
{Upper CI} & 0.99 & 0.91 & 0.92 & 0.92 & 0.95 & 0.88 \\ \hline
\end{tabular}
\end{center}
\end{table}

Table~\ref{tab:parameter_estimates_exp-Gauss} shows the estimated parameters with 90\% CIs for the model with exponential-Gaussian gauge functions used for pairs $(3, 7)$, $(4, 7)$, and $(5, 7)$. The full bootstrap distributions of all parameter estimates are provided in Figure~S.4 and Figure~S.5 in Appendix~B.2. The parameter estimates are similar to the ones obtained with the original model for the pairs where the Gaussian gauge function was used for adjacent flow-connected pairs and for flow-unconnected pairs. The estimates of the $\gamma$ parameters are $\geq 1$ suggesting joint extreme river flows for these pairs. However, the CIs also include values of $\gamma<1$. The CIs are particularly wide for $\gamma_{\{5,7\}}$ and $\rho_{\{5,7\}}$, which suggests that different combinations of $\gamma_{\{5,7\}}$ and $\rho_{\{5,7\}}$ can lead to similar likelihoods.

\begin{figure}[b!]
    \centering
    \includegraphics[width=0.32\linewidth]{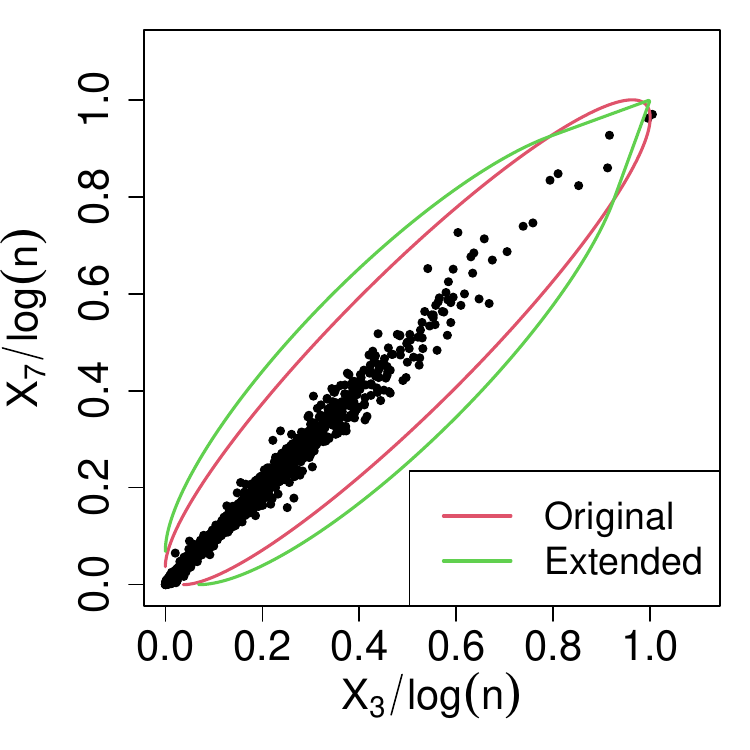}
    \includegraphics[width=0.32\linewidth]{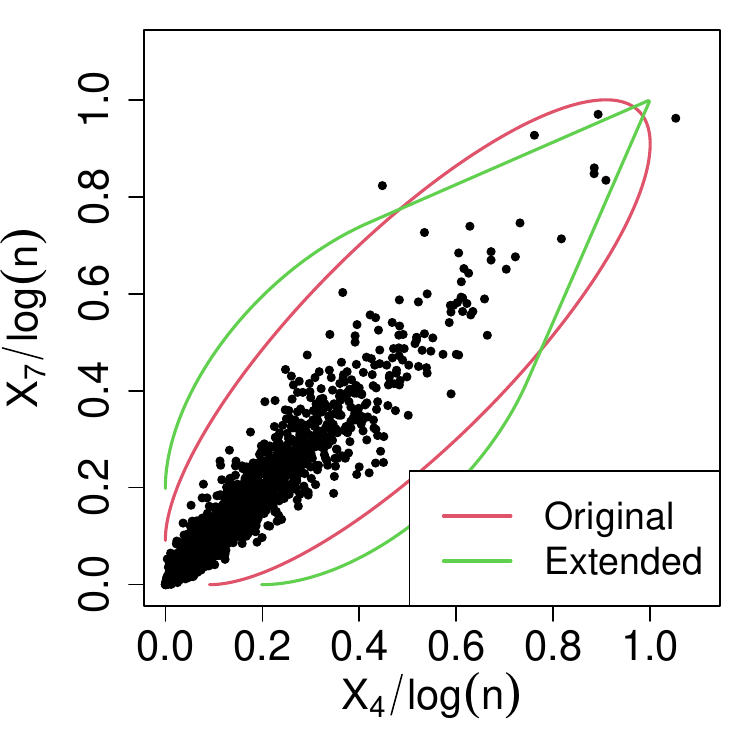}
    \includegraphics[width=0.32\linewidth]{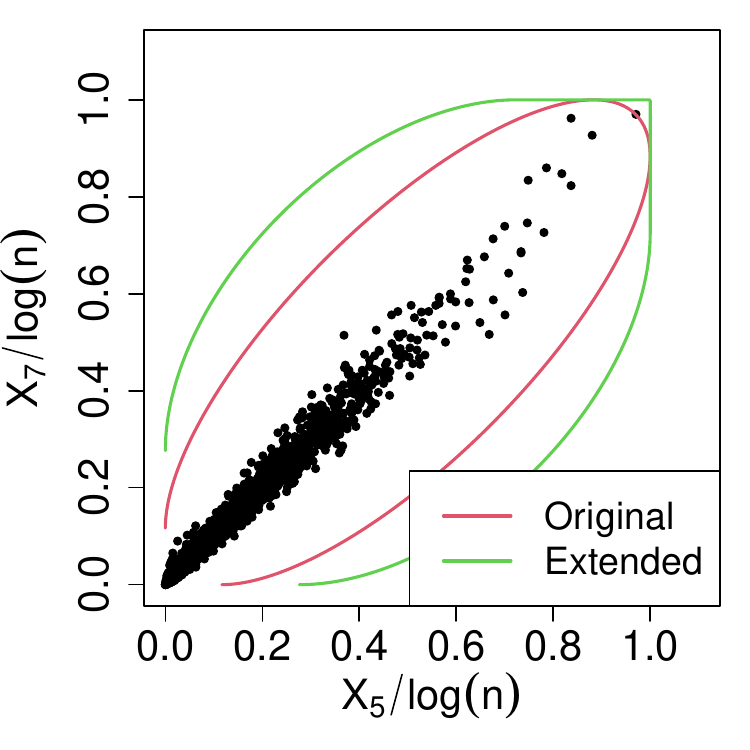}
    \caption{Unit level sets of the bivariate projections of the graphical gauge function with Gaussian gauge function used for all pairs (red line) and with exponential-Gaussian used for pairs $(3, 7)$, $(4, 7)$ and $(5, 7)$ with corresponding scaled sample clouds (black points).}
    \label{fig:bivariate_projections_extended_model}
\end{figure}

The effect of the parameter estimates on the shape of the sample cloud can be seen through the bivariate projections. Figure~\ref{fig:bivariate_projections_extended_model} shows the comparison between the bivariate projections of the graphical gauge function for the original model with Gaussian gauge function used for all flow-connected pairs, and the extended model with exponential-Gaussian used for pairs $(3, 7)$, $(4, 7)$, and $(5, 7)$.
Projections from both models for pair $(3, 7)$ describe the shape of the scaled sample cloud well. The projection from the extended model for pair $(4, 7)$ is significantly influenced by the exponential, which is also reflected in the high $\gamma$ parameter in Table~\ref{tab:parameter_estimates_exp-Gauss}. The projection from the extended model for pair $(5, 7)$ is a lot wider than for the original model. In this case, the Gaussian gauge function appears to describe the shape of the scaled sample cloud better.

\begin{figure}[t!]
    \centering
    \includegraphics[width=0.4\linewidth]{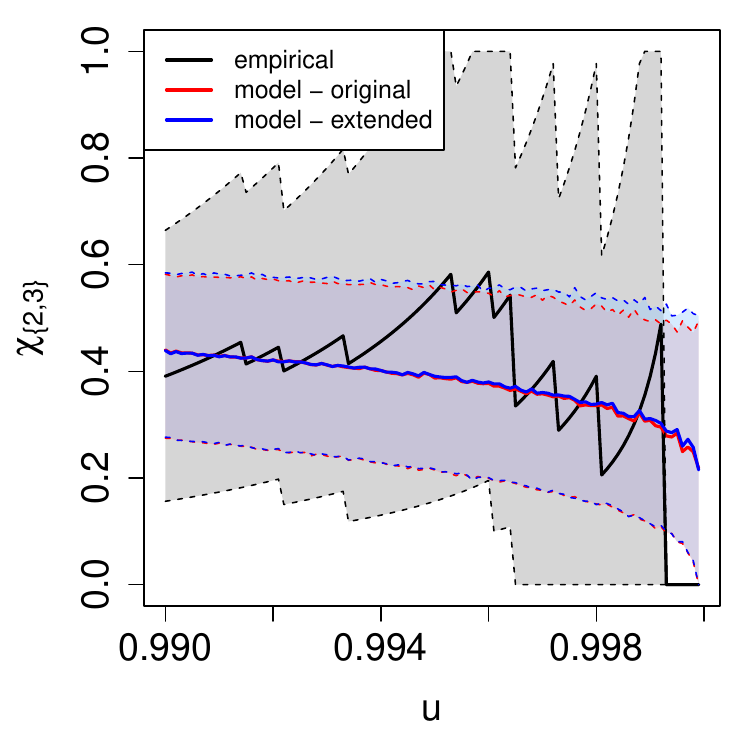}
    \includegraphics[width=0.4\linewidth]{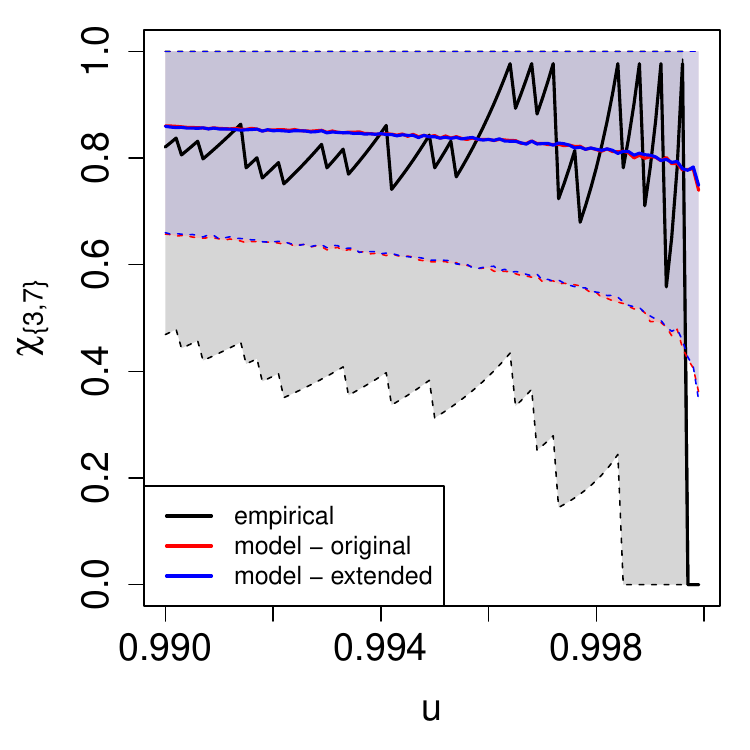}
    \caption{Plot of empirical (solid black) and model-based $\chi_S(u)$ for original (solid red) and extended (solid blue) model with empirical (grey region) and model-based for original (light red region) and for extended (light blue region) 95\% CIs for flow-unconnected $ S = \{2,3\}$ (left) and an adjacent flow-connected $S = \{3,7\}$ pair (right).}
    \label{fig:chi_plot_bivariate_plots_example_comparison}
\end{figure}

Figure~\ref{fig:chi_plot_bivariate_plots_example_comparison} compares the model-based $\chi_S(u)$ for the original and extended model for $u \rightarrow 1$ for a flow-unconnected and an adjacent flow-connected pair. Additional $\chi_S(u)$ plots can be found in Figure~S.10 in Appendix~C.2.3. The model-based $\chi_S(u)$ for both models are very close to each other. The 95\% CIs are wider for the extended model in some cases, for example for flow-connected pairs $(4,8)$ and $(3,4)$. This is likely due to larger parameter space of the extended model, and its ability to capture both joint and separate extremes.

The annual probability of a flood at gauging station (3), (4), (5) and (7) was also calculated using this extended model to see whether the fact that the exponential-Gaussian gauge function used for the pairs $(3,7)$, $(4,7)$ and $(5,7)$ would have an effect on this probability. The exceedance probability per event with 95\% CIs is $2.7 [1.3, 3.7]\cdot 10^{-6}$, which gives an annual exceedance probability of $1.6 [0.8, 2.2]\cdot 10^{-4}$. Interestingly, the probability from the original fit, and the upper CI, are slightly lower than in the original model.

\section{Discussion}
\label{sec:discussion}
In this work, we implemented the first statistical model using the geometric extremal graphical modelling framework introduced by \cite{papastathopoulos_geometric_2026}. We fitted this graphical model to river flow data from 10 gauging stations on a river network near Preston, in the north west of England. This allows extrapolation and estimation of flood risk at 10 locations, which has previously only been done within the geometric framework in a spatial setting or using deep-learning methods. Through numerous diagnostics, the model was found to perform very well. Flood risk estimation in the whole region of space is possible using just one model fit, which is an advantage compared to the conditional extremes approach, where different models each conditioning on a different station have to be fitted and used to calculate probabilities for different regions of space. While in theory probabilities for any combinations of stations could be calculated, in practice river flows corresponding to bankfull could not be obtained for all gauging stations on this river network. This was due to all observations being far from bankfull and the statistical relationship between river height and river flow not being valid for the values near bankfull.

For additional flexibility, an extension to the model was proposed, that allows the adjacent flow-connected pairs to experience the most extreme river flows jointly or separately. This is determined based on the estimate of the parameter and does not have to be assumed a priori, whereas the extremal graphical model of \cite{engelke_graphical_2020} assumes full asymptotic dependence between all stations. However, the differences in extrapolation between the original and extended model can only be seen very far in the tail, beyond the levels that we are interested in. For that reason, the simpler model was preferred in this case.

If the sample cloud of the pairs was more complex, more flexible shapes might be required. In that case, other additively mixed gauge functions \citep{wadsworth_statistical_2024,lee_geometric_2025} or piecewise-linear gauge functions \citep{campbell_piecewise-linear_2024} could be explored within the graphical gauge function.

For estimating a suitable threshold, we used a kernel density estimation based technique. In higher dimensions, other methods will likely be required. It may be possible in this case to also exploit a graphical structure, since the gauge function has an approximate inverse relationship with high quantiles of $R|\{\bm{W}=\bm{w}\}$ \citep{wadsworth_statistical_2024}.

Finally, the simulation of points above the high threshold $r_0(\bm{w})$ could be improved by introducing an angular distribution rather than resampling the angles from the empirical distribution. This would resolve the problem of the trails in the data illustrated in the left hand side plot in Figure~\ref{fig:simulated_points_2_3_and_3_7_original_scale}. This, however, did not seem to impede our analysis.

\section*{Funding}
This paper is based on work completed while KG was part of the EPSRC funded STOR-i Centre for Doctoral Training EP/S022252/1.
JW gratefully acknowledges funding from EPSRC grant EP/X010449/1.
TLT acknowledges the support of the Research Council of Norway through its Centre of Excellence Integreat – The Norwegian Centre for Knowledge-driven Machine Learning, project number 332645.

\section*{Data and code availability}
Data underpinning this publication are openly available from NRFA at \url{https://nrfa.ceh.ac.uk/}.
A sample code to run and reproduce the results can be found at \url{https://github.com/grolmus-k/flood_risk_estimation_via_geometric_extremal_graphical_modelling.git}.

\section*{Supplementary Material}

\subsection*{A: River flow data and pre-processing}
Additional details of the matching of the peaks and seasonality.

\subsection*{B: Histograms of parameter estimates}
Investigation of parameter estimates from bootstrap samples.

\subsection*{C: Diagnostics}
Additional bivariate projections and $\chi_S(u)$ plots.

\subsection*{D: Event set generation for calculating probabilities}
Description of event set generation for flood risk estimation calculation.

\bibliography{bibliography}

@article{wadsworth_statistical_2024,
	title        = {Statistical inference for multivariate extremes via a geometric approach},
	author       = {Wadsworth, Jennifer L and Campbell, Ryan},
	year         = 2024,
	month        = nov,
	journal      = {Journal of the Royal Statistical Society Series B: Statistical Methodology},
	volume       = 86,
	number       = 5,
	pages        = {1243--1265},
	urldate      = {2024-12-06},
}

@article{papastathopoulos_geometric_2026,
  title = {Geometric extremal graphical models and coefficients of extremal dependence on block graphs},
  author = {Papastathopoulos, Ioannis and Wadsworth, Jennifer},
  journal   = {arXiv preprint arXiv:2601.00239},
  year      = {2026},
  archivePrefix = {arXiv},
  primaryClass  = {cs.LG},
}

@article{asadi_extremes_2015,
	title        = {Extremes on river networks},
	author       = {Asadi, Peiman and Davison, Anthony C. and Engelke, Sebastian},
	year         = 2015,
	month        = dec,
	journal      = {The Annals of Applied Statistics},
	volume       = 9,
	number       = 4,
}

@misc{NRFA,
  author       = {{UK Centre for Ecology \& Hydrology}},
  title        = {{National River Flow Archive (NRFA)}},
  year         = {2025},
  howpublished = {{https://nrfa.ceh.ac.uk/}},
  note         = {[Accessed: 11-10-2024]}
}

@article{engelke_graphical_2020,
	title        = {Graphical {models} for {extremes}},
	author       = {Engelke, Sebastian and Hitz, Adrien S.},
	year         = 2020,
	month        = sep,
	journal      = {Journal of the Royal Statistical Society Series B: Statistical Methodology},
	volume       = 82,
	number       = 4,
	pages        = {871--932},
}

@article{coles_modelling_1991,
	title        = {Modelling {extreme} {multivariate} {events}},
	author       = {Coles, Stuart G. and Tawn, Jonathan A.},
	year         = 1991,
	journal      = {Journal of the Royal Statistical Society. Series B (Methodological)},
	volume       = 53,
	number       = 2,
	pages        = {377--392},
}

@article{campbell_piecewise-linear_2024,
  title = {Piecewise-linear modeling of multivariate geometric extremes},
  author = {Campbell, Ryan and Wadsworth, Jennifer},
  journal   = {arXiv preprint arXiv:2412.05195},
  year      = {2024},
  archivePrefix = {arXiv},
  primaryClass  = {cs.LG}
}

@article{farrell_conditional_2024,
  title = {Conditional extremes with graphical models},
  author = {Farrell, Aiden and Eastoe, Emma F. and Lee, Clement},
  journal   = {arXiv preprint arXiv:2411.17013},
  year      = {2024},
  archivePrefix = {arXiv},
  primaryClass  = {cs.LG}
}

@article{nolde_linking_2022,
	title        = {Linking representations for multivariate extremes via a limit set},
	author       = {Nolde, Natalia and Wadsworth, Jennifer L.},
	year         = {2022},
	month        = sep,
	journal      = {Advances in Applied Probability},
	volume       = 54,
	number       = 3,
	pages        = {688--717},
}

@article{lee_geometric_2025,
  title = {Geometric criteria for identifying extremal dependence and flexible modeling via additive mixtures},
  author = {Lee, J. and Wadsworth, J. L.},
  journal   = {arXiv preprint arXiv:2512.24392},
   year = {2025},
  archivePrefix = {arXiv},
  primaryClass  = {cs.LG}
}

@misc{UKHSA2023HECC,
  author       = {{UK Health Security Agency}},
  title        = {{Health Effects of Climate Change (HECC) in the UK: 2023 report}},
  institution  = {UK Health Security Agency},
  year         = {2023},
  howpublished          = {https://www.gov.uk/government/publications/climate-change-health-effects-in-the-uk},
  note         = {[Accessed: 08-09-2025]}
}

@misc{VesuvianoGriffin2025,
  author       = {Vesuviano, G. and Griffin, A.},
  title        = {The {FEH} 2025 statistical method update},
  institution  = {UK Centre for Ecology \& Hydrology},
  address      = {Wallingford, UK},
  year         = {2025},
  howpublished = {https://www.ceh.ac.uk/services/feh-2025-statistical-updates},
  note         = {[Accessed: 08-09-2025]}
}

@book{FEH1999,
  author      = {{Institute of Hydrology}},
  title       = {Flood Estimation Handbook},
  year        = {1999},
  publisher   = {Institute of Hydrology},
  address     = {Wallingford, UK},
  howpublished         = {https://www.ceh.ac.uk/data/software-models/flood-estimation-handbook},
}

@misc{kjeldsen2008,
  author       = {Kjeldsen, T. R. and Jones, D. A. and Bayliss, A. C.},
  title        = {Improving the {FEH} statistical procedures for flood frequency estimation: Science Report: SC050050},
  institution  = {Joint Defra / Environment Agency Flood and Coastal Erosion Risk Management R\&D Programme},
  publisher    = {European Environment Agency (EEA)},
  address      = {Bristol, UK},
  year         = {2008},
  pages        = {137},
  project      = {Science Report SC050050},
  howpublished = {https://nora.nerc.ac.uk/id/eprint/3545/},
  note         = {[Accessed: 22-08-2025]}
}

@article{heffernan_conditional_2004,
    title        = {A {conditional} {approach} for {multivariate} {extreme} {values} (with {discussion})},
	author       = {Heffernan, Janet E. and Tawn, Jonathan A.},
	year         = 2004,
	month        = aug,
	journal      = {Journal of the Royal Statistical Society Series B: Statistical Methodology},
	volume       = 66,
	number       = 3,
	pages        = {497--546},
}

@article{majumder_semiparametric_2025,
   title={Semiparametric estimation of the shape of the limiting bivariate point cloud},
   volume={1},
   number={1},
   journal={Bayesian Analysis},
   publisher={Institute of Mathematical Statistics},
   author={Majumder, Reetam and Shaby, Benjamin A. and Reich, Brian J. and Cooley, Daniel S.},
   year={2025},
   month=jan,
}

@article{simpson_estimating_2024,
    author = {Emma S. Simpson and Jonathan A. Tawn},
    title = {{Estimating the limiting shape of bivariate scaled sample clouds: With additional benefits of self-consistent inference for existing extremal dependence properties}},
    volume = {18},
    journal = {Electronic Journal of Statistics},
    number = {2},
    publisher = {Institute of Mathematical Statistics and Bernoulli Society},
    pages = {4582 -- 4611},
    year = {2024},
    howpublished = {https://doi.org/10.1214/24-EJS2300}
}

@article{murphy-barltrop_deep_2024,
  title = {Deep learning of multivariate extremes via a geometric representation},
  author = {Murphy-Barltrop, Callum J. R. and Majumder, Reetam and Richards, Jordan},
  journal   = {arXiv preprint arXiv:2406.19936},
  archivePrefix = {arXiv},
  primaryClass  = {cs.LG}
}

@article{de_monte_generative_2025,
  title = {Generative modelling of multivariate geometric extremes using normalising flows},
 author = {De Monte, Lambert and Huser, Raphaël and Papastathopoulos, Ioannis and Richards, Jordan},
  journal   = {arXiv preprint arXiv:2505.02957},
   year = {2025},
  archivePrefix = {arXiv},
  primaryClass  = {cs.LG}
}

@article{campbell_analysing_2026,
  title = {Analysing extreme rainfall via a geometric framework},
  author = {Campbell, Ryan and Grolmusova, Kristina and Kakampakou, Lydia and Lee, Jeongjin},
  journal   = {arXiv preprint arXiv:2603.18149},
   year = {2026},
  archivePrefix = {arXiv},
  primaryClass  = {cs.LG}
}

@article{keef_estimating_prob_2013,
	title        = {Estimating the probability of widespread flood events},
	author       = {Keef, Caroline and Tawn, Jonathan A. and Lamb, Rob},
	year         = 2013,
	month        = feb,
	journal      = {Environmetrics},
	volume       = 24,
	number       = 1,
	pages        = {13--21},
}

@article{kakampakou_geometric_2025,
  title     = {Geometric modelling of spatial extremes},
  author = {Kakampakou, Lydia and Wadsworth, Jennifer L.},
  journal   = {arXiv preprint arXiv:2511.08192},
  year      = {2025},
  archivePrefix = {arXiv},
  primaryClass  = {cs.LG}
}

@article{huser_modeling_2019,
    author = {Huser, Raphaël G. and Wadsworth, Jennifer L.},
    year = {2019},
    title = {Modeling spatial processes with unknown extremal dependence class},
    volume = {114},
    number = {525},
    pages = {434-444},
    journal = {{Journal of the American Statistical Association}},
}

@article{wadsworth_higher-dimensional_2022,
	title = {Higher-dimensional spatial extremes via single-site conditioning},
	volume = {51},
	pages = {100677},
	journaltitle = {Spatial Statistics},
	journal = {Spatial Statistics},
	author = {Wadsworth, J. L. and Tawn, J. A.},
	urldate = {2026-01-23},
    year = {2022},
	date = {2022-10},
}

@book{lauritzen_graphical_1996,
    author = {Lauritzen, Steffen L.},
    booktitle = {Graphical models},
    isbn = {0198522193},
    address = {Oxford},
    publisher = {Oxford University Press},
    title = {Graphical models},
    year = {1996},
}

@article{towe_modelbased_2019,
	title = {Model‐based inference of conditional extreme value distributions with hydrological applications},
	volume = {30},
	number = {8},
	urldate = {2026-03-03},
	journal = {Environmetrics},
	author = {Towe, R. P. and Tawn, J. A. and Lamb, R. and Sherlock, C. G.},
	month = dec,
	year = {2019},
	pages = {e2575},
}

@article{keef_spatial_dependence_2009,
	title = {Spatial dependence in extreme river flows and precipitation for {Great} {Britain}},
	volume = {378},
	number = {3-4},
	journal = {Journal of Hydrology},
	author = {Keef, Caroline and Svensson, Cecilia and Tawn, Jonathan A.},
	month = nov,
	year = {2009},
	pages = {240--252},
}

@article{keef_spatial_risk_2009,
	title = {Spatial risk assessment for extreme river flows},
	volume = {58},
	pages = {601--618},
	number = {5},
	journal = {Journal of the Royal Statistical Society Series C: Applied Statistics},
	author = {Keef, Caroline and Tawn, Jonathan and Svensson, Cecilia},
    year = 2009,
}

@article{rentschler_flood_2022,
	title = {Flood exposure and poverty in 188 countries},
	volume = {13},
	issn = {2041-1723},
	pages = {3527},
	number = {1},
	journaltitle = {Nature Communications},
	journal = {Nat Commun},
	author = {Rentschler, Jun and Salhab, Melda and Jafino, Bramka Arga},
	urldate = {2026-04-27},
	date = {2022-06-28},
    year = 2022,
}

@article{schneeberger_generation_2018,
	title = {Generation of spatially heterogeneous flood events in an {A}lpine region — {A}daptation and application of a multivariate modelling procedure},
	volume = {5},
	pages = {5},
	number = {1},
	journaltitle = {Hydrology},
	journal = {Hydrology},
	author = {Schneeberger, Klaus and Steinberger, Thomas},
	urldate = {2026-05-05},
	date = {2018-01-04},
    year = {2018},
}

@article{olcese_developing_2024,
	title = {Developing a fluvial and pluvial stochastic flood model of {S}outheast {A}sia},
	volume = {60},
	pages = {e2023WR036580},
	number = {6},
	journaltitle = {Water Resources Research},
	journal = {Water Resources Research},
	author = {Olcese, Gaia and Bates, Paul D. and Neal, Jeffrey C. and Sampson, Christopher C. and Wing, Oliver E. J. and Quinn, Niall and Murphy‐Barltrop, Callum J. R. and Probyn, Izzy},
    year = {2024},
}

@article{quinn_spatial_2019,
	title = {The spatial dependence of flood hazard and risk in the {U}nited {S}tates},
	volume = {55},
	pages = {1890--1911},
	number = {3},
	journaltitle = {Water Resources Research},
	journal = {Water Resources Research},
	author = {Quinn, Niall and Bates, Paul D. and Neal, Jeff and Smith, Andy and Wing, Oliver and Sampson, Chris and Smith, James and Heffernan, Janet},
    year = {2019},
}

@article{olcese_use_2022,
	title = {Use of hydrological models in global stochastic flood modeling},
	volume = {58},
	pages = {e2022WR032743},
	number = {12},
	journaltitle = {Water Resources Research},
	journal = {Water Resources Research},
	author = {Olcese, Gaia and Bates, Paul D. and Neal, Jeffrey C. and Sampson, Christopher C. and Wing, Oliver E. J. and Quinn, Niall and Beck, Hylke E.},
    year = {2022},
}

@article{wing_toward_2020,
	title = {Toward global stochastic river flood modeling},
	volume = {56},
	pages = {e2020WR027692},
	number = {8},
	journaltitle = {Water Resources Research},
	journal = {Water Resources Research},
	author = {Wing, Oliver E. J. and Quinn, Niall and Bates, Paul D. and Neal, Jeffrey C. and Smith, Andrew M. and Sampson, Christopher C. and Coxon, Gemma and Yamazaki, Dai and Sutanudjaja, Edwin H. and Alfieri, Lorenzo},
    year = {2020},
}

@article{kakampakou_spatial_2024,
	title = {Spatial extremal modelling: {A} case study on the interplay between margins and dependence},
	volume = {13},
	shorttitle = {Spatial Extremal Modelling},
	pages = {e70021},
	number = {4},
	journaltitle = {Stat},
	journal = {Stat},
	author = {Kakampakou, Lydia and Simpson, Emma S. and Wadsworth, Jennifer L.},
    year = {2024},
}
\bibliographystyle{dcu}

\pagebreak
\appendix

\end{document}


\maketitle

\appendix
\renewcommand{\thesection}{\Alph{section}}
\setcounter{section}{0}
\renewcommand{\thefigure}{S.\arabic{figure}}
\setcounter{figure}{0}
\renewcommand{\theequation}{S.\arabic{equation}}
\setcounter{equation}{0}
\renewcommand{\thetable}{S.\arabic{table}}
\setcounter{table}{0}

\section{River flow data and pre-processing}
\subsection{Matching the peak flows}
\label{sec:supp:matching_the_peak_flows}
Here we explain the procedure of \cite{asadi_extremes_2015} in detail. Our dataset consists of 10 time series corresponding to the GDF at all of the considered gauging stations. This dataset will be referred to as the original dataset and will not be modified during the pre-processing. 
Each event within an individual time series is ranked based on the value of the river flow, where the highest river flow has the highest rank. This results in a new dataset, which contains the 10 time series of river flows as well as 10 time series of ranks. This dataset will be referred to as the ranked dataset and will be modified during the pre-processing.
The highest ranked event is chosen from the ranked dataset. If there are multiple events, the earliest one is chosen. A window of $\pm p$ days is constructed around the event. Within this time window, the highest river flows are identified for each of the time series of river flows in the original dataset. The highest ranked event is modified such that for each gauging station the highest river flow from the time window is considered as the river flow for the highest ranked event. This event is added to a new dataset, which will be referred to as the matched dataset. A window of $\pm 2p$ around the highest ranked event is deleted in the ranked dataset and new highest ranked event is selected. This process continues until there are no events left in the ranked dataset. The matched dataset is the new dataset of unique flood events.

\subsection{Seasonality}
\label{sec:supp:seasonality}
\begin{figure}[H]
    \centering
    \includegraphics[width=0.49\linewidth]{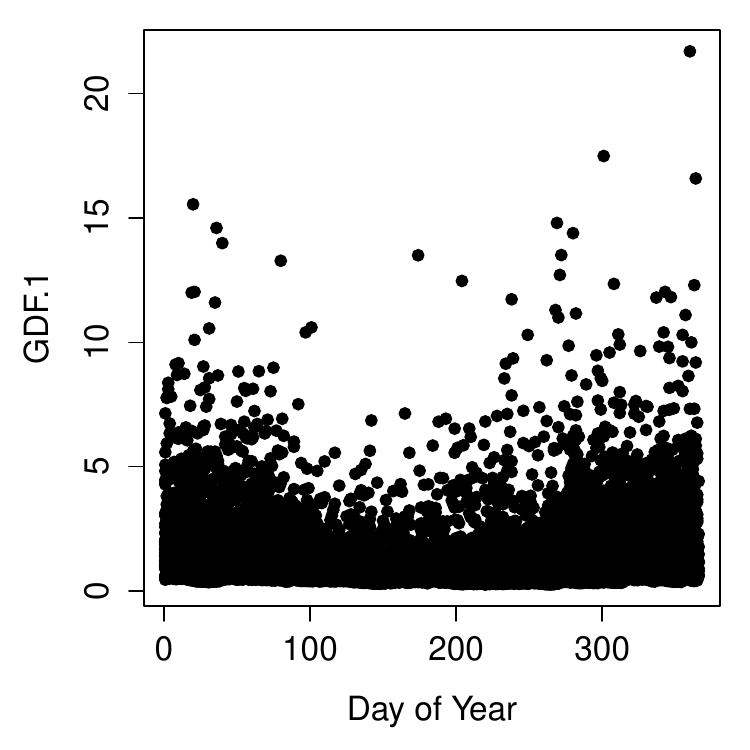}
    \includegraphics[width=0.49\linewidth]{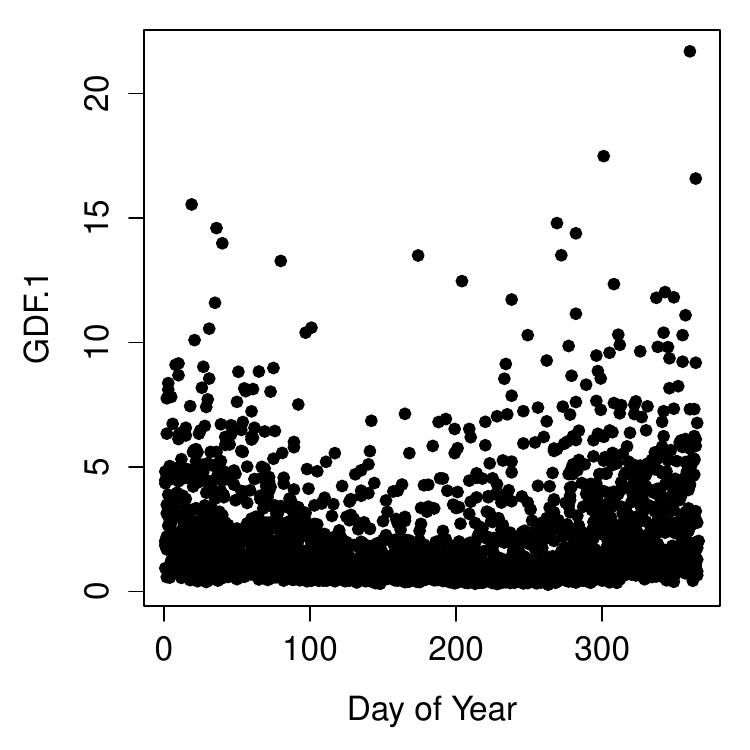}
    \caption{Plot of GDF against the day of the year using original (left) and matched data (right) for station (1).}
    \label{fig:supp:seasonality}
\end{figure}

Figure~\ref{fig:supp:seasonality} shows an example plot of GDF against the day of the year using original and matched data for gauging station (1). There appears to be a seasonal pattern in the data with higher and lower river flows in the winter and summer months, respectively. We have considered addressing this seasonality in two different ways: splitting the dataset into summer and winter seasons which has been done for example by \cite{schneeberger_generation_2018}, and performing the analysis on these separately, or modelling the marginal non-stationarity using methods discussed by \cite{kakampakou_spatial_2024}, for example. However, splitting the dataset into summer and winter would leave very few data points (around 100) for fitting of the extreme value model increasing the uncertainty of the model fit.
When it comes to modelling the marginal non-stationarity, many approaches rely on pre-processing the data at each gauging station to remove seasonal patterns and/or trends. This results in the summer values being more extreme after pre-processing and therefore more summer values contributing to the fitting of the model. In flood risk estimation, we are particularly interested in river flows that are above a certain high value, which does not change depending on the season. Many of the extra summer extreme values will not be close to this value in the original dataset making them less interesting from flooding perspective. Finally, \cite{kakampakou_spatial_2024} also illustrated that decoupling dependence and margins is a non-trivial task, and different pre-processing methods for marginal non-stationarity can lead to different conclusions from the model.

\newpage
\section{Histograms of parameter estimates}

\subsection{Original model}
\label{sec:supp:histograms_param_estimates_original_model}
\begin{figure}[H]
    \centering
    \includegraphics[width=0.32\linewidth]{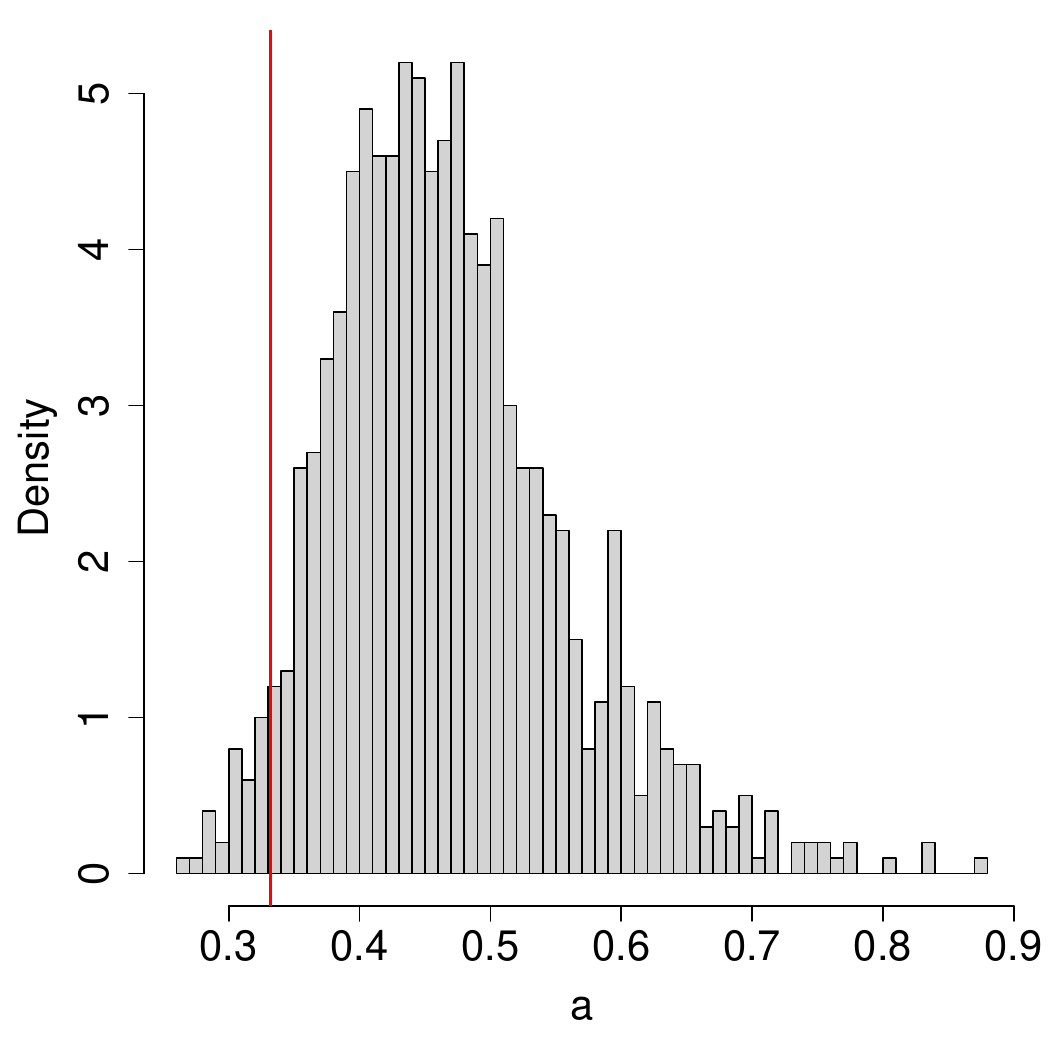}\\
    \includegraphics[width=0.32\linewidth]{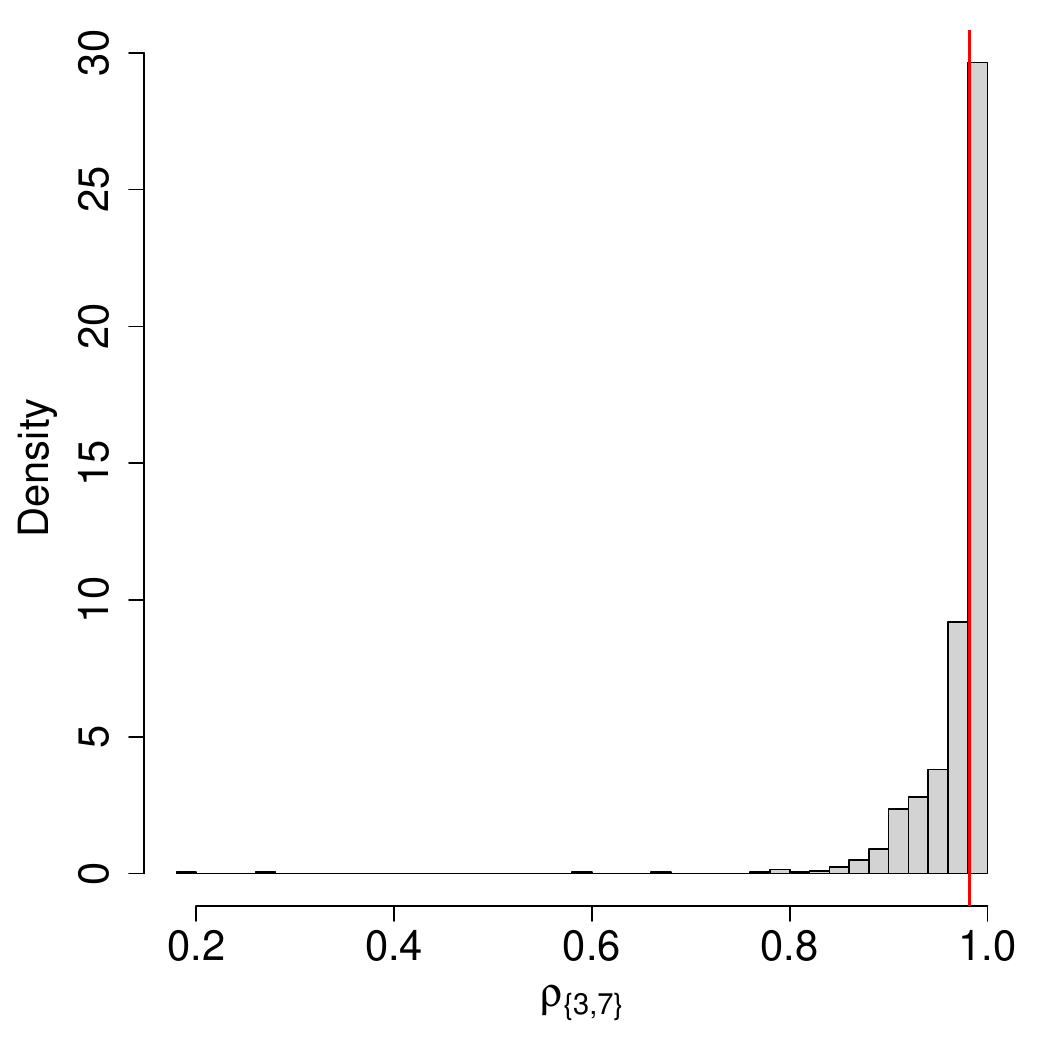}
    \includegraphics[width=0.32\linewidth]{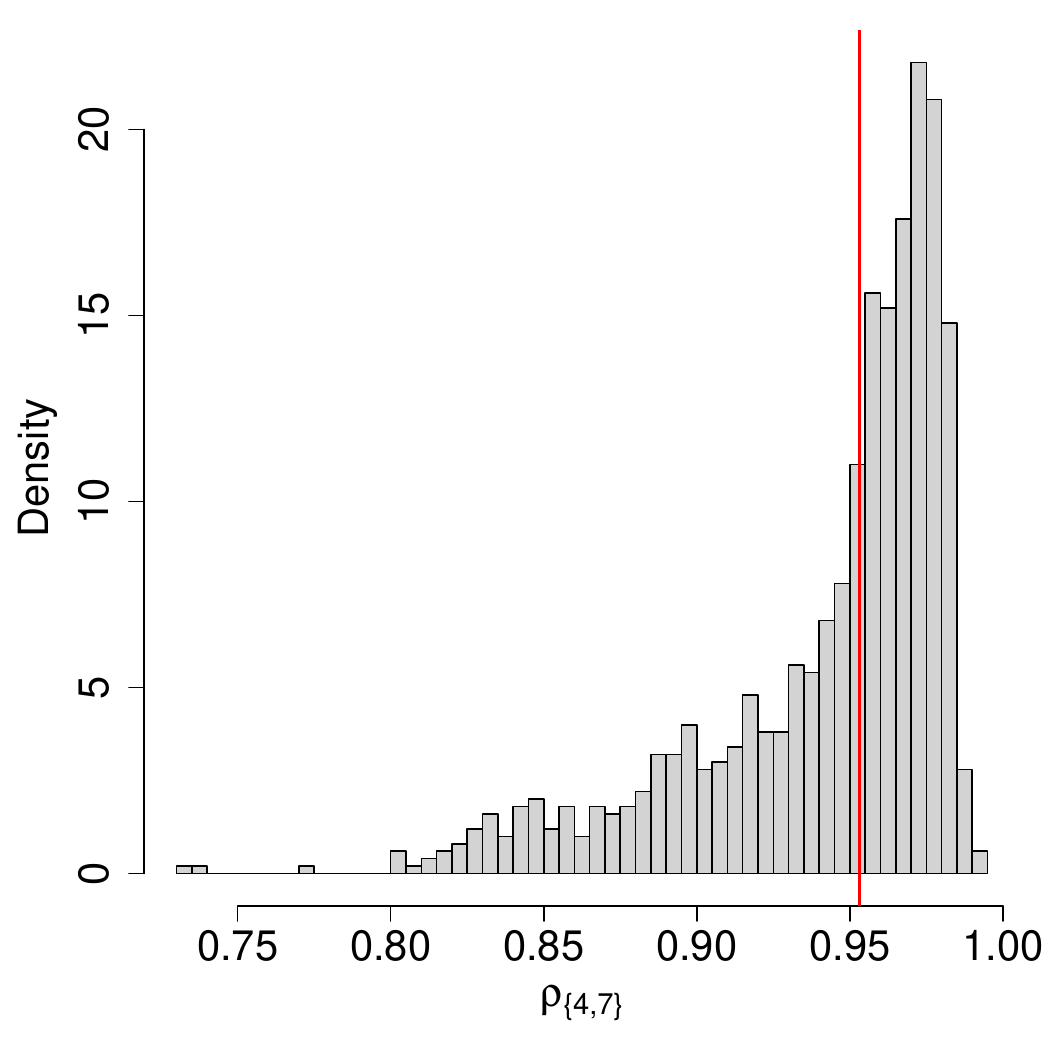}
    \includegraphics[width=0.32\linewidth]{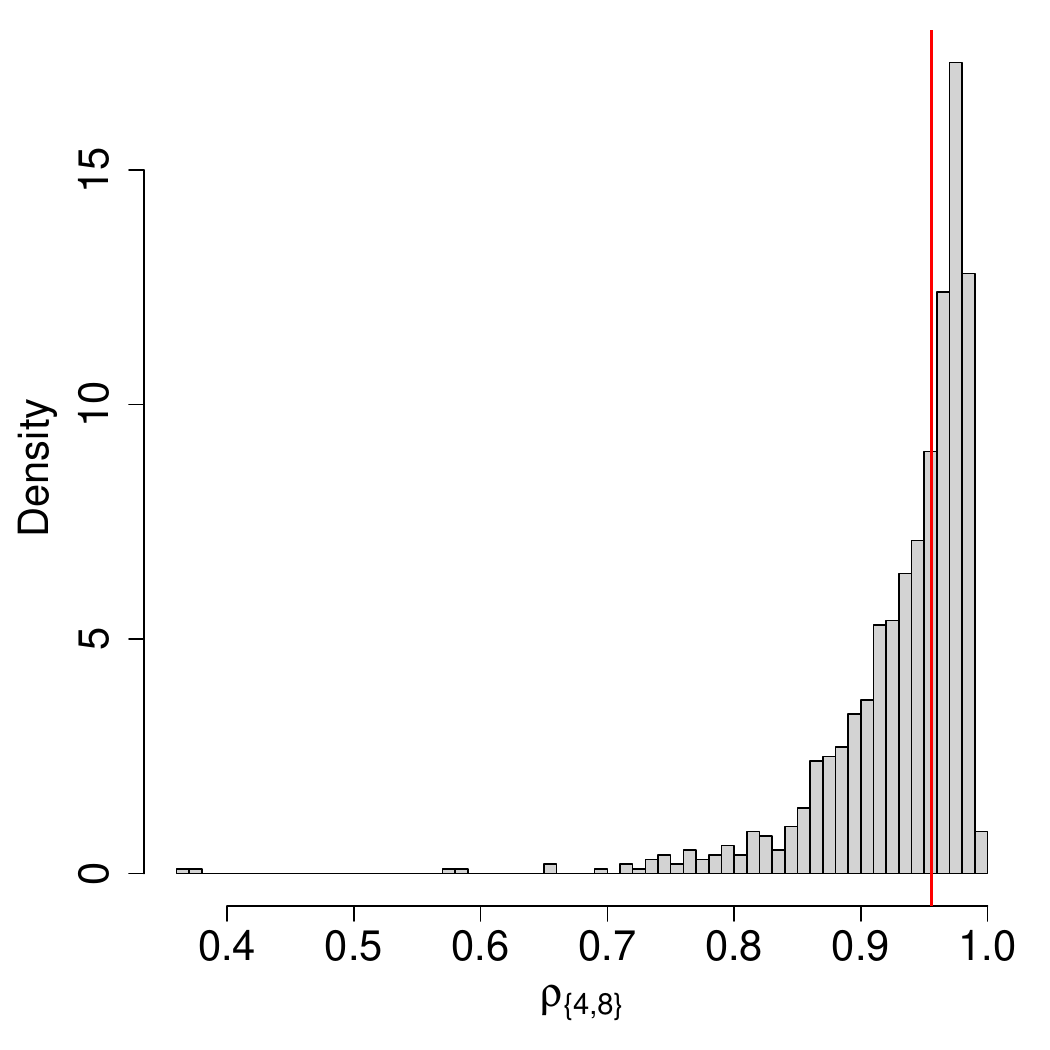}
    \includegraphics[width=0.32\linewidth]{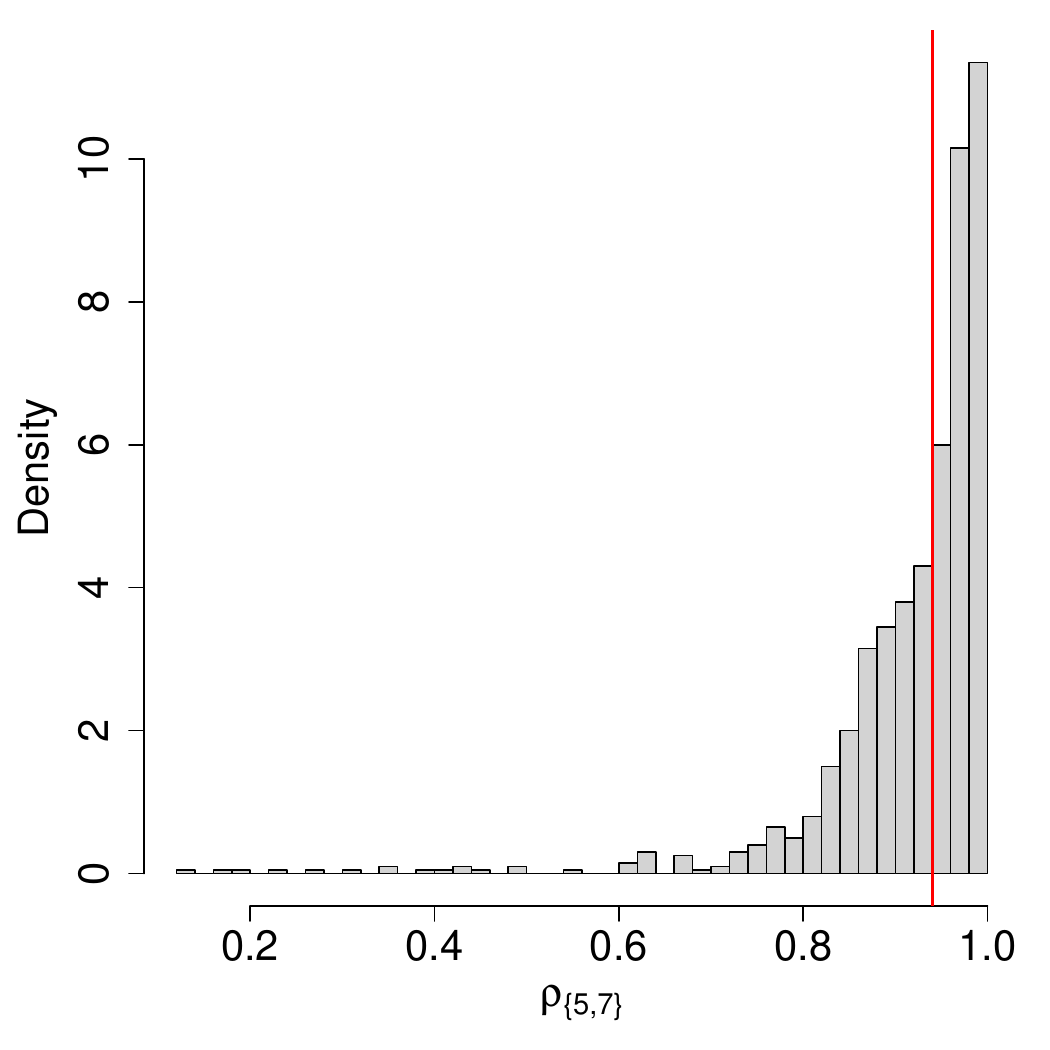}
    \includegraphics[width=0.32\linewidth]{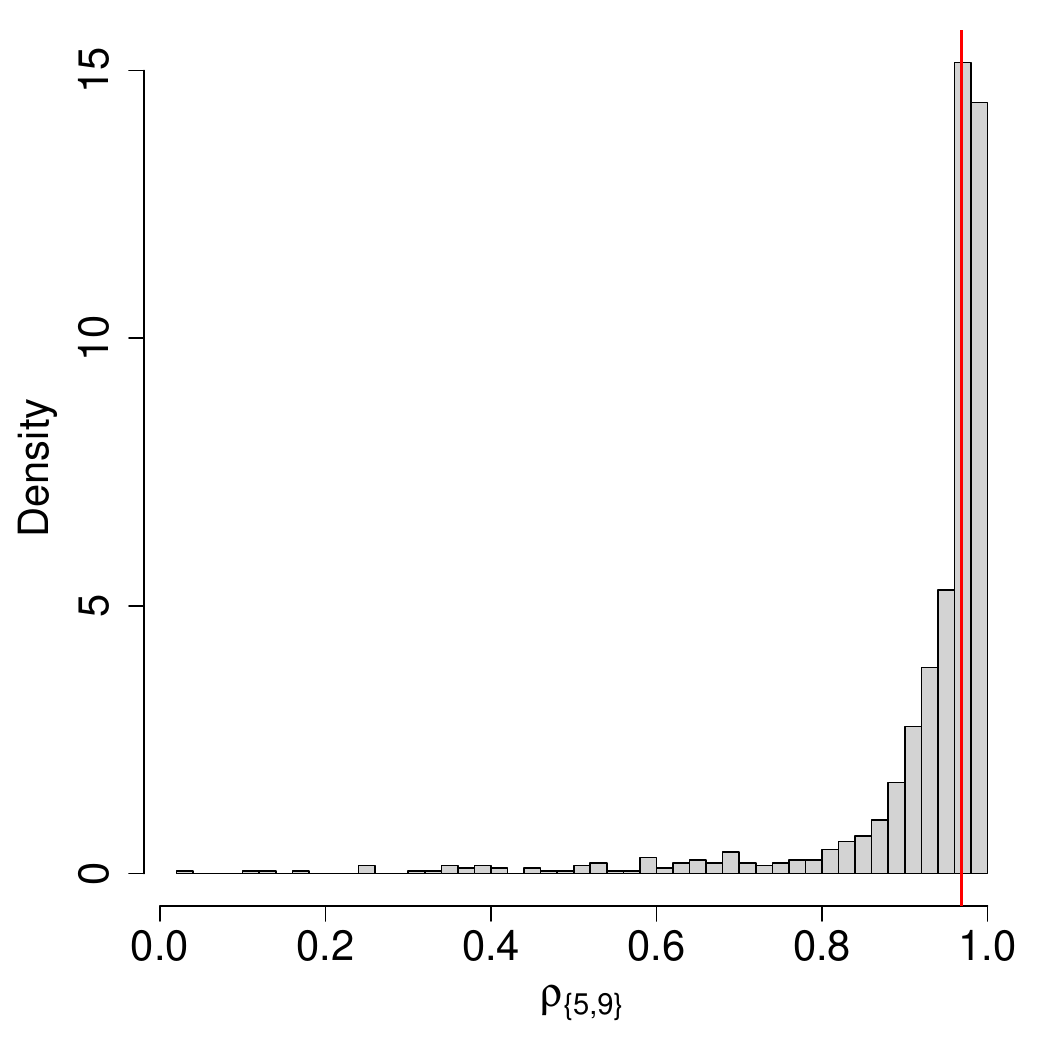}
    \includegraphics[width=0.32\linewidth]{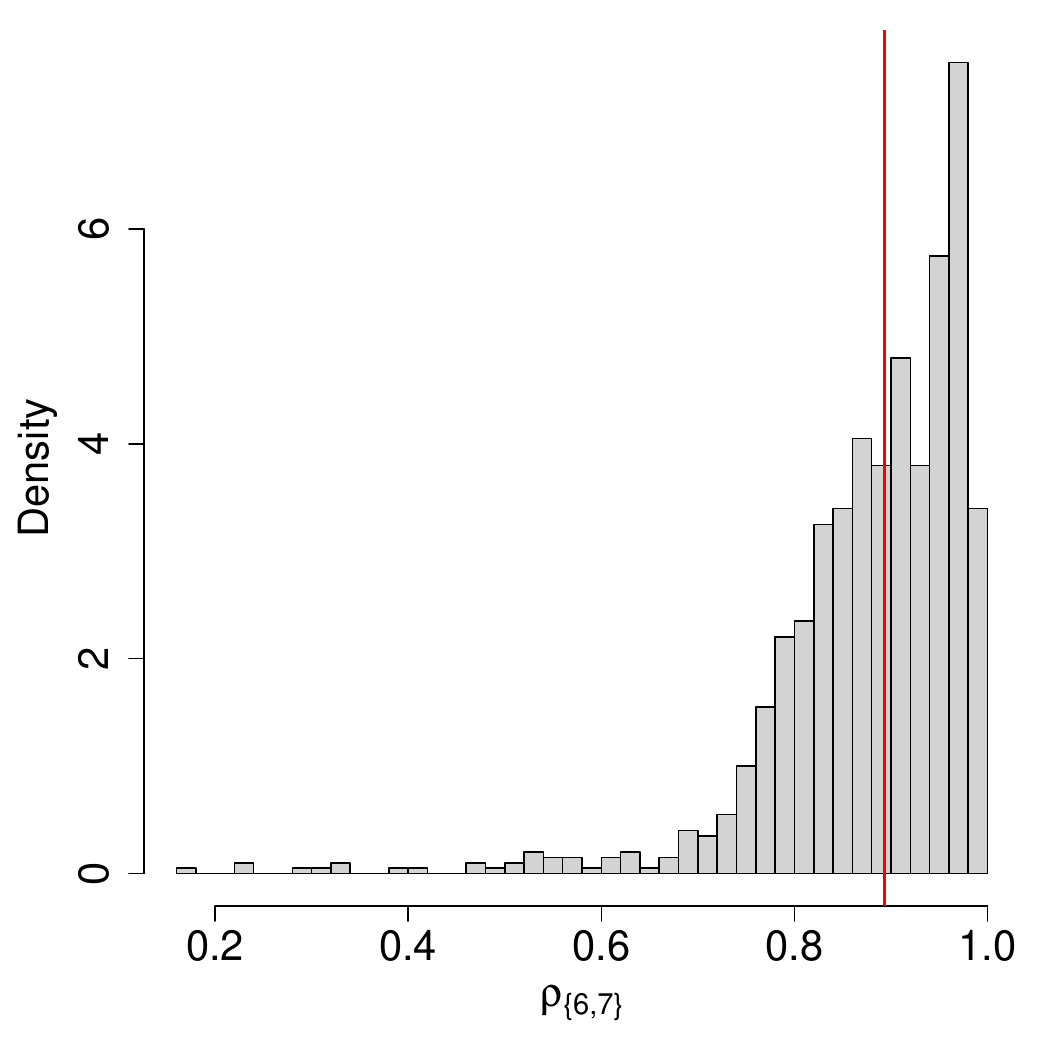}
    \caption{Histograms of $a$ and $\rho_{\{i,j\}}$ for adjacent flow-connected pairs $(i,j)$ from block-bootstrap samples of the fit from the truncated gamma model with the graphical gauge function for the Preston river network, with Gaussian gauge function used for all adjacent flow-connected pairs. Vertical red line shows the maximum likelihood estimate obtained using the matched dataset.}
    \label{fig:histograms_param_estimates_original_model_a_rho}
\end{figure}

\begin{figure}[H]
    \centering
    \includegraphics[width=0.32\linewidth]{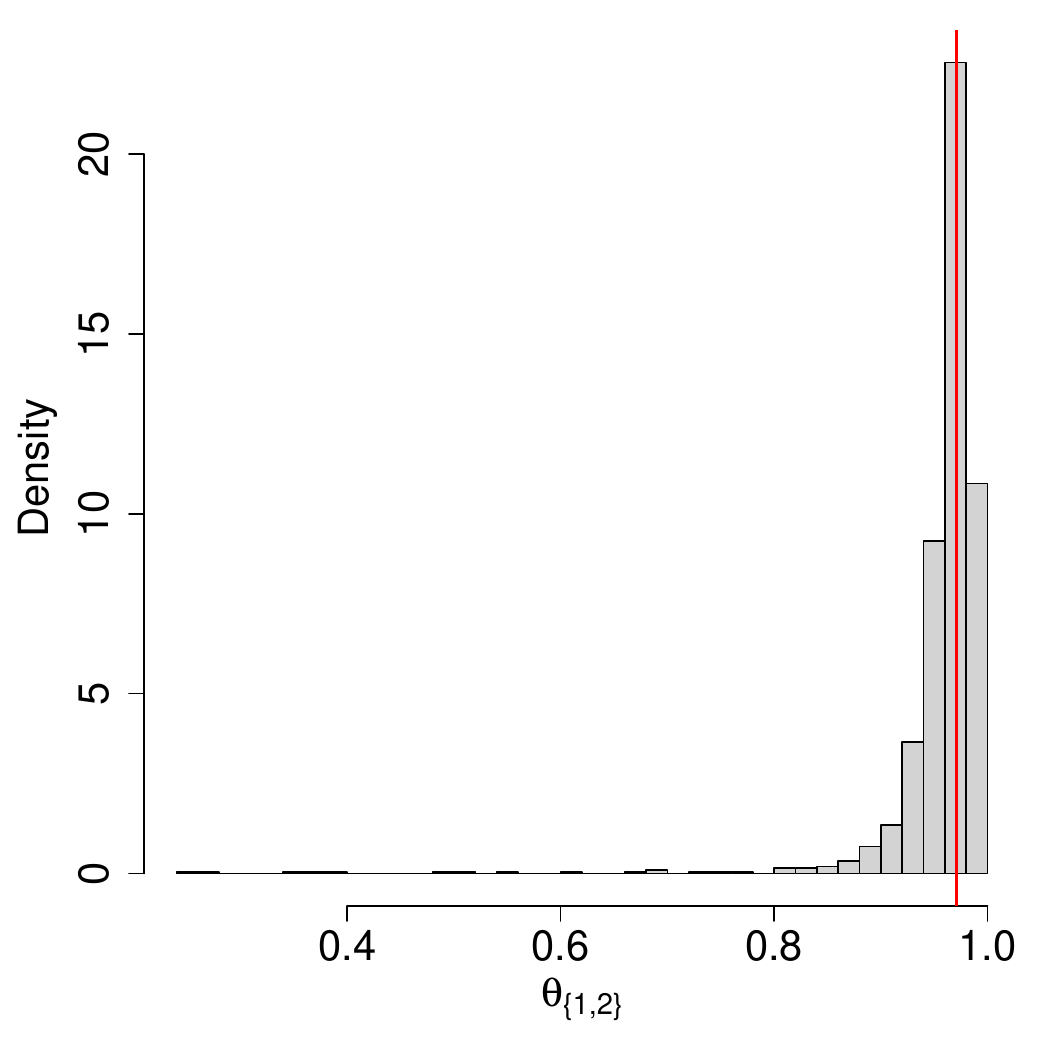}
    \includegraphics[width=0.32\linewidth]{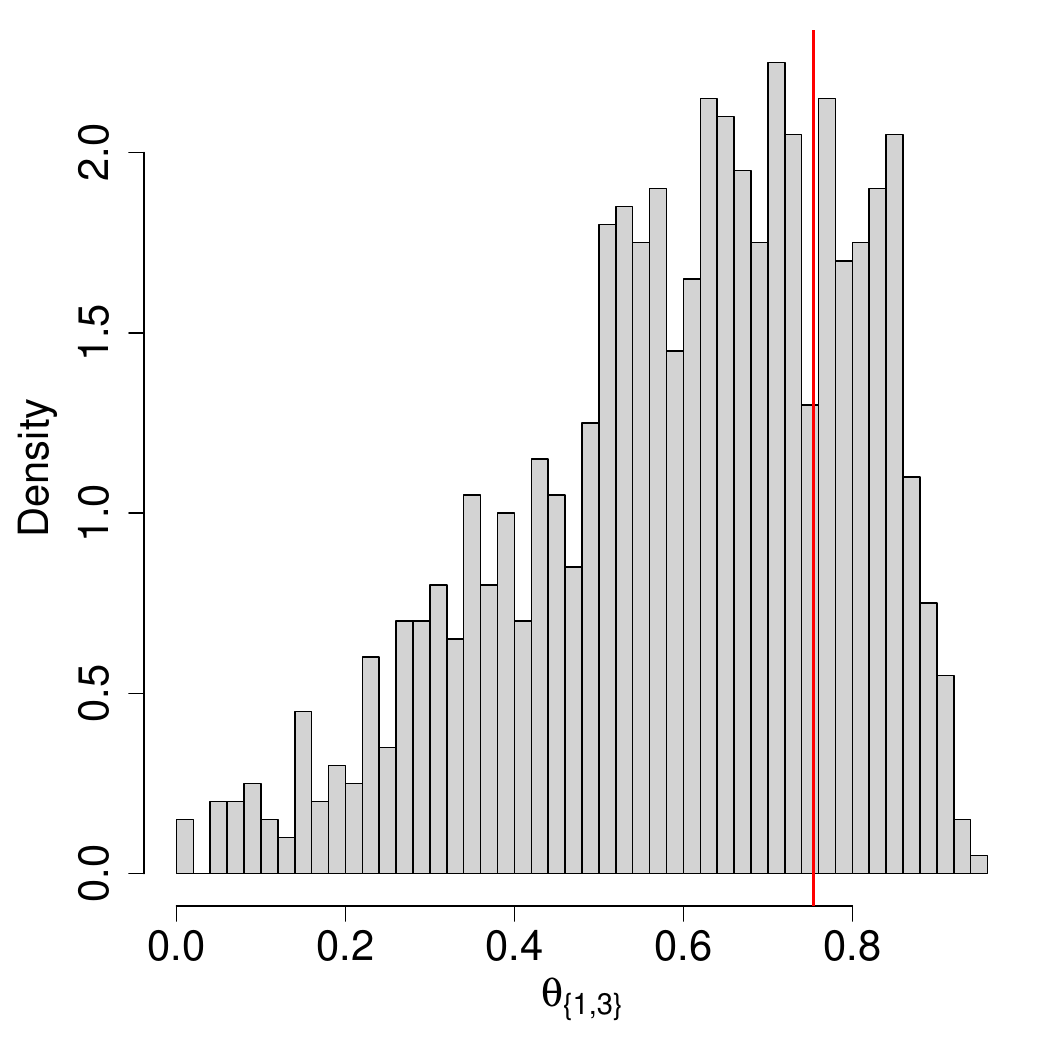}
    \includegraphics[width=0.32\linewidth]{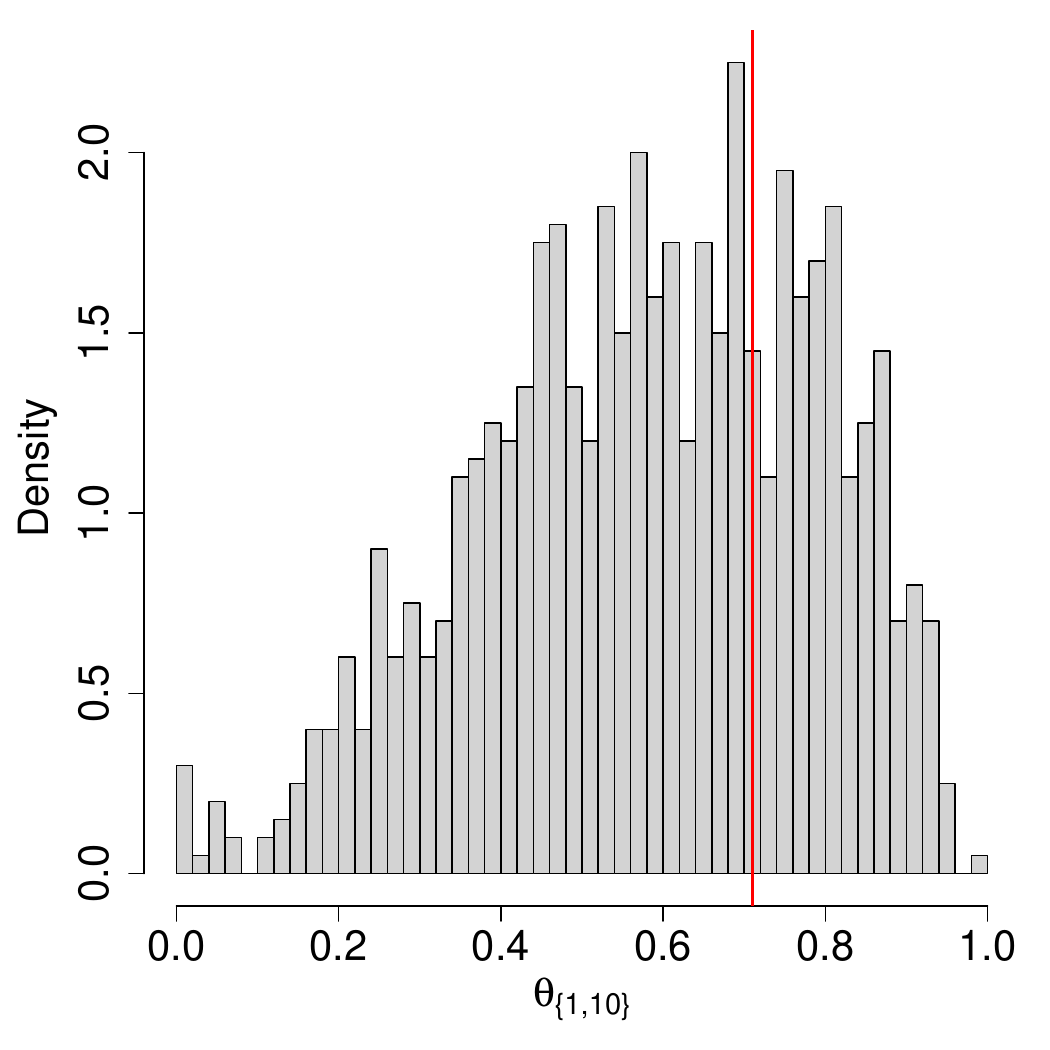}
    \includegraphics[width=0.32\linewidth]{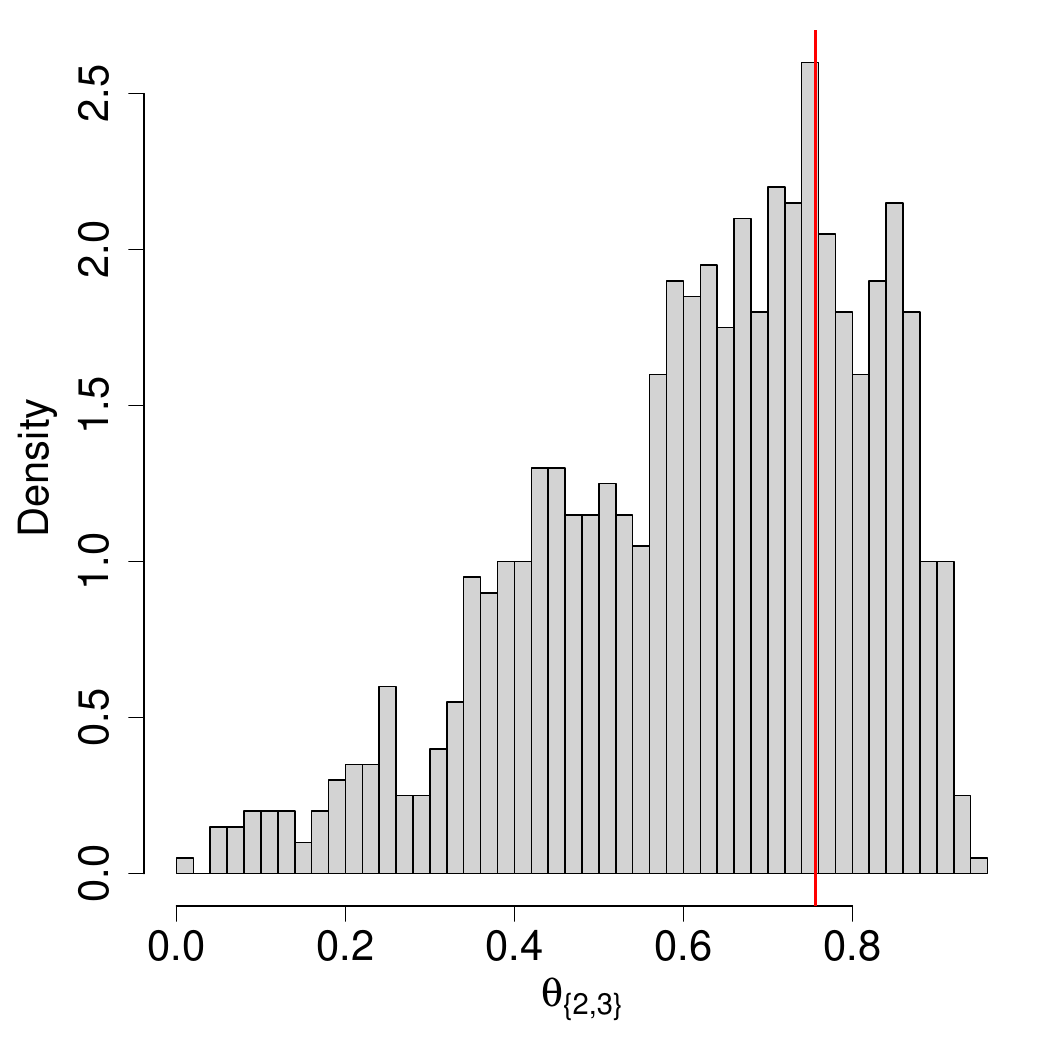}
    \includegraphics[width=0.32\linewidth]{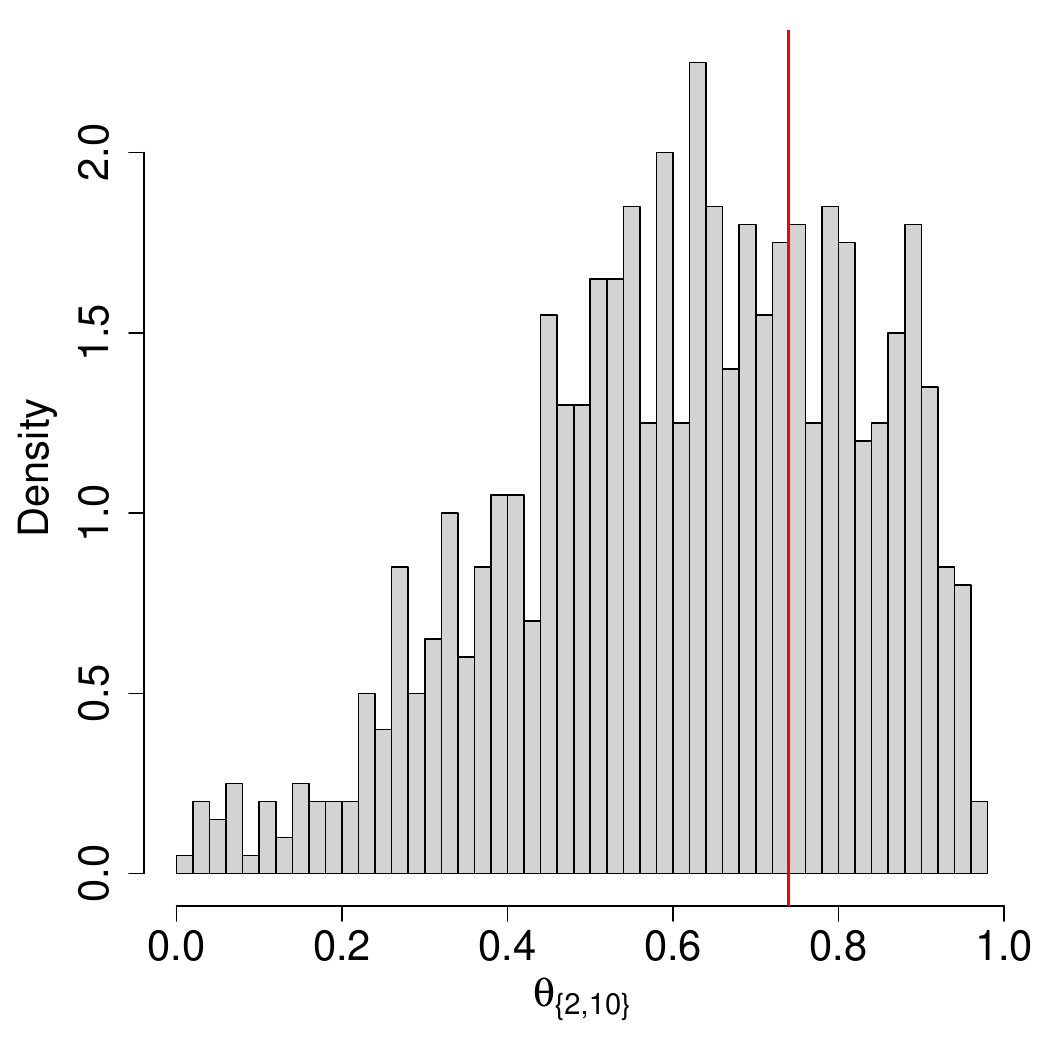}
    \includegraphics[width=0.32\linewidth]{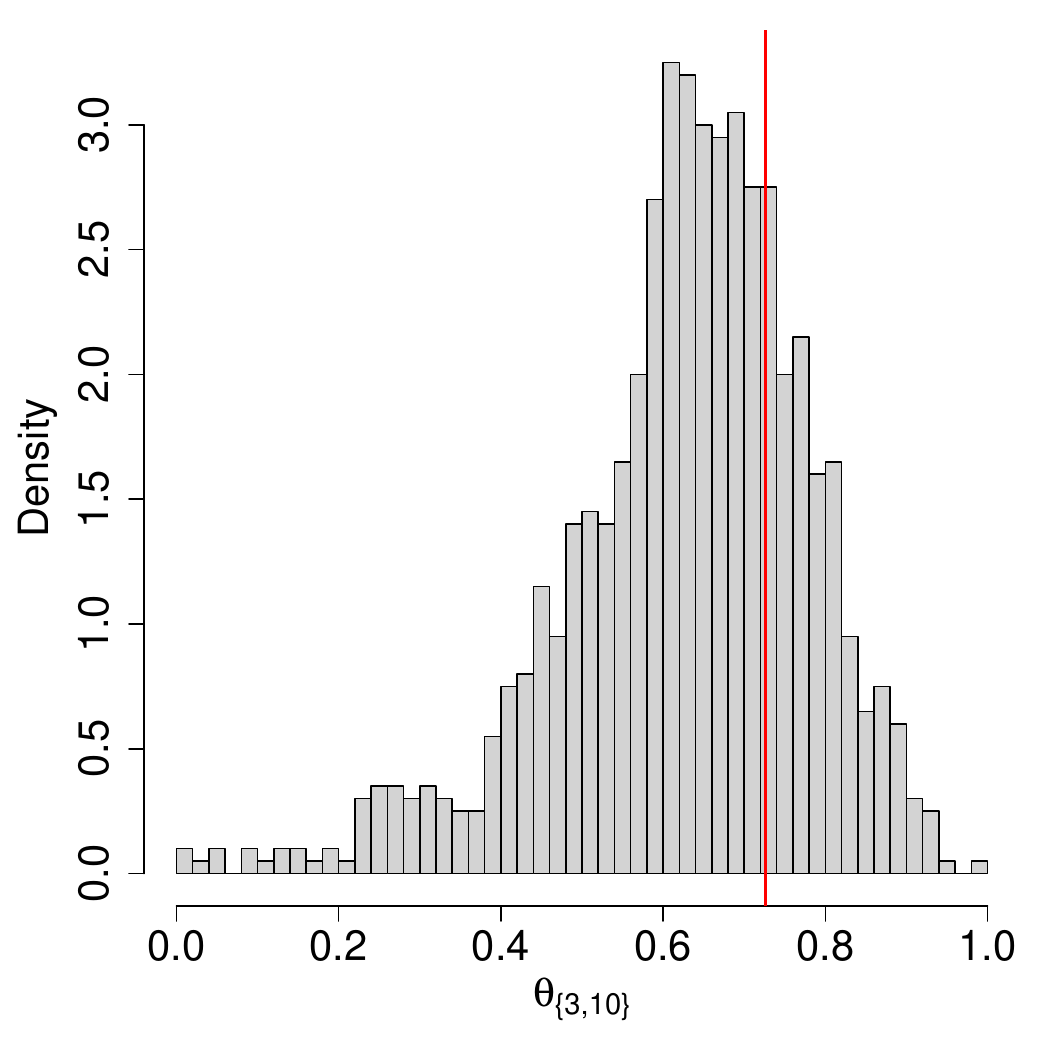}
    \caption{Histograms of $\theta_{\{i,j\}}$ for flow-unconnected pairs $(i,j)$ from block-bootstrap samples of the fit from the truncated gamma model with the graphical gauge function for the Preston river network, with Gaussian gauge function used for all adjacent flow-connected pairs. Vertical red line shows the maximum likelihood estimate obtained using the matched dataset.}
    \label{fig:histograms_param_estimates_original_model_theta}
\end{figure}

Figure~\ref{fig:histograms_param_estimates_original_model_a_rho} and Figure~\ref{fig:histograms_param_estimates_original_model_theta} show the histograms of parameters from block-bootstrap samples of the fit from the truncated gamma model with the graphical gauge function for the Preston river network with Gaussian gauge function used for all adjacent flow-connected pairs. 

\subsection{Extended model}
\label{sec:supp:histograms_param_estimates_extended_model}
\begin{figure}[H]
    \centering
    \includegraphics[width=0.32\linewidth]{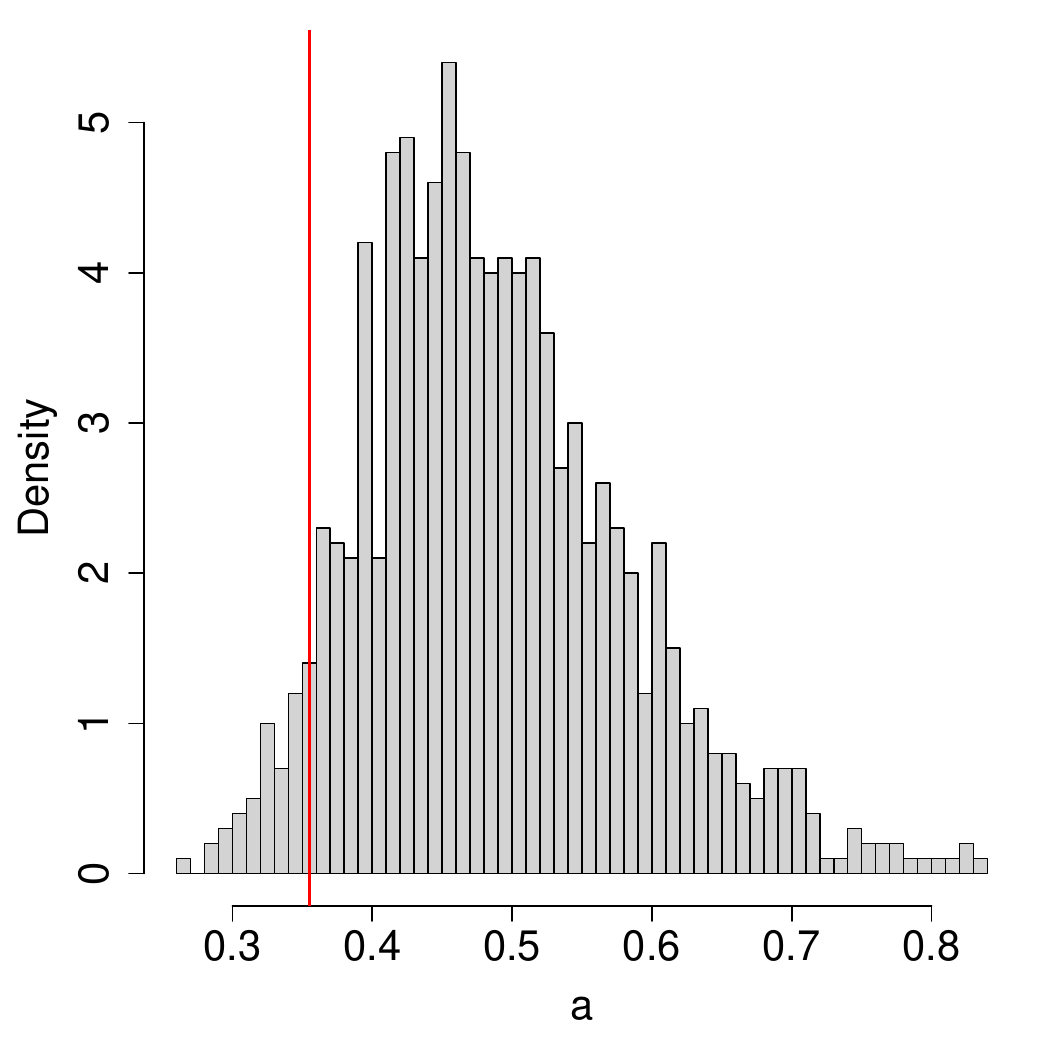}\\
    \includegraphics[width=0.32\linewidth]{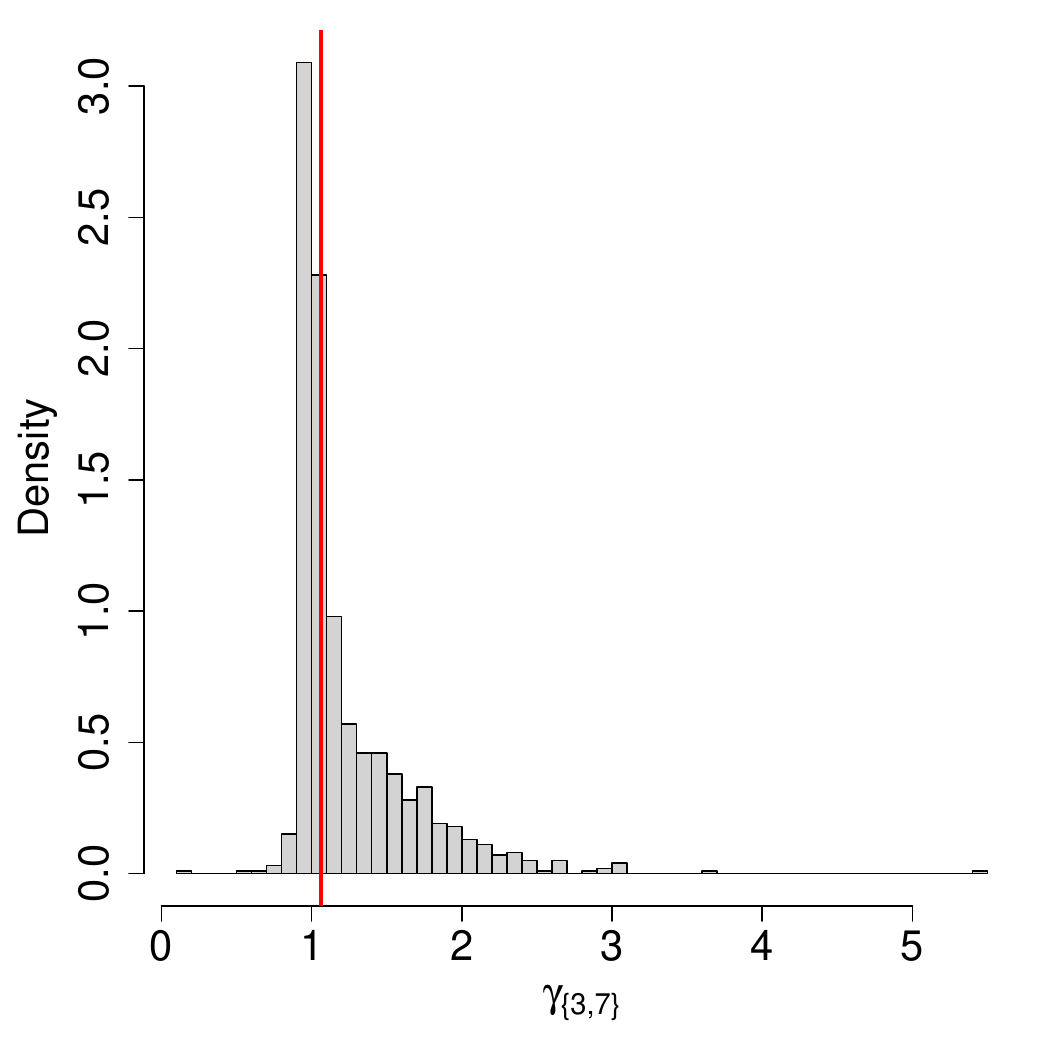}
    \includegraphics[width=0.32\linewidth]{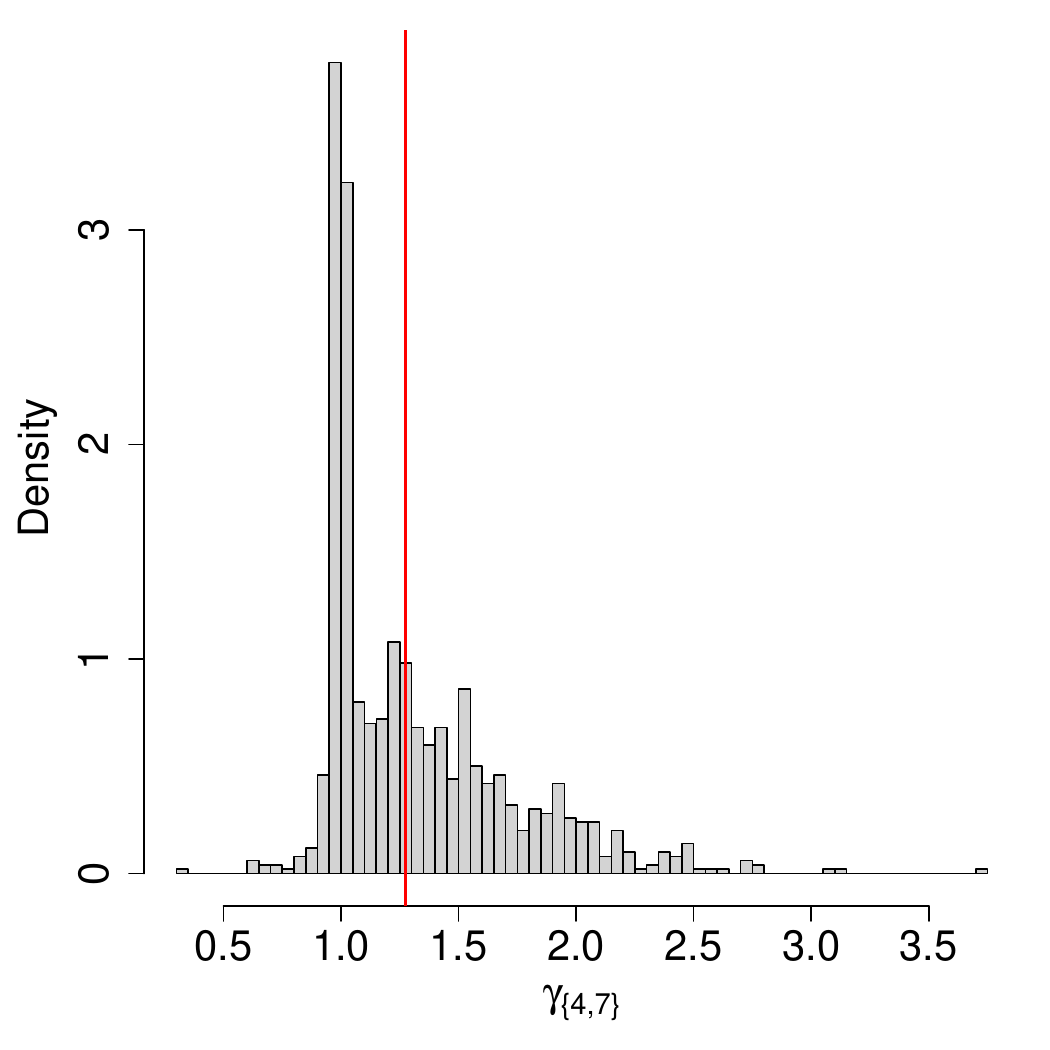}
    \includegraphics[width=0.32\linewidth]{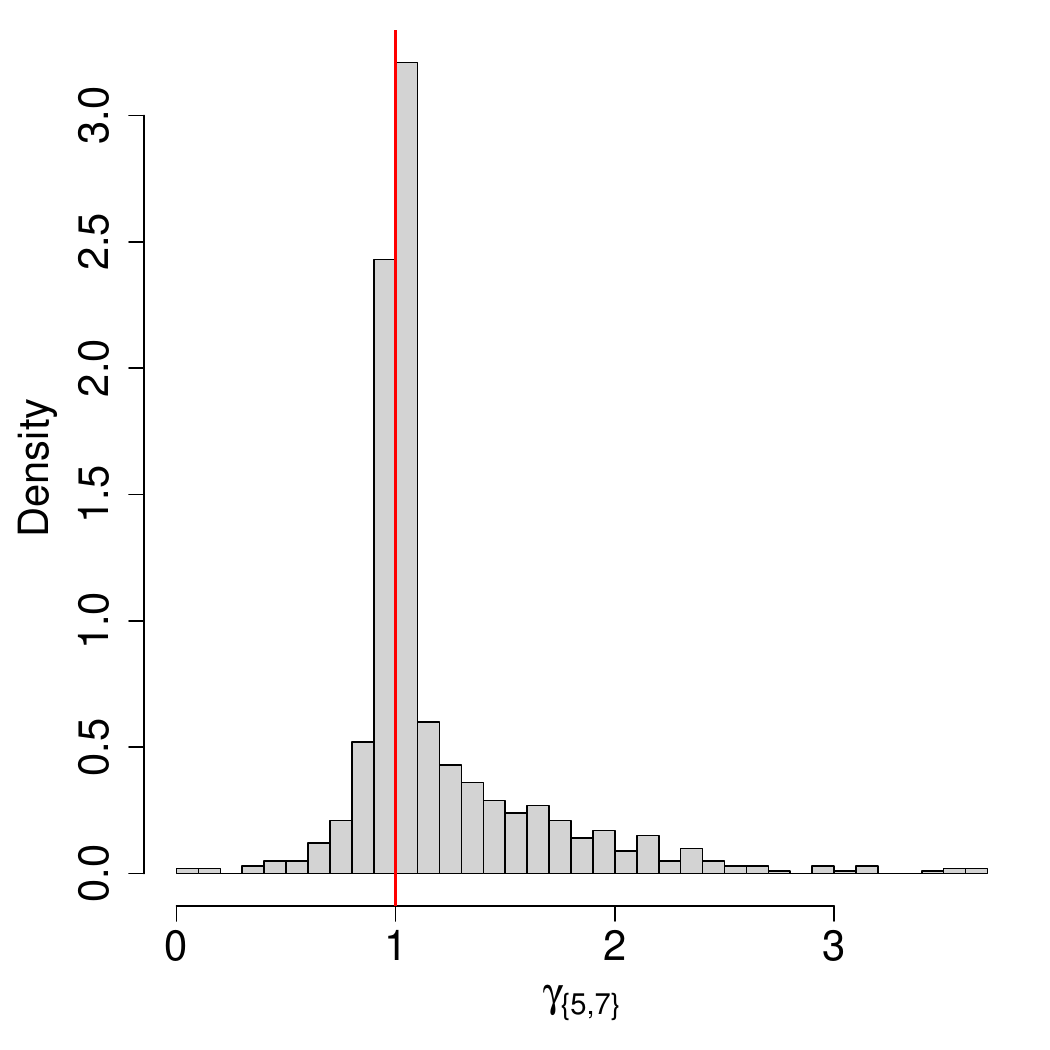}
    \includegraphics[width=0.32\linewidth]{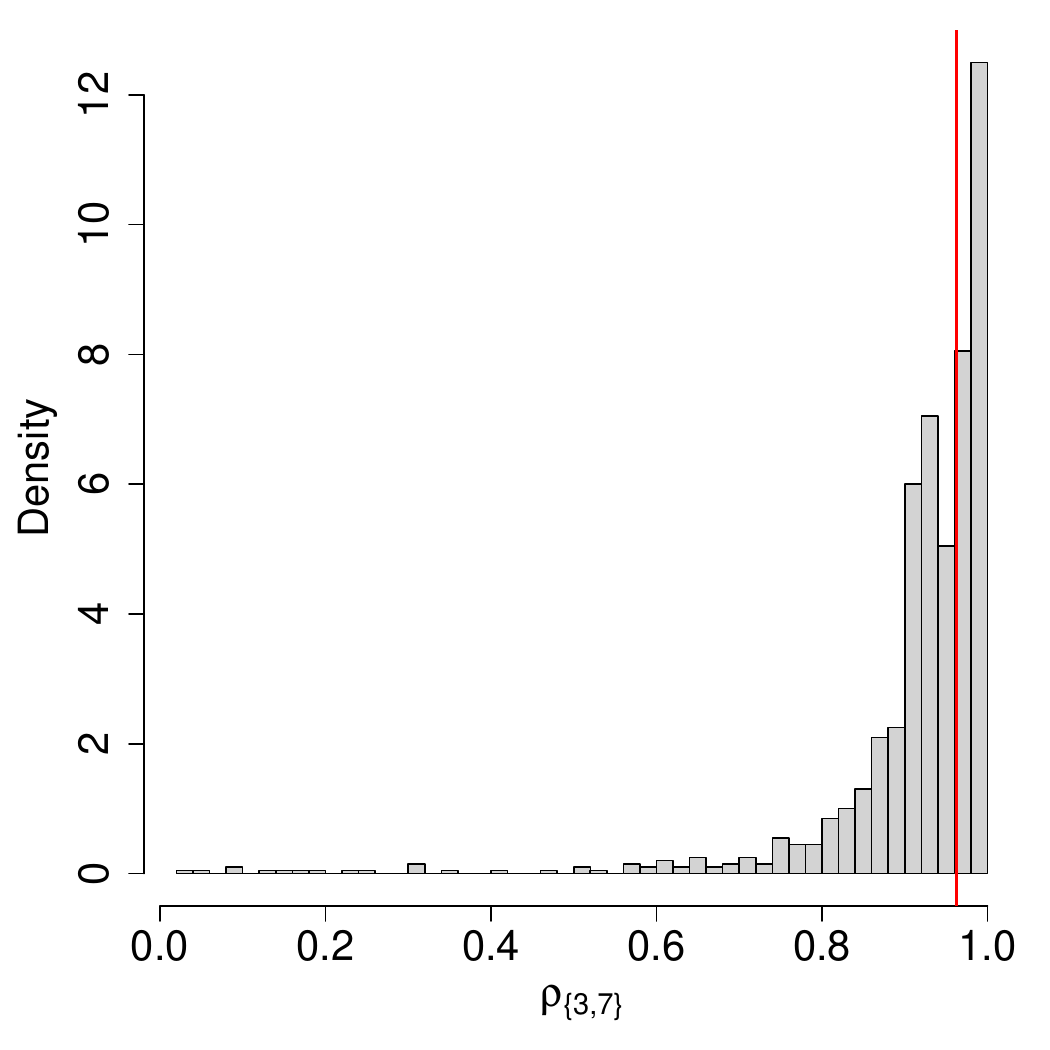}
    \includegraphics[width=0.32\linewidth]{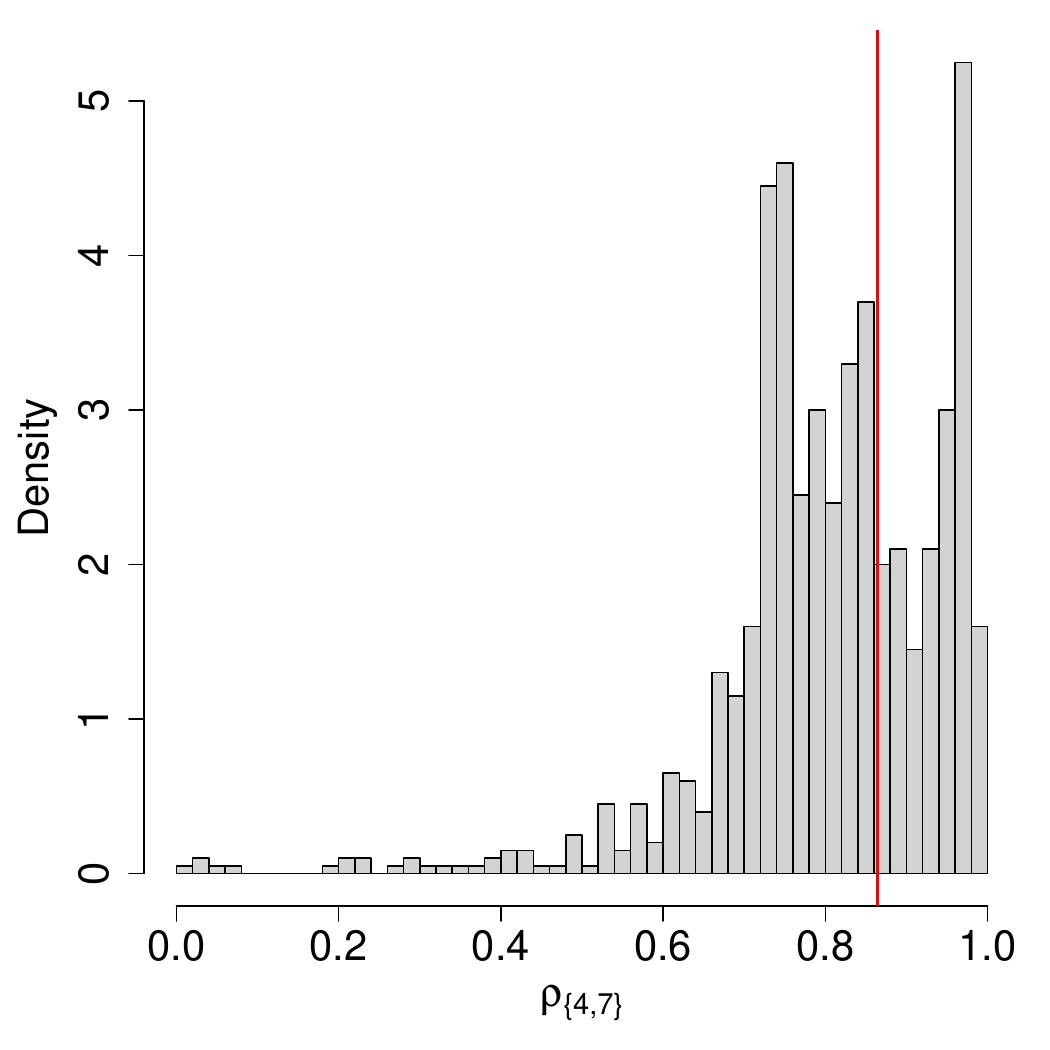}
    \includegraphics[width=0.32\linewidth]{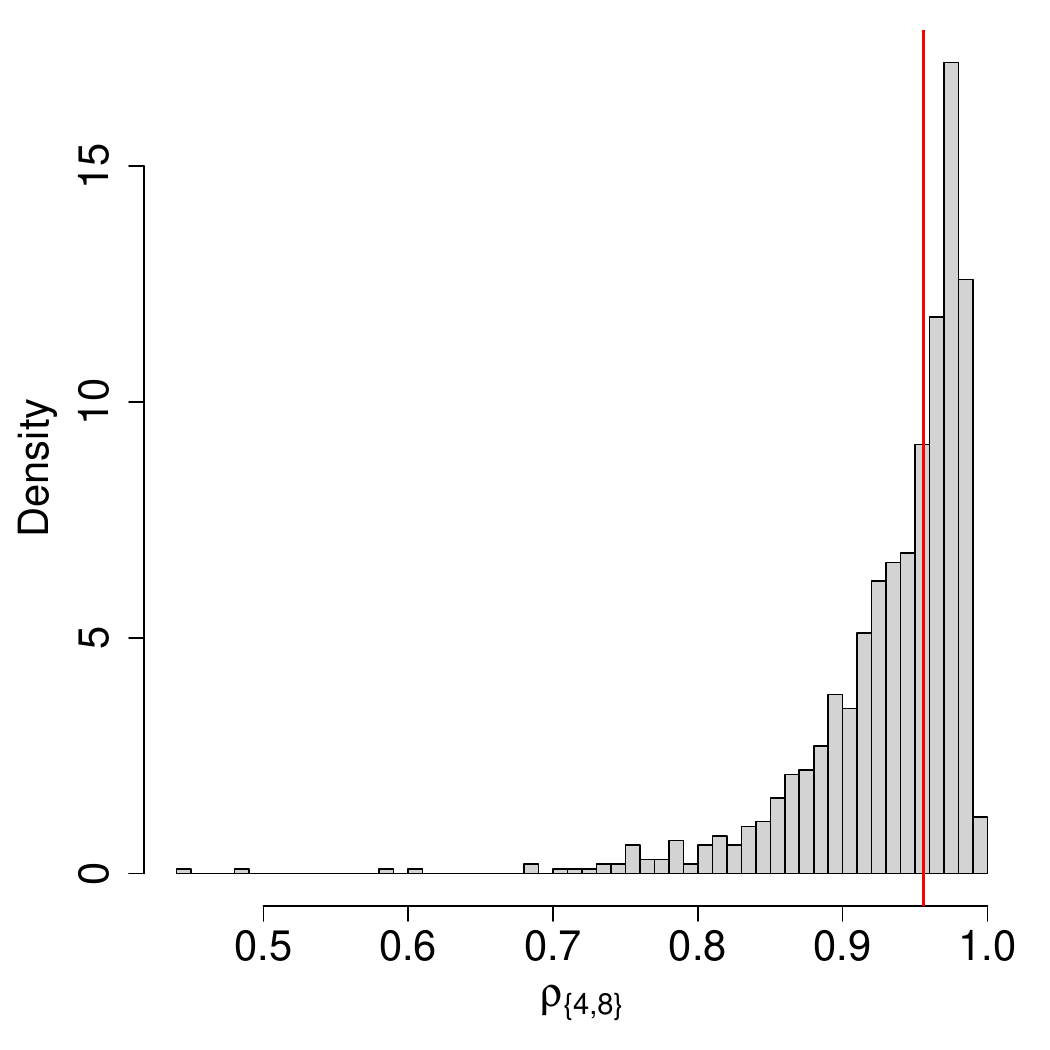}
    \includegraphics[width=0.32\linewidth]{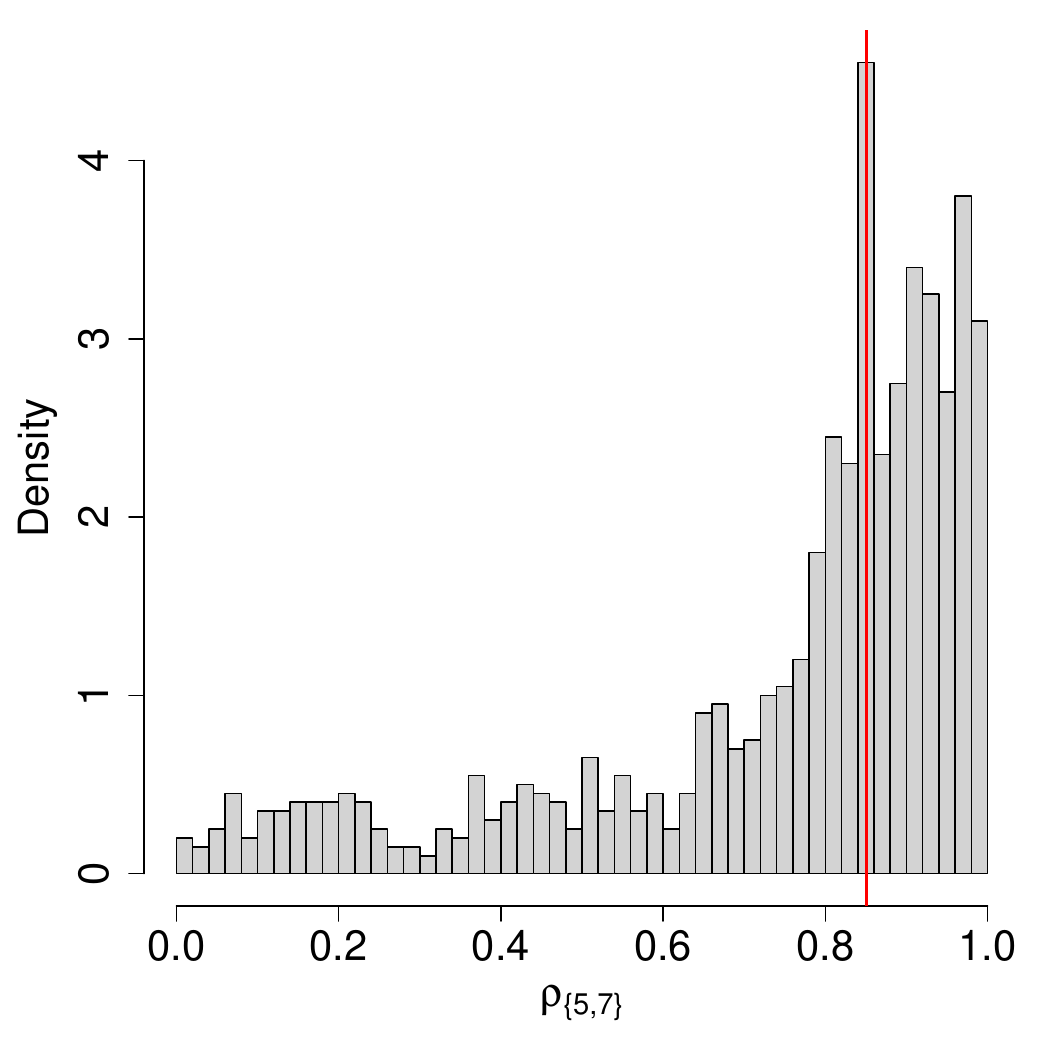}
    \includegraphics[width=0.32\linewidth]{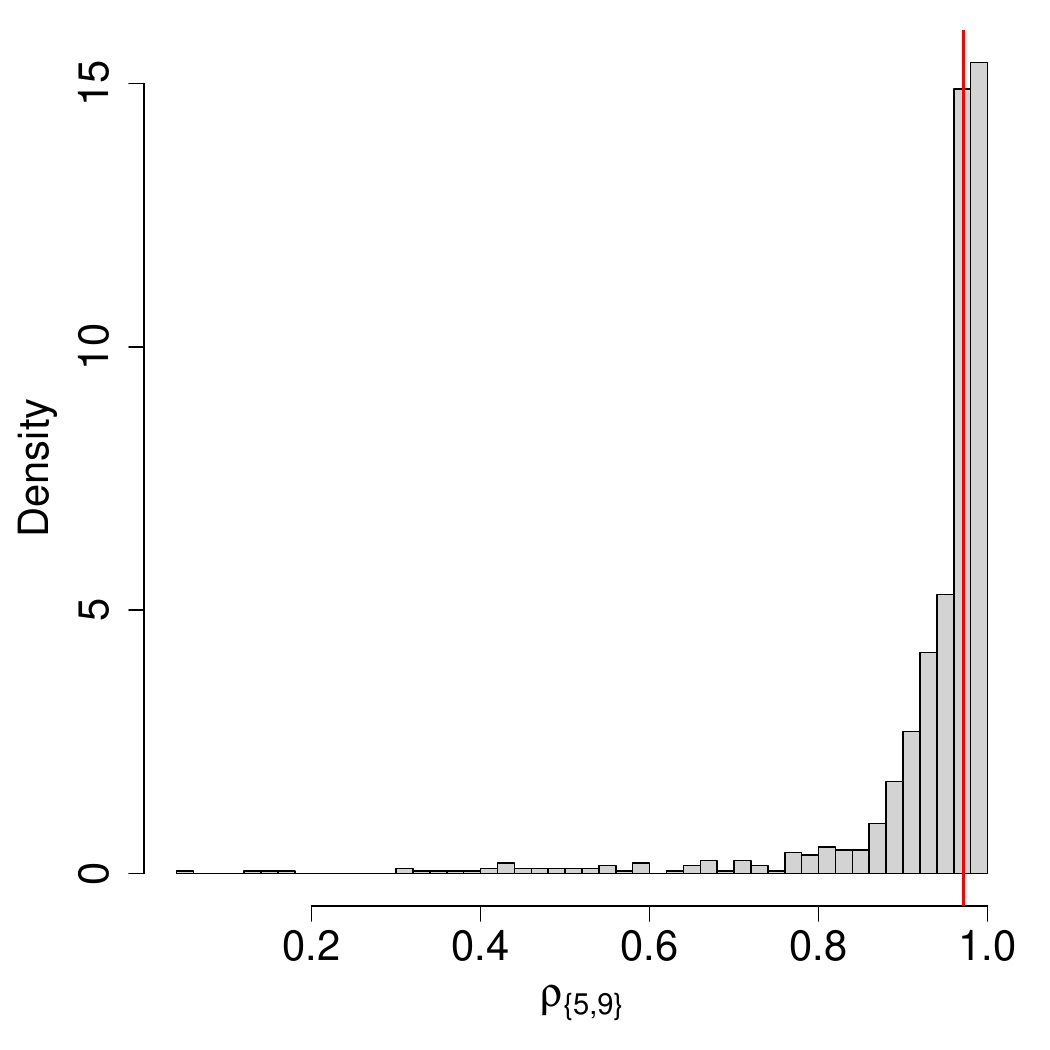}
    \includegraphics[width=0.32\linewidth]{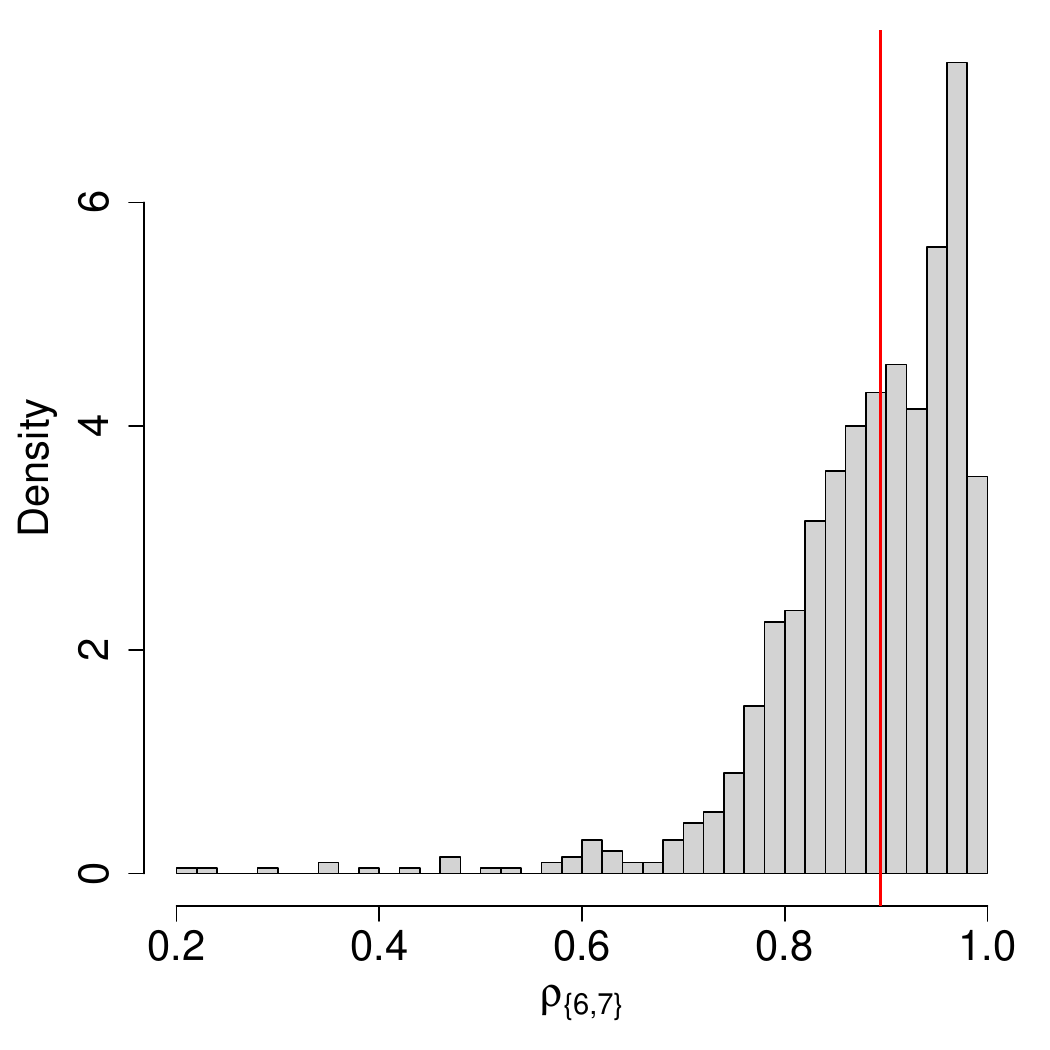}
    \caption{Histograms of $a$, $\gamma_{\{3,7\}}$, $\gamma_{\{4,7\}}$, $\gamma_{\{5,7\}}$ and $\rho_{\{i,j\}}$ for flow-connected pairs $(i,j)$ from block-bootstrap samples of the fit from the truncated gamma model with the graphical gauge function for the Preston river network, with exponential-Gaussian gauge function used for flow-connected pairs $(3,7)$, $(4,7)$ and $(5,7)$. Vertical red line shows the maximum likelihood estimate obtained using the matched dataset.}
    \label{fig:histograms_param_estimates_extended_model_a_rho}
\end{figure}

\begin{figure}[H]
    \centering
    \includegraphics[width=0.32\linewidth]{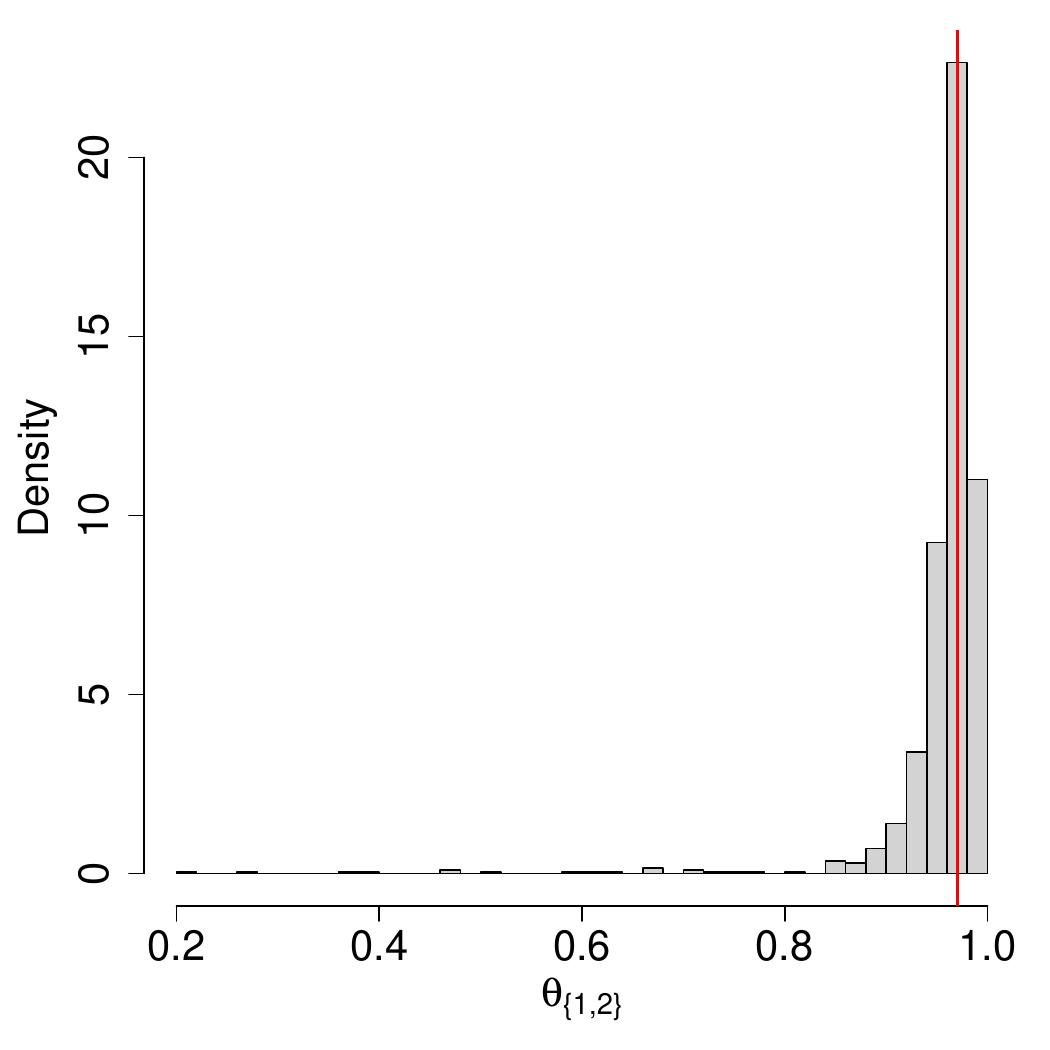}
    \includegraphics[width=0.32\linewidth]{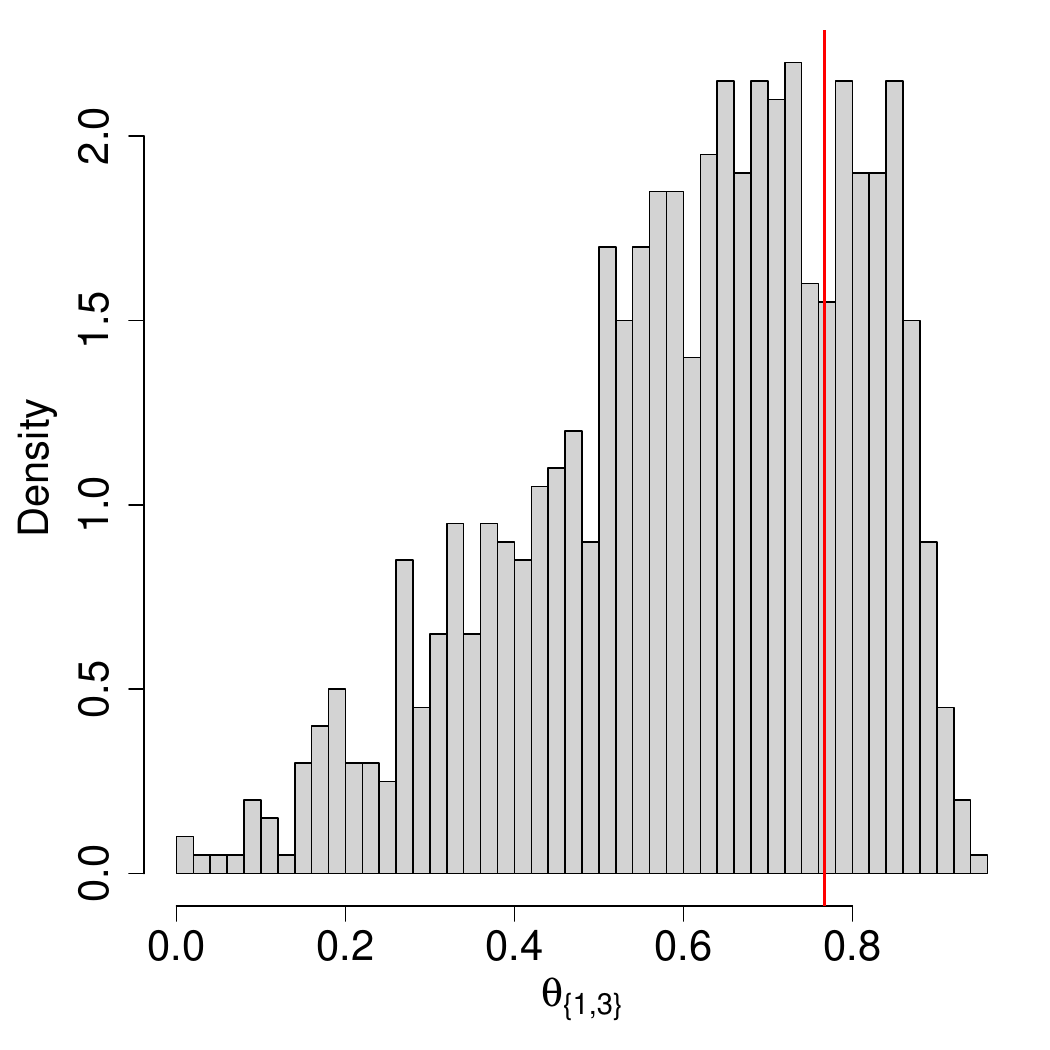}
    \includegraphics[width=0.32\linewidth]{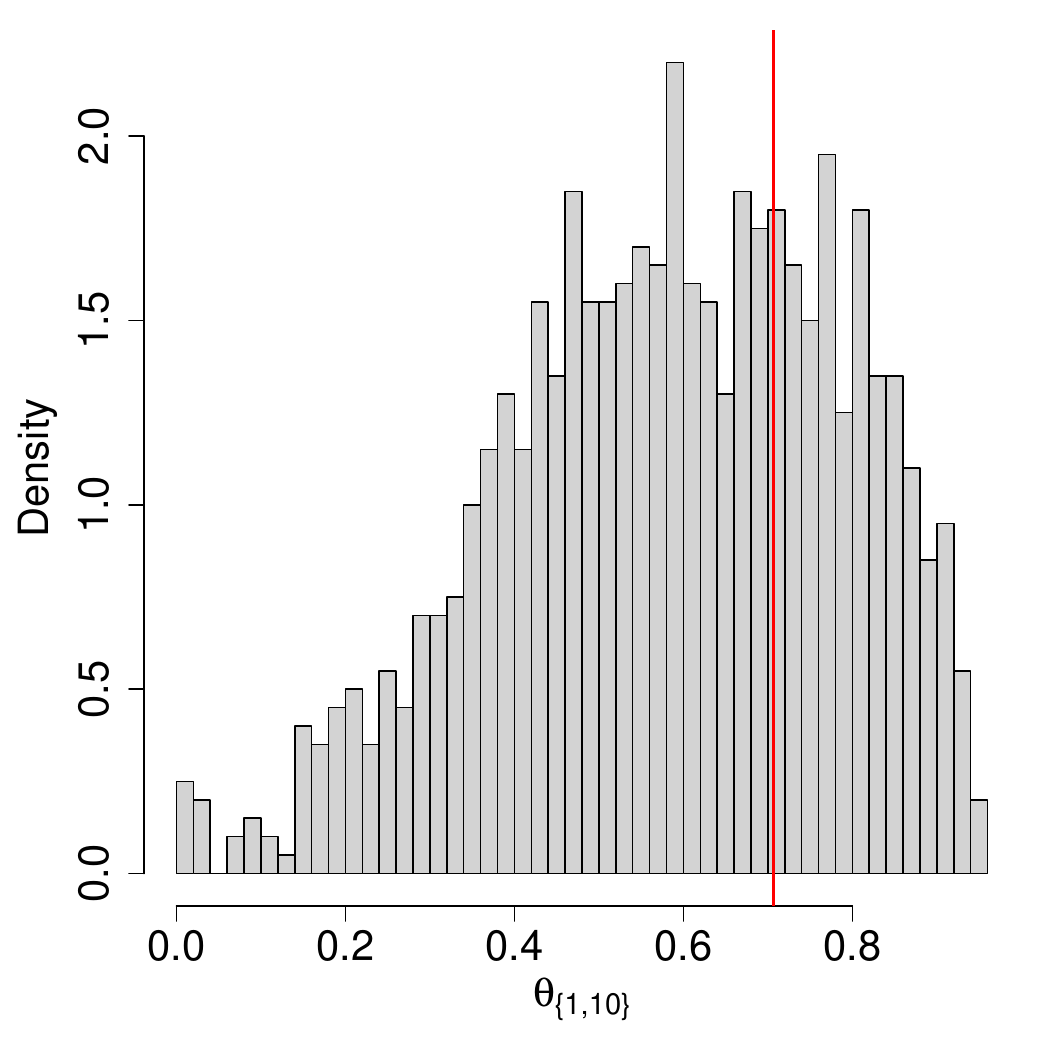}
    \includegraphics[width=0.32\linewidth]{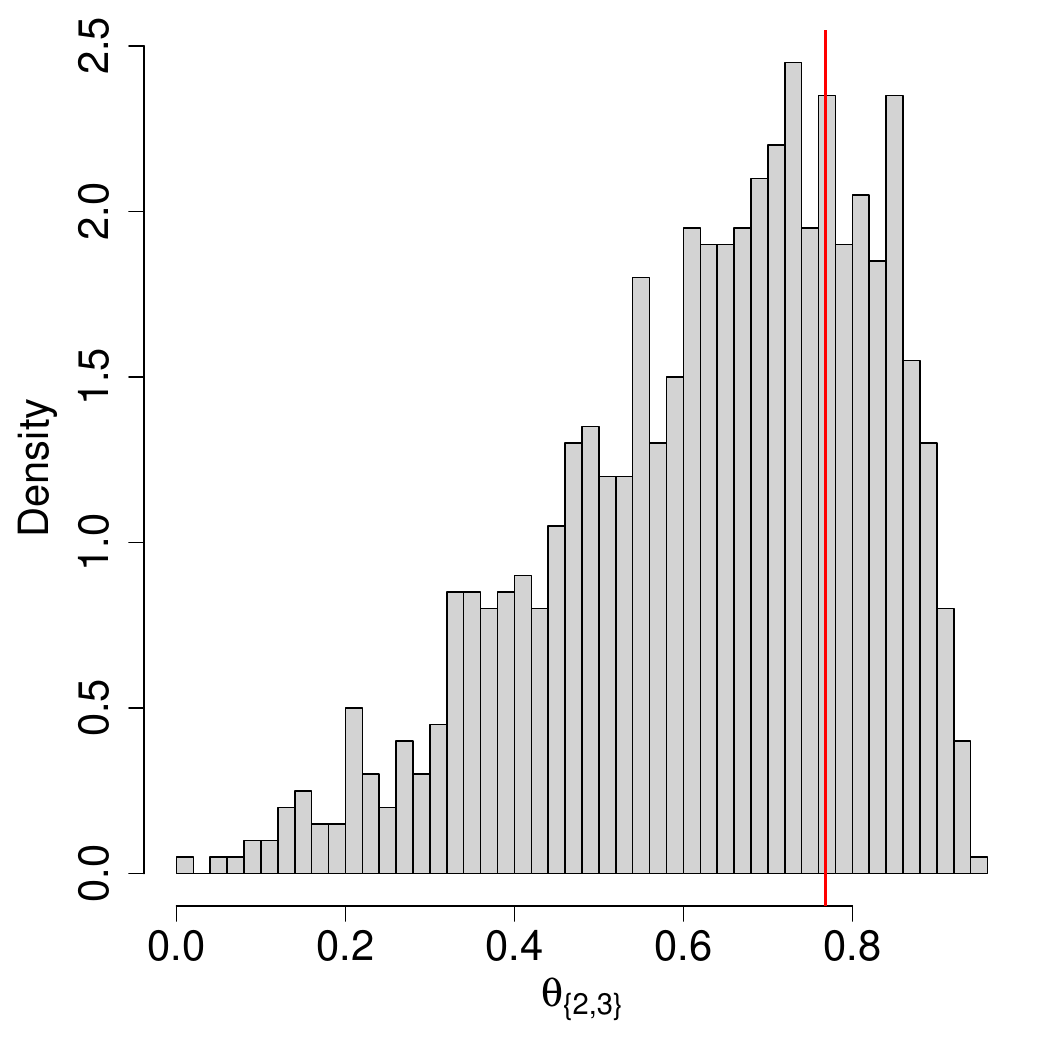}
    \includegraphics[width=0.32\linewidth]{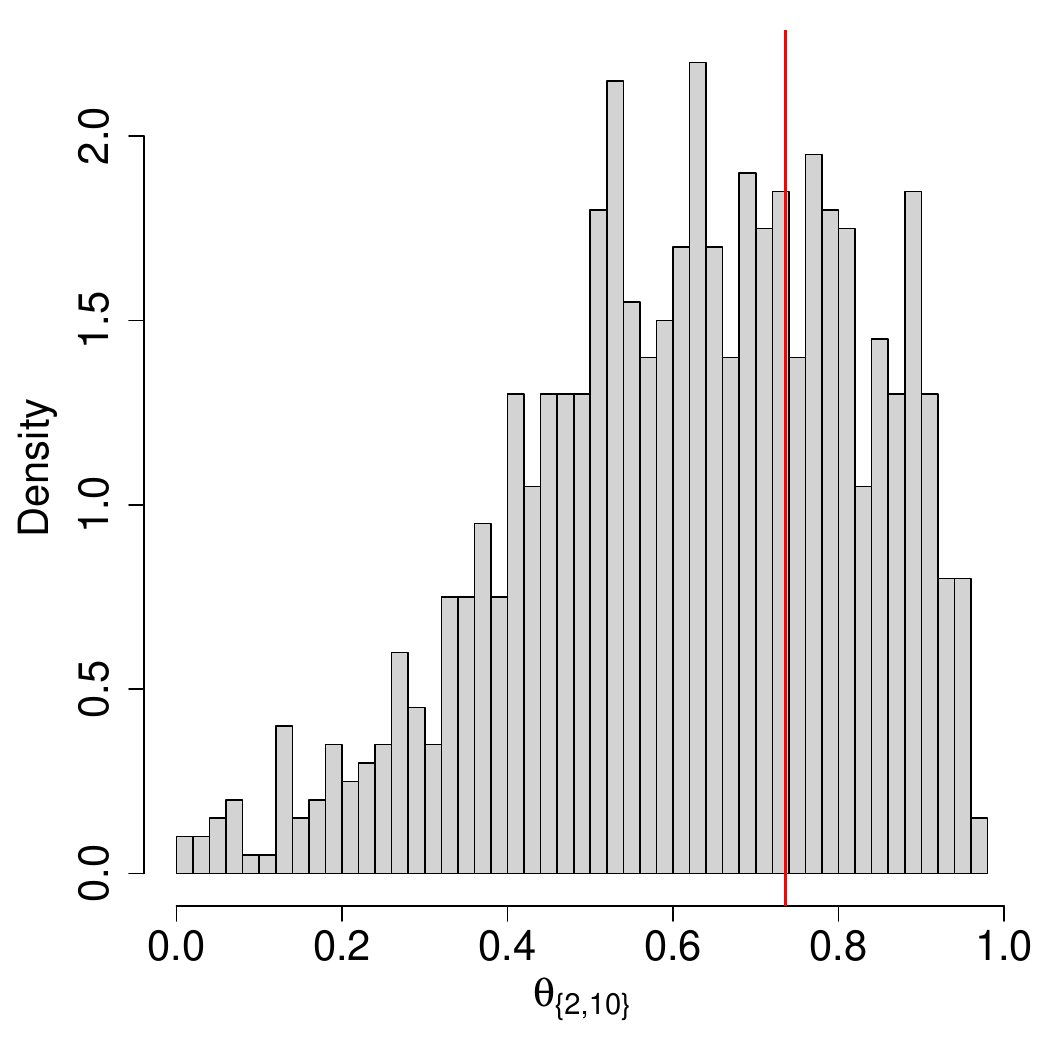}
    \includegraphics[width=0.32\linewidth]{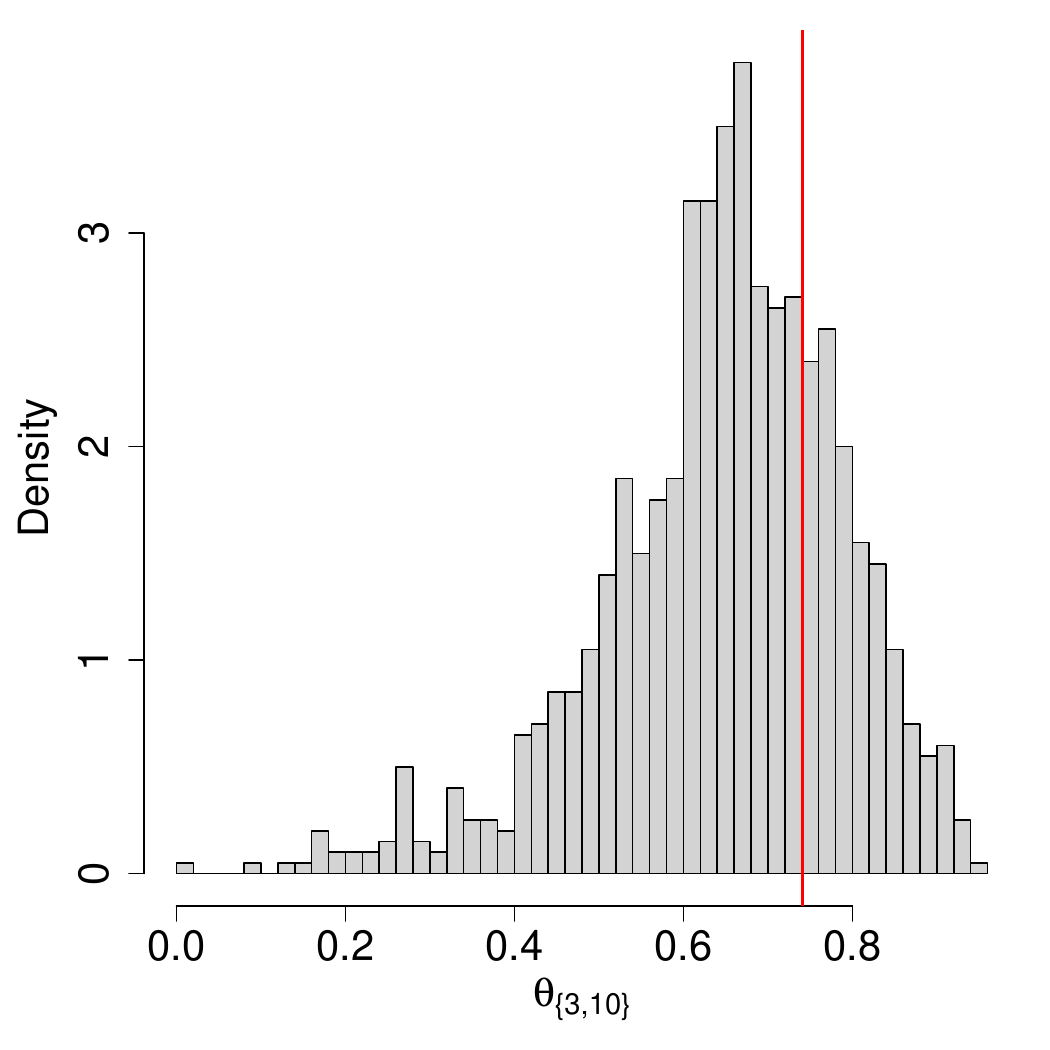}
    \caption{Histograms of $\theta_{\{i,j\}}$ for flow-unconnected pairs $(i,j)$ from block-bootstrap samples of the fit from the truncated gamma model with the graphical gauge function for the Preston river network, with exponential-Gaussian gauge function used for flow-connected pairs $(3,7)$, $(4,7)$ and $(5,7)$. Vertical red line shows the maximum likelihood estimate obtained using the matched dataset.}
    \label{fig:histograms_param_estimates_extended_model_theta}
\end{figure}

Figure~\ref{fig:histograms_param_estimates_extended_model_a_rho} and Figure~\ref{fig:histograms_param_estimates_extended_model_theta} show the histograms of parameters from block-bootstrap samples of the fit from the truncated gamma model with the graphical gauge function for the Preston river network with exponential-Gaussian gauge function used for flow-connected pairs $(3,7)$, $(4,7)$ and $(5,7)$.

\subsection{Investigating an influential point for pair $(6,7)$ and its impact on the fit}
\label{sec:supp:outlying_point_6_7}

\begin{figure}[H]
    \centering
    \includegraphics[width=0.5\linewidth]{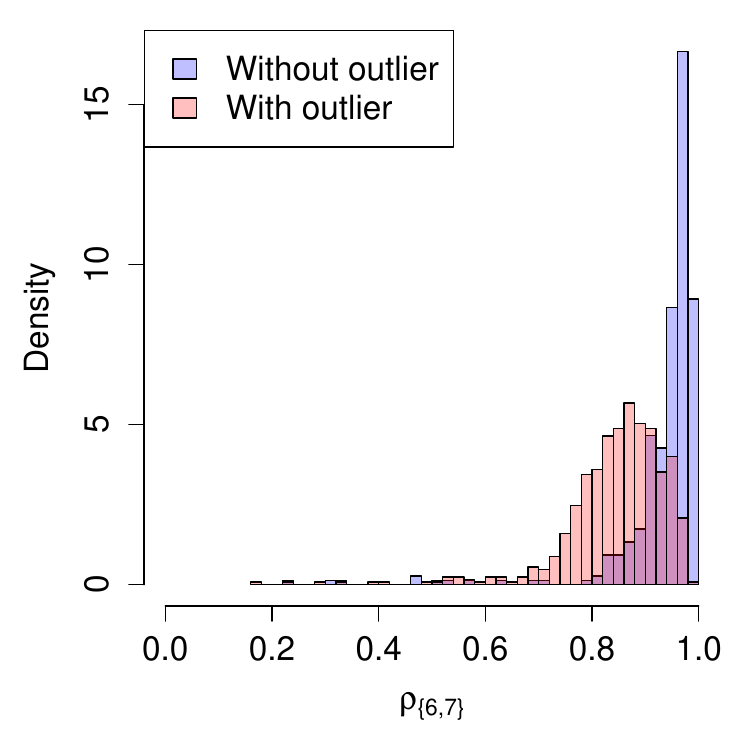}
    \caption{Histogram of the estimate of $\rho_{\{6,7\}}$ for bootstrap samples that contain the outlier (red) and that do not contain the outlier (blue).}
    \label{fig:hist_mle_6_7_outlier}
\end{figure}

The shape of the limit set for the pair $(6,7)$ in Figure~\ref{fig:bivariate_projections_adjacent_flow_connected_rest} is a little wider than the scaled sample cloud. We investigated whether this could be due to the slightly outlying point at (0.69, 0.96). We looked at the 1000 bootstrap samples and determined whether this point is present in a particular sample or not, and then plotted a histogram of the estimates of the parameter $\rho_{\{6,7\}}$ for the two cases, which can be found in Figure~\ref{fig:hist_mle_6_7_outlier}. There were 375 samples without and 625 with this particular point. The estimates of the parameter $\rho_{\{6,7\}}$ were in general higher for the case where the outlying point was not present, supporting the belief that the wider fit is likely due to this point. 

\newpage
\section{Diagnostics}
\subsection{Bivariate projections}
\label{sec:supp:bivariate_projections}
\begin{figure}[H]
    \centering
    \includegraphics[width=0.32\linewidth]{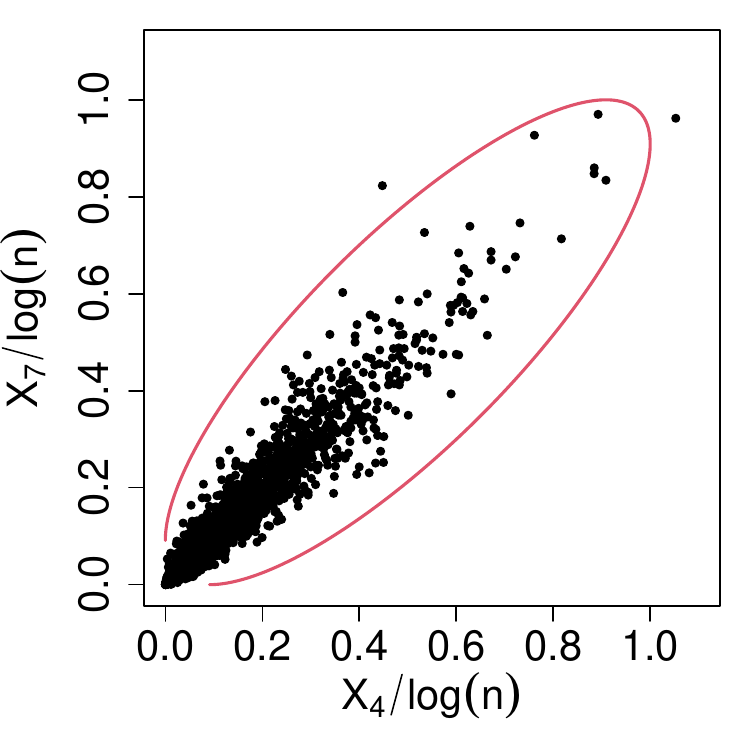}
    \includegraphics[width=0.32\linewidth]{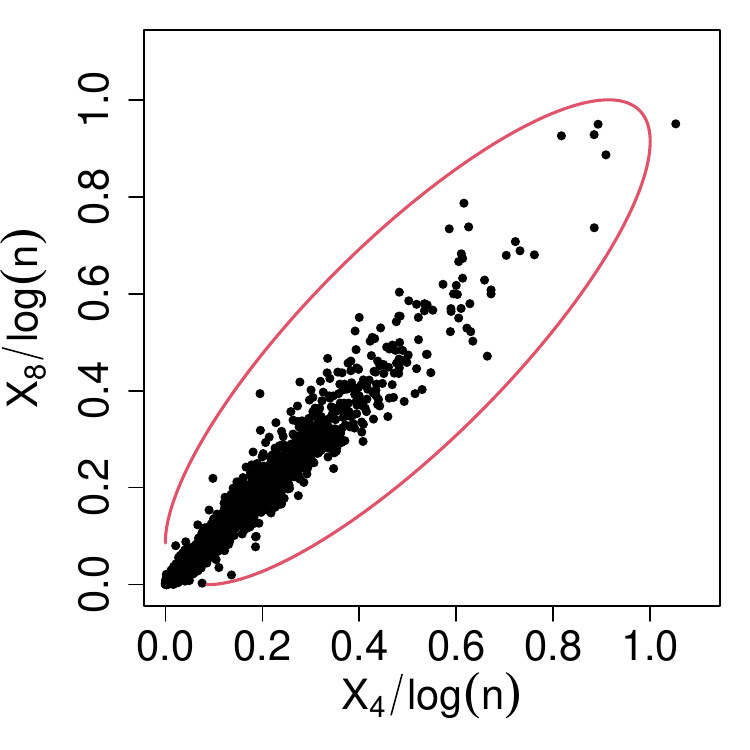}
    \includegraphics[width=0.32\linewidth]{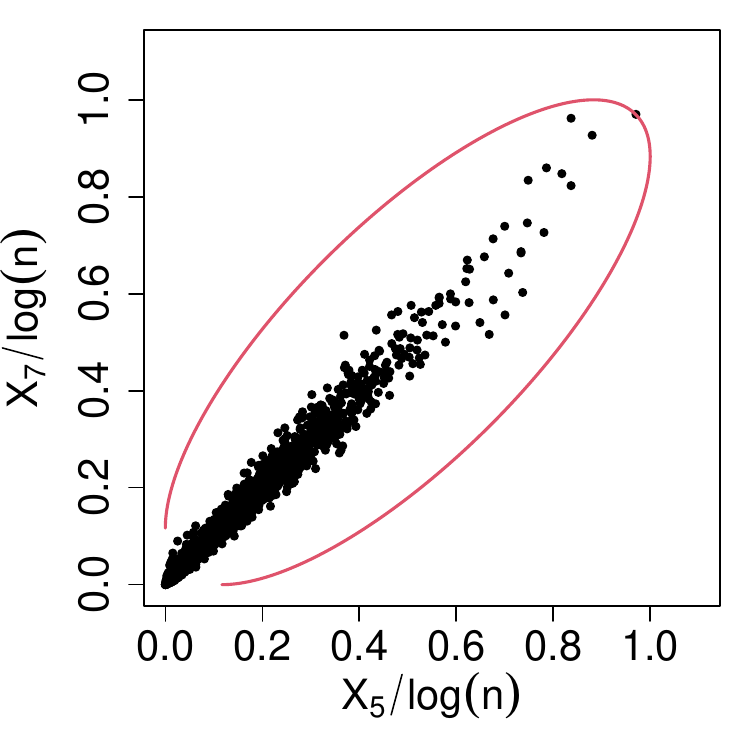}
    \includegraphics[width=0.32\linewidth]{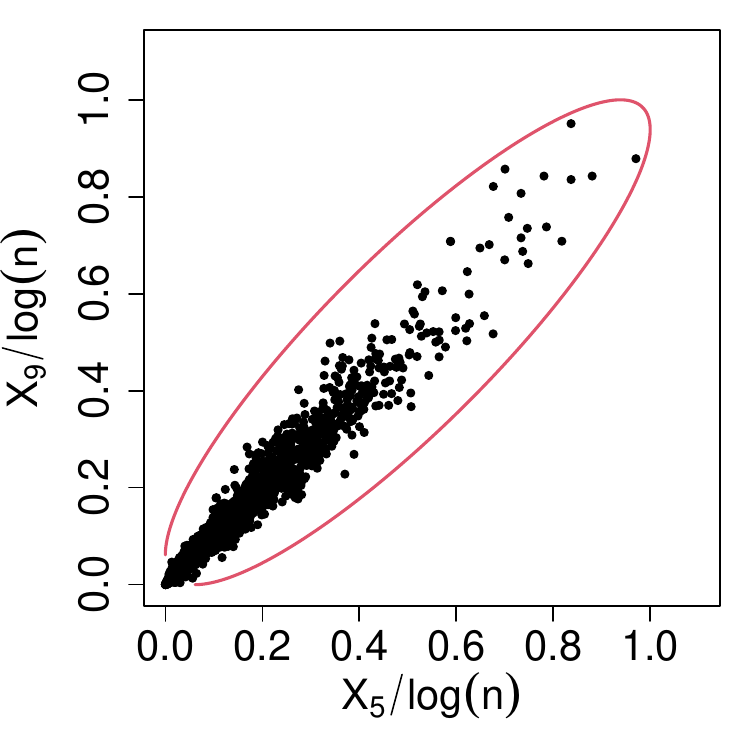}
    \includegraphics[width=0.32\linewidth]{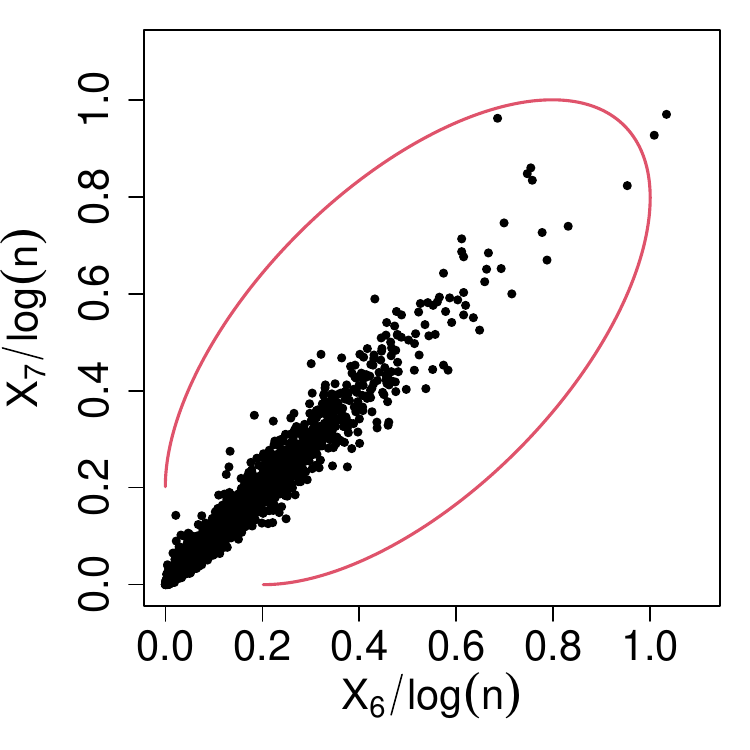}
    \caption{Plots showing the bivariate projections of the limit set (red line) fitted to all 10 stations on the Preston river network with corresponding scaled sample clouds (black points) for adjacent flow-connected pairs.}
    \label{fig:bivariate_projections_adjacent_flow_connected_rest}
\end{figure}

Figure~\ref{fig:bivariate_projections_adjacent_flow_connected_rest} shows the bivariate projections for the rest of the adjacent flow-connected pairs.

\subsection{$\chi_S(u)$ plots}
\subsubsection{Original model}
\label{sec:supp:chi_original_model}
\begin{figure}[H]
    \centering
    \includegraphics[width=0.4\linewidth]{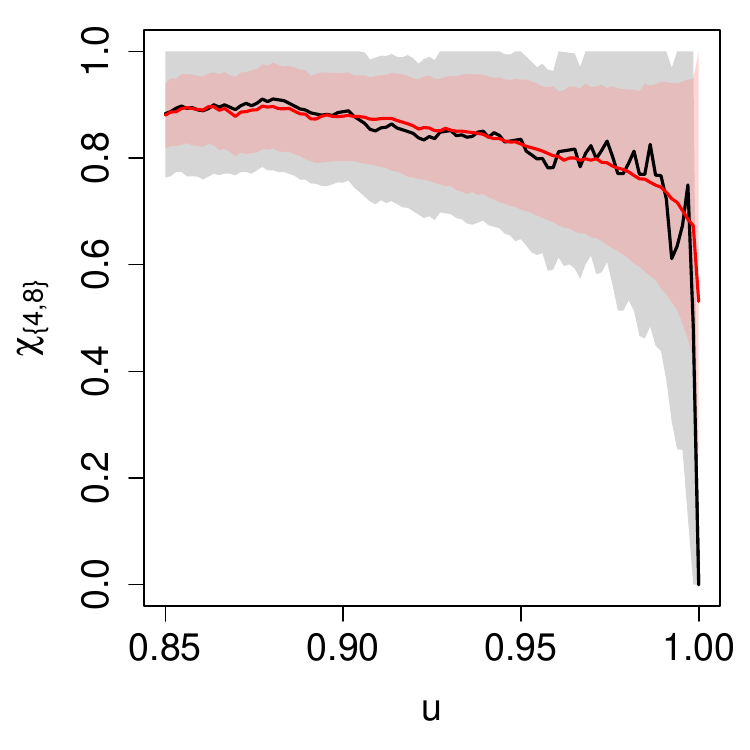}
    \includegraphics[width=0.4\linewidth]{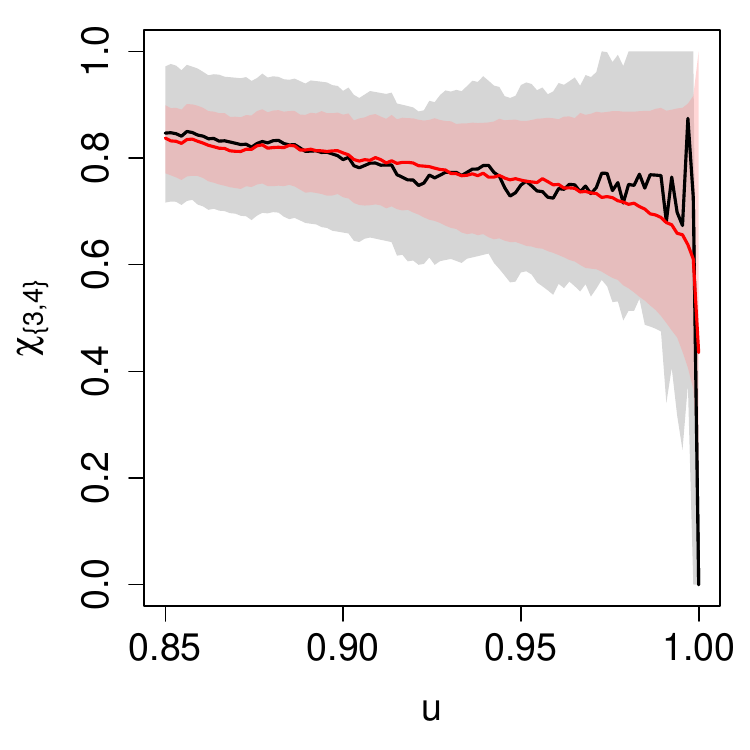}
    \includegraphics[width=0.4\linewidth]{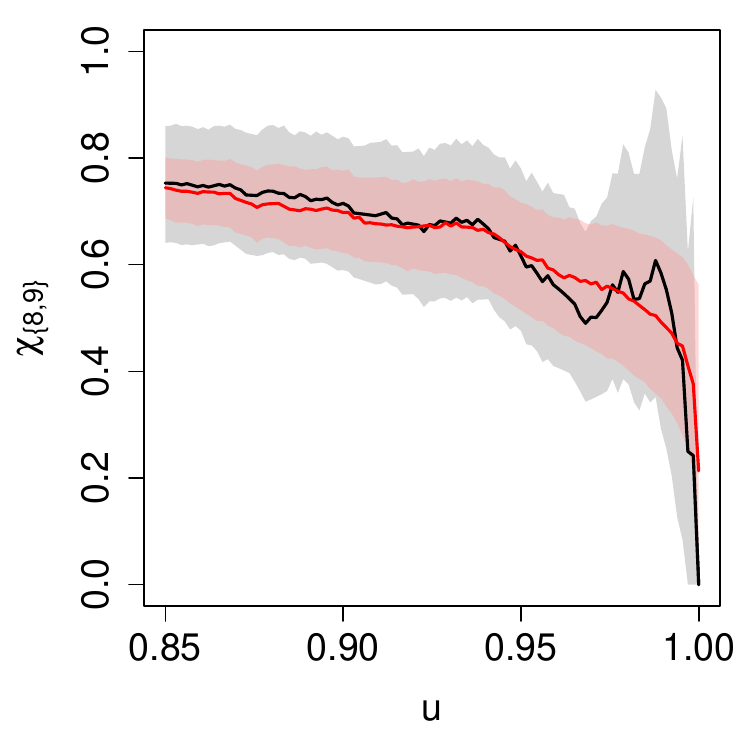}
    \includegraphics[width=0.4\linewidth]{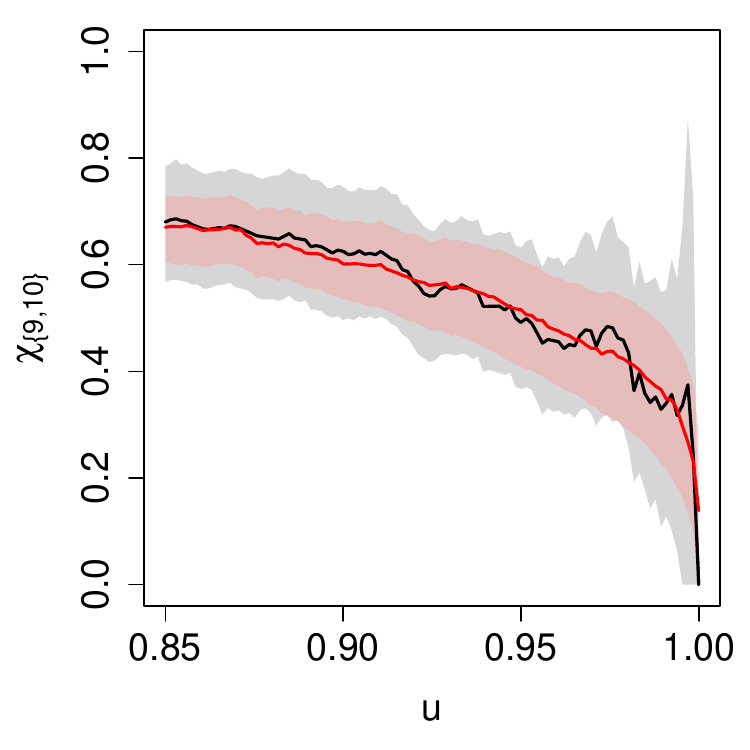}
    \caption{Plot of empirical (solid black) and model-based $\chi_S(u)$ (solid red) with empirical (grey region) and model-based (light red region) 95\% CIs for flow-connected $S = \{4,8\}, \{3,4\}$ (top) and flow-unconnected pairs $ S = \{8,9\}, \{9,10\}$ (bottom).}
    \label{fig:chi_plot_bivariate_plots}
\end{figure}

Figure~\ref{fig:chi_plot_bivariate_plots} shows the comparison of bivariate empirical and model-based $\chi_S(u)$ for flow-connected $S = \{4,8\}, \{3,4\}$ and flow-unconnected pairs $S = \{8,9\}, \{9,10\}$. Pair $(4,7)$ is adjacent flow-connected, and $(3,4)$ is non-adjacent flow-connected. Pair $(8,9)$ is a flow-unconnected pair with both gauging stations being on the tree of flow-connected stations, whereas for pairs $(9,10)$ one of the stations is on the tree and the other one is not.

\subsubsection{Impact of correction coefficient on $\chi_S(u)$ plots}
\label{sec:supp:correction_coefficient_chi_plot}

\begin{figure}[H]
    \centering
     \includegraphics[width=0.4\linewidth]{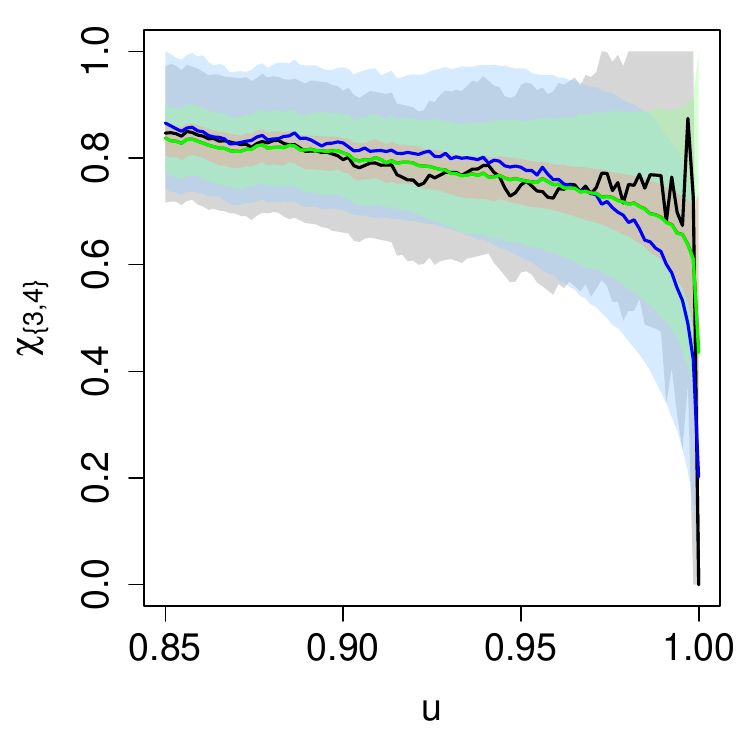}
    \caption{Plot of empirical (solid black) and model-based $\chi_S(u)$ without $C_{\mathrm{corr}}$ (solid blue), with $C_{\mathrm{corr}}$ assuming standard exponential distribution (solid red) and with $C_{\mathrm{corr}}$ with fitted exponential distribution (solid green) for $S=\{3,4\}$. The corresponding empirical and model-based 95\% CIs are shown as grey and light blue, red and green regions, respectively.}
    \label{fig:chi_plot_3_4_model_based_chi_comparison_with_CI}
\end{figure}

Figure~\ref{fig:chi_plot_3_4_model_based_chi_comparison_with_CI} shows an example of a plot of empirical and model-based $\chi_S(u)$, illustrating three different model-based 95\% confidence intervals (CIs). The widest CI is that from the model without the correction coefficient $C_{\mathrm{corr}}$, while the narrowest one is that obtained with $C_{\mathrm{corr}}$ and assuming a standard exponential distribution in the block bootstrap. Finally, when $C_{\mathrm{corr}}$ is used with a fitted exponential distribution in the block bootstrap, we obtain the middle width.

\subsubsection{Extended model}
\label{sec:supp:chi_extended_model}
\begin{figure}[H]
    \centering
    \includegraphics[width=0.4\linewidth]{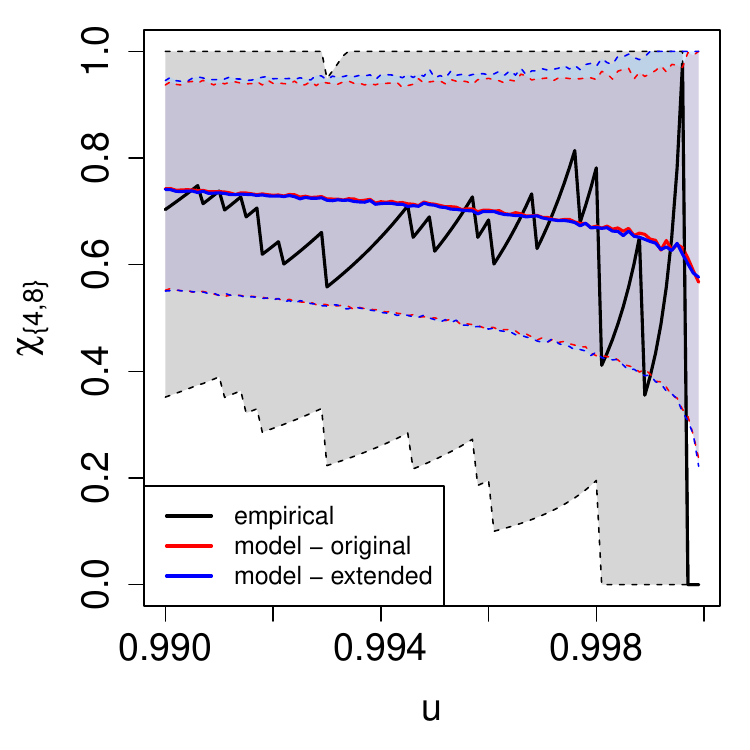}
    \includegraphics[width=0.4\linewidth]{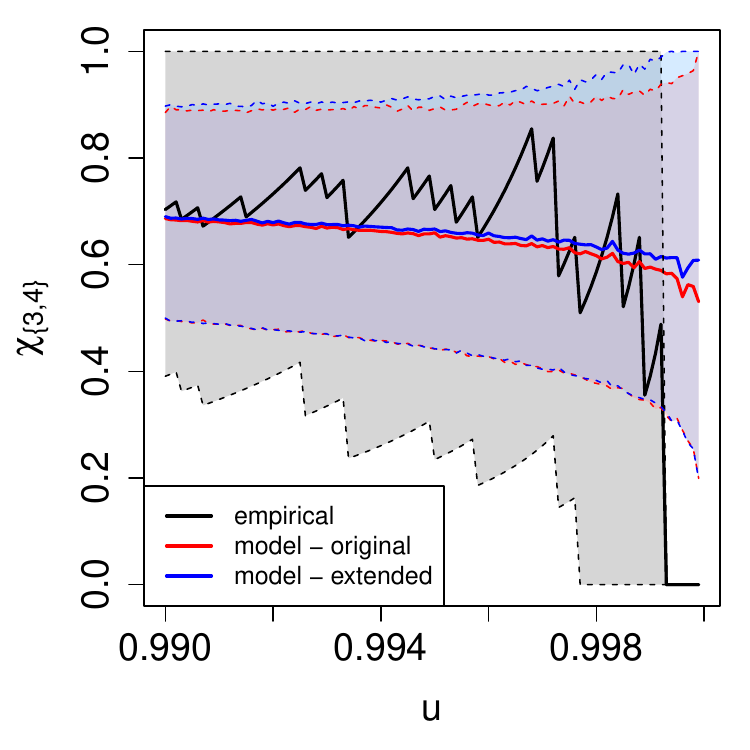}
    \includegraphics[width=0.4\linewidth]{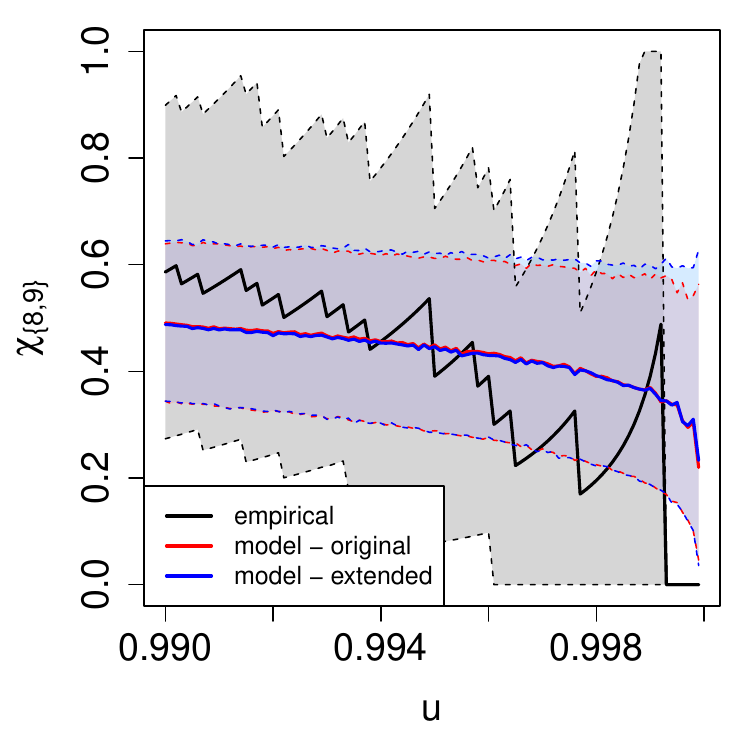}
    \includegraphics[width=0.4\linewidth]{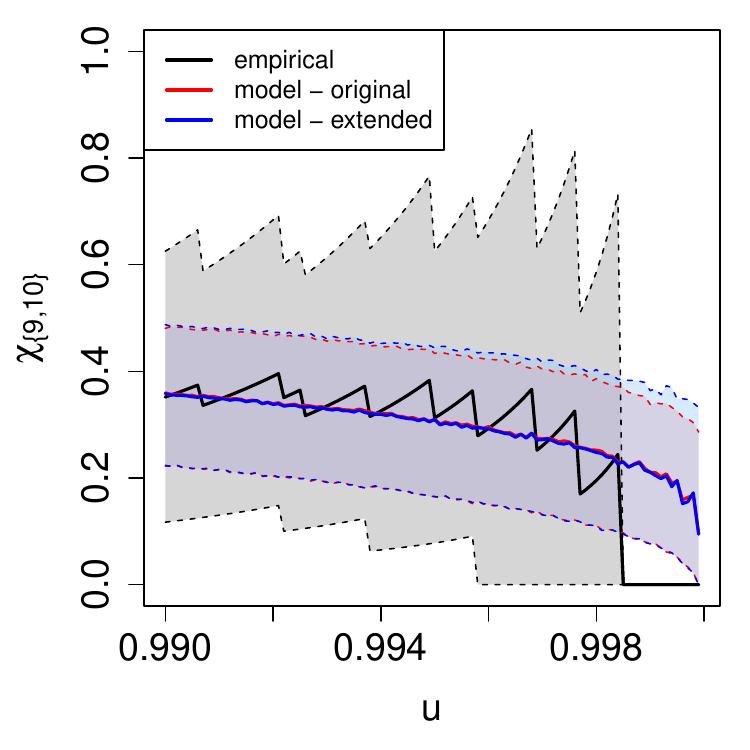}
    \caption{Plot of empirical (solid black) and model-based $\chi_S(u)$ for original (solid red) and extended (solid blue) model with empirical (grey region) and model-based for original (light red region) and for extended (light blue region)  95\% CIs for flow-connected $S = \{4,8\}, \{3,4\}$ (top) and flow-unconnected pairs $ S = \{8,9\} \{9,10\}$ (bottom).}
    \label{fig:chi_plot_bivariate_plots_comparison_rest}
\end{figure}

Figure~\ref{fig:chi_plot_bivariate_plots_comparison_rest} shows the comparison of bivariate empirical and model-based $\chi_S(u)$ for the original and extended model for flow-connected $S = \{4,8\}, \{3,4\}$ and flow-unconnected pairs $S = \{8,9\}, \{9,10\}$. Pair $(4,7)$ is adjacent flow-connected, and $(3,4)$ is non-adjacent flow-connected. Pair $(8,9)$ is a flow-unconnected pair with both gauging stations being on the tree of flow-connected stations, whereas for pairs $(9,10)$ one of the stations is on the tree and the other one is not. The model-based $\chi_S(u)$ estimates are very similar for both models with slightly higher upper CI for the extended model as $u \rightarrow 1$.

\newpage
\section{Event set generation for calculating probabilities}
\label{sec:supp:event_set_generation}
In the event set generation approach, the desired number of years is selected. For each year, we resample the number of events per year from the empirical distribution of number of events per year for the years with no missing data. In our dataset, there were 29 years with no missing data and the average number of events per year $\bar{n}_{\mathrm{event/year}}$ was $60.5$, with range $56 - 64$. This procedure gives the total number of events that need to be simulated. In order to simulate an event, we resample an event from the matched dataset. If this event is a non-exceedance, we add this event to our new dataset. If this event is an exceedance, we simulate a new point above the high radial threshold $r_0(\bm{w})$ from the fitted model with the graphical gauge function using the methods described in Section~3.2.2. We repeat this procedure until the desired number of events is reached. The annual exceedance probability is calculated by counting the number of points that fall into the region of interest $B$ and dividing by the total number of years. In this case $B =(0,\infty)^2\times(v_{\mathrm{exp},3},\infty)\times(v_{\mathrm{exp},4},\infty)\times(v_{\mathrm{exp},5},\infty)\times(0,\infty)\times(v_{\mathrm{exp},7},\infty)\times(0,\infty)^3$. We also calculate the correction coefficient to account for the difference between the marginal distribution of the sampled points and the exponential distribution. This is done by obtaining the annual exceedance probability using the exponential distribution, which is approximately $\exp(-v_{i,\mathrm{exp}}) \bar{n}_{\mathrm{event/year}}$, for $i=3,4,5,7$. We then divide this by a fraction, where the numerator is the number of points in the region where $X_i>v_{i,\mathrm{exp}}$ and the denominator is the total number of years. This approach allows us to generate a new long dataset with extreme values that have not been observed before. The downside of this is that if $B$ is very far in the tail, a very high number of years is necessary to observe a few events in $B$, which is computationally expensive. In this case, to obtain around 100 events in $B$, we needed 1,000,000 years worth of data, which we found to be challenging for a standard computer.

\bibliography{bibliography}
\bibliographystyle{dcu}